\documentclass[10pt,aps,physrev,raggedbottom,longbibliography,nobalancelastpage,reprint,citeautoscript,letterpaper,superscriptaddress]{revtex4-2}

\usepackage{multirow}
\usepackage{bbold}
\usepackage{graphicx}
\usepackage{subfigure}

\usepackage[usenames,dvipsnames]{color}
\usepackage{graphicx,microtype}
\usepackage{amsmath}
\usepackage[bookmarks=false,colorlinks]{hyperref}
\usepackage{lmodern}
\usepackage[all]{hypcap} 
\hypersetup{linkcolor=magenta,citecolor=MidnightBlue,filecolor=Plum,urlcolor=MidnightBlue}

\usepackage[scaled=0.875]{helvet}


\usepackage{ccaption}
\usepackage{ragged2e}
  \captionnamefont{\small \sffamily \bfseries }
  \captiontitlefont{\small }
  \captiondelim{~\textbar~}
  \captionstyle{\justifying}

\newcommand{\supf}{\textcolor{DarkOrchid}{Fig.}\,S}
\newcommand{\supt}{\textcolor{DarkOrchid}{Table}\,S}

\newcommand{\supd}{\textcolor{DarkOrchid}{Supplementary Discussion} }

\begin{document}
\preprint{APS/123-QED}

\title{Accelerated Discovery of Topological Conductors for Nanoscale Interconnects}

\author{Alexander C.\ Tyner}
\affiliation{NORDITA, KTH Royal Institute of Technology and Stockholm University, Stockholm, Sweden}
\affiliation{Department of Physics, University of Connecticut, Storrs, CT, USA}

\author{William Rogers}
\affiliation{Graduate Program in Applied Physics, Northwestern University, Evanston, IL, 60208, USA}

\author{Po-Hsin Shih}
\affiliation{IBM Thomas J. Watson Research Center, Yorktown Heights, NY, USA}

\author{Yi-Hsin Tu}
\affiliation{Industry Academia Innovation School, National Yang Ming Chiao Tung University, Hsinchu, Taiwan}

\author{Gengchiau Liang}
\affiliation{Industry Academia Innovation School, National Yang Ming Chiao Tung University, Hsinchu, Taiwan}

\author{Hsin Lin}
\affiliation{Institute of Physics, Academia Sinica, Taipei, Taiwan}

\author{Ching-Tzu Chen}
\affiliation{IBM Thomas J. Watson Research Center, Yorktown Heights, NY, USA}

\author{James M.\ Rondinelli}
\affiliation{Department of Materials Science and Engineering, Northwestern University, Evanston, IL, USA}

\date{\today}

\begin{abstract} 
The sharp increase in resistivity of copper interconnects at ultra-scaled dimensions threatens the continued miniaturization of integrated circuits.
Topological semimetals (TSMs) with gapless surface states (Fermi arcs) provide conduction channels resistant to localization. 
Here we develop an efficient computational framework to quantify $0$~K surface-state transmission in nanowires derived from Wannier tight-binding models of topological conductors that faithfully reproduce relativistic density functional theory results. 
Sparse matrix techniques enable scalable simulations incorporating disorder and surface roughness, allowing systematic materials screening across sizes, chemical potentials, and transport directions. 
A dataset of 3000 surface transmission values reveals TiS, ZrB$_2$, and nitrides $A$N where $A=(\mathrm{Mo},\, \mathrm{Ta},\, \mathrm{W})$ as candidates with conductance matching or exceeding copper and benchmark TSMs NbAs and NbP. 
This dataset further supports machine learning models for rapid interconnect compound identification. 
Our results highlight the promise of topological conductors in overcoming copper’s scaling limits and provide a roadmap for data-driven discovery of next-generation interconnects.
\end{abstract}

\maketitle

\section{Introduction} 
As the dimensions of logic and memory components in modern integrated circuits continue to shrink, copper interconnects between devices must also be proportionally scaled down to maintain density. 
At nanoscale dimensions, the electrical resistivity of copper increases sharply, leading to higher energy consumption and $RC$ delays.
This effect becomes significant when the thickness of copper interconnects falls below the electron mean free path, $\lambda= 39$\,nm \cite{Gall:2022,chen2020topological,zou2021review,gall2021materials,moon2023materials}, where surface and grain boundary scattering dominate electron transport. 
Degradation in performance results in slower, less energy-efficient computation, underscoring the need for alternative materials to replace copper.
Topological semimetals (TSMs)~\cite{burkov2016topological,RevModPhys.93.025002,hu2019transport,gao2019topological,yan2017topological} and topological metals (TMs)\cite{Zhu2016,kumar2019extremely,han2023topological} have emerged as a promising materials class to replace copper \cite{chen2020topological,kim2024addressing,soulie2024selecting,Kiani2025}, owing to their topologically protected surface states that enable robust, low-resistance conducting channels (see \supd IV).
The TSM/TM classification is broad, encompassing Weyl \cite{PhysRevX.5.031013,yan2017topological}, multifold \cite{robredo2024multifold,wang2025exhaustive}, triple point \cite{Zhu2016}, and nodal line \cite{lou2018experimental,Fang2015,fang2016topological,yu2017topological} systems among others. Common among each these classes are nodal points or lines in the Brilloun zone where bands cross near the Fermi energy. In the case of a nodal point semimetal, the band crossing locations act as monopoles and antimonopoles of Berry curvature.
Importantly, the non-trivial Berry flux gives rise to chiral edge states, commonly known as Fermi arcs states, between the projection of the nodal points on the surface. 
Prior computational works have investigated the utility of the Fermi arcs states as conduction channels in the ultra-thin limit for Weyl semimetal NbAs \cite{kumar2024surface,cheon2025surface} and multifold fermion semimetal CoSi~\cite{lien2023unconventional}. 
These works showed that conduction is increasingly dominated by Fermi arcs as the sample thickness is decreased, leading to a decreasing resistance-area ($RA$) product with decreasing thickness. 
\begin{figure*}
    \centering
    \includegraphics[width=18cm]{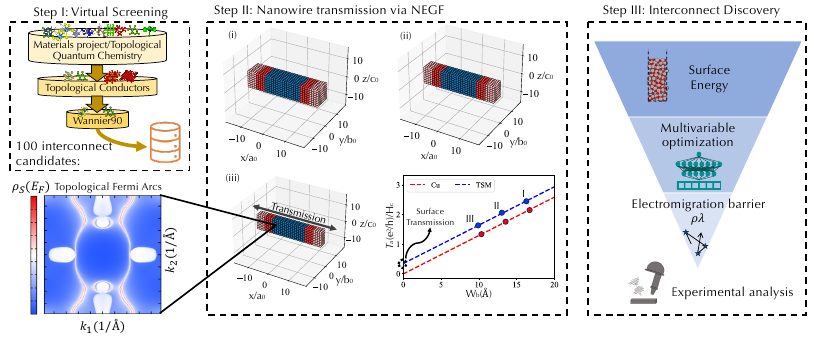}
    \caption{\sffamily  \textbf{Accelerated topological semimetal interconnect discovery workflow.}
    Step I (top left)  details filtering material candidates from existing databases of thermodynamic and topological properties. 
    The current search is limited to binary and selected ternary and elemental conductors, considering those found on the DFT convex hull within the Materials Project \cite{jain2013commentary} or for which examples of experimental synthesis are available. 
    After filtering candidates we perform density-functional theory calculations and construct a Wannier tight-binding model for each compound. 
    Step II (center):  Tight-binding models are used to construct nanowires and transmission is computed via the non-equilibrium Greens function (NEGF) formalism as a function of thickness along an axis perpendicular to the transmission direction as indicated in (i). 
    In the geometry shown, transmission is computed along [100], $T_{a}$, arising from the Fermi arcs on the (010) surface (example spectra density shown in lower left) as the transverse width of the nanowire along the [010] direction, $W_b$, decreases moving from (i) -- (iii). (The vertical height of the nanowire along [001] is fixed.)
    The zero-thickness intercept is obtained by linearly fitting the transmission as a function of the transverse width to obtain the surface transmission $T_s$, which is equivalent to $T_a(W_{b} \rightarrow 0)$, for the given geometry (bottom right, center panel). 
    Step III (right) Higher-fidelity simulations and targeted property evaluations are performed to optimize candidate materials against additional interconnect-application constraints. 
    Based on these refined criteria, the most promising candidates are identified and recommended for experimental validation.
    }
    \label{fig:Fig1}
\end{figure*}

Although isolated case studies have demonstrated the promise of topological conductors as next-generation interconnects, they have also shown that not all are created equal in this context. A broader and systematic exploration is needed to fully assess and optimize their potential.
Recent high-throughout screenings of inorganic materials for non-trivial topology, such as those compiled in the topological quantum chemistry database~\cite{bradlyn2019disconnected,vergniory2019complete}, 
have significantly expanded the catalog of known TSMs,  providing a reasonable starting point targeted screening efforts aimed at identifying TSMs with favorable surface transport properties for interconnect applications.
Nevertheless, such a screening is yet to be undertaken due to the computational demands associated with determination of nanowire surface conductance via first-principles. 
This expense arises due to the need to consider nanowire geometries of sufficient size to project the conductance performance at the relevant length scale for the interconnect technologies. 
Such nanowires can contain hundreds of atoms, creating a bottleneck not only for convergence within density functional theory (DFT), but particularly for computation of transmission via the non-equilibrium Green's function (NEGF) formalism~\cite{groth2014kwant}. 
To make matters worse, these nanowire computations must be performed multiple times, considering a variety of surfaces, transmission directions, and disorder.

We reduce this computational expense by constructing nanowires from Wannier tight-binding models of candidate TSMs  to obtain a high-fidelity representation of the electronic structure of each compound using automated and efficient schemes.
Once constructed, we use the hopping parameters provided by the Wannier tight-binding to construct nanowire geometries, which are represented as sparse matrices for which computationally advantageous methods can be used to tackle large system sizes efficiently. 
Furthermore, the computational efficiency of this protocol allows us to investigate the effects of disorder and surface roughness on the bulk and surface transmission. These computations are performed in the $0K$ limit, however we expect the results to be relevant for down-selection of optimal room-temperature interconnects due to the low electron-phonon coupling computed for topological surface states in prior works\cite{heid2017electron,Pan2012,benedek2020origin}. 
Our analysis reveals TiS\cite{Li2018,xu2020centrosymmetric}, ZrB$_{2}$\cite{Feng2018,lou2018experimental} and nitrides $A$N\cite{Zhu2016} where $A=(\mathrm{Mo},\, \mathrm{Ta},\, \mathrm{W})$ as previously known but unidentified interconnect candidates, displaying surface transmission competitive with TSMs that have been studied in depth previously.
Last, we discuss other materials properties requiring optimization to ensure manufacturability and reliability during device operation.

\section{Computational Workflow}
\subsection{Candidate selection}
The first stage in our workflow involves a virtual screening of publicly available inorganic structure databases (\autoref{fig:Fig1}).  
We limit our search primarily to binary and elemental topological conductors including a small number of experimentally realized ternaries that are computationally tractable. 
This constraint in structure is motivated by the high memory demands associated with the final step of our workflow—nanowire transmission calculations—which remain a computational bottleneck.
The virtual screening process aims to curate a focused dataset of promising compounds for further investigation, thereby reducing the dimensionality of the candidate space \cite{Zhang2025}.

\begin{figure*}
    \centering
    \includegraphics[width=0.98\textwidth]{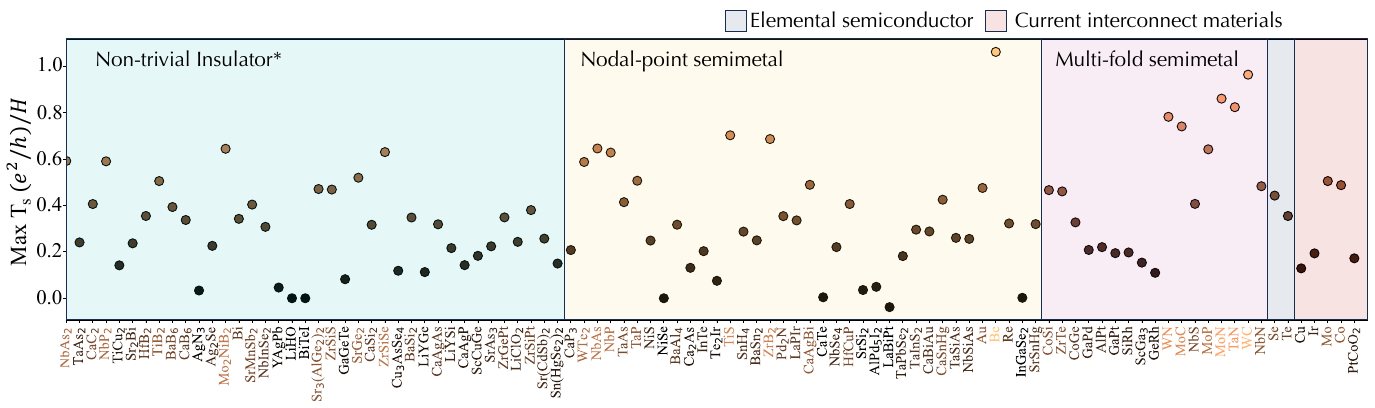}
    \caption{\sffamily  \textbf{Maximum calculated nanowire surface transmission values.} 
    The maximum $T_s$ values obtained at the Fermi energy for the compounds screened via the process in \autoref{fig:Fig1}, Step II. Materials sorted by their topological quantum chemistry classification. The general term, non-trivial insulator$^{*}$ is used to group materials with occupied topological bands. An asterisk is included to emphasize that these systems need not, and in most cases do not, have a direct band gap at the Fermi level. }
    \label{fig:SurfaceT}
\end{figure*}

In practice, candidate identification is carried out through a combination of manual searches of online databases, including the Topological Materials Database \cite{Vergniory2022}, Topological Materials Arsenal \cite{arsenal},  and Materiae \cite{Zhang2019}, as well literature reviews. 
We prioritize compounds that have either been experimentally synthesized or exhibit evidence of non-trivial bulk topology in prior computational studies.
The full list of selected compounds is provided in \supt1, identified by Materials Project ID \cite{jain2013commentary}.
While this list may not encompass all viable topological conductors, it serves as a practical starting point for deploying our discovery workflow.
For compounds that have not been experimentally realized, we require that they lie on the  convex hull, which is readily accessible via the Materials Project \cite{jain2013commentary} (v2023.11.1). 
Additionally, we limit our current screening to non-magnetic systems, deferring the inclusion of ferromagnetic and antiferromagnetic candidates for future work. 
All necessary files to reproduce the data generated in this study are available at Ref.\ \cite{Data}. 
A breakdown of the space group distribution and elemental composition within the dataset is provided in  \supf1. 
Last, we also include current elemental interconnect candidates PtCoO$_{2}$, Mo, Ir, and Cu to provide a benchmark for the workflow. 

\begin{figure*}[t]
    \centering
    \includegraphics[width=.98\textwidth]{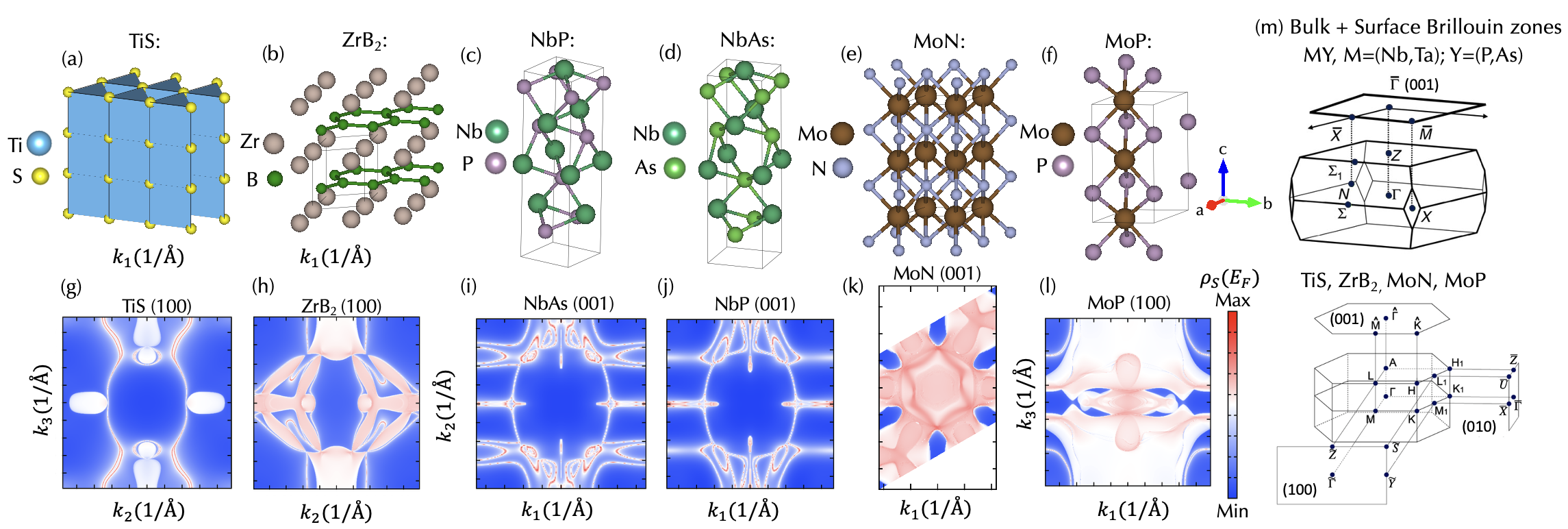}
    \caption{\sffamily  \textbf{Crystal structure and spectral density dispersions for promising topological semimetals.}
    Constant energy surface spectral density at the Fermi energy centered about $\bar{\Gamma}$ for TiS, ZrB$_{2}$ and MoN, identified in this work, compared with topological interconnect candidates NbPn, where Pn=(P,\,As), and MoP reported to support enhanced surface transmission. 
    Spectral density figures labeled by surface for which transmission is maximal. Colorbar represents the magnitude of the surface spectral density at the $E_F$ (red, maximal; blue, minimal).
    }
    \label{fig:Fig3}
\end{figure*}

\begin{figure}
    \centering
    \includegraphics[width=0.95\linewidth]{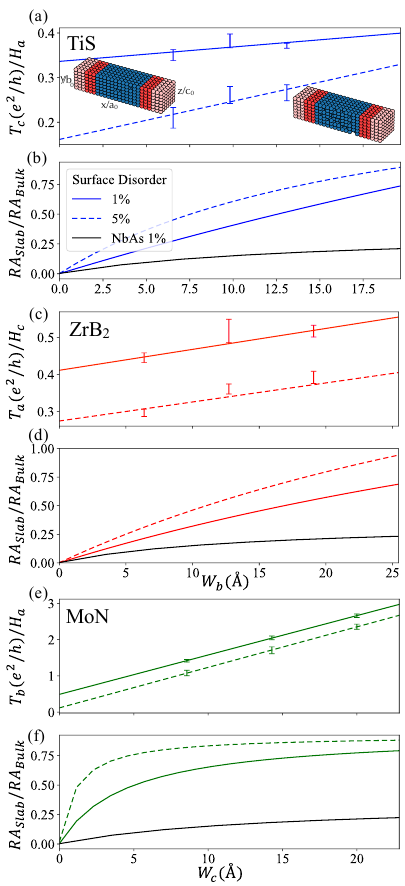}
    \caption{\sffamily \textbf{Surface disorder effects on transmission and resistance-area product scaling.} Nanowire surface transmission as a function of surface vacancy density in (a) TiS, (c) ZrB$_{2}$ and (e) MoN at the Fermi energy. In both systems we utilize the geometry which yielded maximal surface transmission in the initial screening. 
    Ratio of slab to bulk resistance area for (b) TiS, (d) ZrB$_{2}$ and (f) MoN as a function of the slab surface roughness compared to NbAs with $1\%$ surface vacancy density utilizing the geometry which yielded maximal surface transmission. 
    Inset of (a) displays example of nanowire with unit cells removed from the surface to simulate disorder. Example configurations show $1\%$ and $10\%$ surface vacancy density for clarity. }
    \label{fig:Fig4}
\end{figure}

\subsection{Electronic structure simulations}

The next stage in our computational workflow involves constructing a Wannier tight-binding model for each compound using the conventional unit cell as the structural basis \cite{Pizzi2020,vitale2020automated}. 
We deliberately choose the conventional unit cell over the primitive unit cell to ensure compatibility with downstream nanowire transmission simulations, which are performed along the principal crystallographic axes of the conventional unit cell. 
This choice facilitates a direct mapping between bulk electronic structure and transport directionality, enabling consistent and physically meaningful comparisons across different materials.
We perform density functional theory (DFT) computations with spin-orbit coupling (SOC) using the Quantum Espresso software package \cite{QE-2020,Perdew1996}.

\subsection{Nanowire surface transmission}
We next construct a nanowire of dimension $(L\times W \times H)$ conventional unit cells using the Kwant software package \cite{groth2014kwant}. 

With this geometry, the transmission $T$ is  always simulated along $L$ to identify contributions due to the surface perpendicular to $W_i$. 
Since we consider surface orientations for each principal axes, $i = a, b, \textrm{and } c$, 
we append a subscript to the dimension to denote the corresponding crystal orientation. The magnitude of individual lattice parameters are denoted $a_{0}, b_{0}, \textrm{and } c_{0}$ for clarity. 
For example, $(L_{a}\times W_{b} \times H_{c})$ describes the nanowire shown in the center panel of \autoref{fig:Fig1}, for which transmission along the $[100]$ direction is due to states on the $(010)$ surface in a laboratory frame. 
To isolate the contribution of surface states to transmission, we employ a procedure in which leads are attached to the compound under investigation, followed by a series of simulations where the size of the nanowire is progressively reduced in a direction transverse to the transmission axis as described below.
In the example presented in \autoref{fig:Fig1} Step II(i), we begin with an isolated nanowire of dimensions, $(L_{a}\times W_{b} \times H_{c})$= $(16 \times 8 \times 5)$. 
Leads, formed from the same real-space tight-binding model, and extending semi-infinitely along the $[100]$ direction are then attached. 
The setup appears as shown in \autoref{fig:Fig1} under step II(i), where the scattering region is shown in blue, the leads are shown in light-red, and red indicates the region in which the scattering region and leads overlap. 
We then compute the transmission, normalized by the height of the scattering region ($H_{c}$).
For each configuration, we also vary the electron chemical potential between $[-0.2\,\mathrm{eV},0.2\,\mathrm{eV}]$ in increments of $0.1\,\mathrm{eV}$ when computing $T$ transmission as the position of the Fermi energy $E_F$ may be unknown due to unintentional defects. 

In the next step, we decrease the number of unit cells along $W_{b}$ and recompute the transmission for each value of the chemical potential, recognizing that as $W_{b}\rightarrow0$ the total transmission is linear with respect $W_{b}$ as
\begin{equation}\label{eq:G}
T(W_i)=T_{S}+W_i\,g_\mathrm{bulk},
\end{equation}
where $T$ is the total transmission, $T_{S}$ is the transmission due to the surface perpendicular to the $i$ crystal axis and $W_i\,g_\mathrm{bulk}$ gives the bulk transmission as shown in the bottom right of Step II in \autoref{fig:Fig1}.
Provided that the thickness $W_b$ is greater than the critical dimension below which quantum confinement completely gaps out the bulk states, then each time $W_{b}$ is reduced, the number of bulk conducting channels should decrease in a linear manner, while conducting channels on the surface perpendicular to $W_{b}$ remain unaffected. The zero-thickness limit of the linear fit to $T(W_i)$ gives the contribution from the surface states to the transmission. 
This process is repeated for all permutations of transmission direction and surface orientation.
The results are then plotted and inspected to identify optimal compounds and transmission directions. 

\subsection{High-fidelity simulations}

After the initial screening, materials displaying promising surface transmission are further refined by creating a Wannier tight-binding model using hydrogenic orbitals, excluding $f$-electrons if present. We ensure that the refined Wannier models precisely match the DFT band structure along the high-symmetry path in the Brillouin zone within $1\,eV$ of the Fermi energy and then repeat the computations of surface transmission to ensure accuracy. Detailed quantitative analysis of this approach is provided in the \supd III.

\subsection{Machine learning model}
We use data generated from the surface transmission workflow to train a a random forest regression model that predicts transmission across the broader chemical space of binary of compounds.
Each compound is labeled by its maximal surface transmission value, $T_S$, for a given transmission direction/surface combination.
To represent each compound, we employ Magpie descriptors \cite{ward2016general}, accessed via the MatMiner package \cite{ward2018matminer}, which encode a wide range of compositional and elemental properties.
Additionally, we use the topogivity index \cite{ma2023topogivity} as a  feature, which assigns a scalar value to each element, averaged across constituent atoms in a compound, with higher averages indicating a greater likelihood of non-trivial ground state topology. 
The feature captures elemental contributions such as spin-orbit coupling strength and reflects the tendency of certain elements to favor topologically nontrivial phases.
To identify the most important features and reduce dimensionality, we apply SHAP (SHapley Additive exPlanations) analysis using the Python SHAP library \cite{janzing2020}.
This interpretability framework allows us to extract physical insights from the trained model, which is implemented using the scikit-learn library \cite{pedregosa2011}.

\section{Results and Discussion}

\subsection{Maximum surface transmission compounds} 
The maximal transmission direction/surface combination at the Fermi energy for each screened binary and elemental compound is shown in \autoref{fig:SurfaceT}. Compounds are grouped according to their topological classification via analysis of elementary band representations\cite{bradlyn2019disconnected,zhang2019catalogue}. 
The general term non-trivial insulator is used to specify a broad group containing all systems supporting occupied bands with non-trivial topology, but following the formalism of Ref.\ \onlinecite{bradlyn2019disconnected}, an asterisk is included to emphasize that these systems need not, and in most cases do not, have a direct band gap at the Fermi level. Additionally, we separate nodal-point semimetals, which support crossings between two bands, from multi-fold semimetals, which support crossings between $N>2$ bands at high-symmetry locations in the Brillouin zone. 
From this data we immediately find that Cu, the current industry standard for interconnect applications, has minimal surface transmission. 
We also identify two previously known but unidentified promising interconnect candidates, TiS, ZrB$_{2}$ (\autoref{fig:SurfaceT}), in addition to known monopnictide $M$Pn TSMs, where $M=(\mathrm{Nb,Ta})$ and $\mathrm{Pn}=(\mathrm{P,As})$ \cite{PhysRevB.92.115428,PhysRevB.92.235104}. Additionally, we find that mononitrides MoN, WN, and TaN are promising candidates from the family of topological triple point metals\cite{Zhu2016} for which MoP has been identified previously as an optimal interconnect\cite{han2023topological,kumar2019extremely}.
Monosulfide TiS exhibits $P\bar{6}m2$ symmetry (space group 187) with TiS$_6$ triangular prismatic coordination. 
The trigonal prisms are edge and face sharing within the $a-b$ plane and along the $c$ axes of the hexagonal structure (\autoref{fig:Fig3}). 
Nominally, the electronic structure is characterized by a singlet state with electrons from Ti$^{2+}$ ($d^2$ electronic configuration) imposed by the $D_{3h}$ crystal field and is obtained by filling the lowest energy $a_1^\prime$ symmetry orbital with $d(z^2)$ character (\supf3). 
The fully occupied S $3p$ orbitals are found below the Ti $3d$ states.
Moderate mixing occurs with the $e^\prime$ orbitals with $d(x^2-y^2)$ and $d(xy)$ character owing to $\pi$-bonding interactions \cite{Huisman:1971} and flattening of the TiS$_6$ trigonal pyramids \cite{Georgescu:2022}. 
Thus, the low energy electronic structure comprises a mixture of the  $a_1^\prime$ and $e^\prime$ orbitals, making it a self-doped nodal-point semimetal. 
We further calculate the resistivity scaling coefficient, $\rho_0\lambda$, where $\rho_0$ is the bulk resistivity and accessible from first-principles \cite{Gall:2016}. In a trivial metal, the system with the lowest value of $\rho_0\lambda$ is expected to support the lowest nanowire resistivity. For details see \supd VI.
This descriptor has been widely used for identifying promising trivial conductors for interconnect applications \cite{Gall:2022}, and is \,$37 \times 10^{-16}\,\Omega\,\mathrm{m}^2$ and $41 \times 10^{-16}\,\Omega\,\mathrm{m}^2$ for the $[100]$ and $[001]$ transport directions of TiS. This value is computed accounting for spin-orbit coupling and is comparable to the best-of-class anisotropic conductors such as delafossite PtCoO$_2$ ($\rho_0\lambda=13.3 \times 10^{-16}\,\Omega\,\mathrm{m}^2$ \cite{Kumar:2022}).
To the best of our knowledge, no synthesis or transport measurements have been reported for TiS; moreover, the compound is thermodynamically stable, as it appears on the Ti–S convex hull reported by the Materials Project \cite{jain2013commentary}. 
Furthermore, TiS is unlikely to oxidize upon contact with SiO$_2$, which we use as a simplified proxy for the low-$k$ dielectric SICOH. 
The formation of a tie-line between TiS and SiO$_2$ suggests that the compound can exist in stable equilibrium with the dielectric and is unlikely to undergo chemical reactions; however, oxidation in the presence of O$_{2(g)}$ could favor oxysulfide formation.
ZrB$_{2}$ crystallizes in the hexagonal AlB$_2$ structure type with $P6/mmm$ symmetry (space group 191), featuring hexagonal layers of Zr atoms interleaved between borophene sheets.
Although the nominal electronic configuration of Zr$^{4+}$ ($d^0$) and $sp^2$-hybridized B orbitals for its constituents suggests an insulating state, delocalized Zr–B $dp\pi$ bonding interactions result in a semimetallic character. 
This leads to slight doping of the Zr 4$d$ states and several band crossings near $E_F$, as evident in the band dispersions (\supf4).
Previous studies have classified ZrB$_{2}$ as a nodal-line semimetal \cite{PhysRevMaterials.2.014202,lou2018experimental}, and have demonstrated low resistivity in the thin-film limit \cite{shappirio1984resistivity,zhang2020application}, consistent with our nanowire transmission analysis.
Here we calculate  $\rho_0\lambda = 12 \times 10^{-16}\,\Omega\,\mathrm{m}^2$ and $\rho_0\lambda = 11 \times 10^{-16}\,\Omega\,\mathrm{m}^2$ for the $[100]$ and $[001]$ transport directions of ZrB$_{2}$. These values are reduced with respect to TiS, a reflection of the greater Fermi surface area due to additional bulk bands near the Fermi energy as illustrated in the supplementary information \cite{suppl}.
Similar to TiS, ZrB$_{2}$ lies on the convex hull and is in phase equilibrium with SiO$_2$, indicating its chemical stability and compatibility with dielectric environments. 

The $A$N mononitrides where $A=(\mathrm{Mo},\, \mathrm{Ta},\, \mathrm{W})$ exhibit $P\bar{6}m2$ symmetry with with $A$N$_6$ triangular prismatic coordination. The trigonal prisms share edges and faces within the $a-b$ plane and stack along the $c$ axis of the hexagonal structure. Nominally, the electronic structure is characterized by $A^{3+}$ transition metals within a $D{3h}$ crystal field. Strong hybridization occurs between the N $2p$ states and $4d$ ($5d$) states of Mo (Ta,W), leading to a complex low-energy electronic band structure dominated by $d$ (\supf2).
Prior work has studied this family of materials and identified it as hosting a topological triple point near the Fermi energy \cite{Zhu2016} with ultra-high conductivity shown for MoP\cite{kumar2019extremely},  which falls in this family and has been proposed as an emerging interconnect materials\cite{han2023topological}. 
We calculate $\rho_0\lambda = 2.8 \times 10^{-16}\,\Omega\,\mathrm{m}^2$ and $\rho_0\lambda = 2.1 \times 10^{-16}\,\Omega\,\mathrm{m}^2$ for the $[100]$ and $[001]$ transport directions of MoN. These values reflect the large conductivity offered by the large Fermi surface area as a topological metal (\supd II).

\autoref{fig:Fig3} compares the surface spectral density of TiS, ZrB$_{2}$ and MoN with that of MoP and the $M$Pn compounds at the Fermi energy.
The analyzed surfaces correspond to the (100) orientation for TiS, ZrB$_{2}$ and MoP; the (001) orientation for MoN and the $M$Pn compounds. 
A common feature for all the TSMs is the presence of a network of Fermi arcs beginning and terminating at the projection of the bulk band crossing locations; in contrast to the Dirac cone surface states of topological insulators. This is notable as the open-line character of Fermi-arcs is expected to enhance protection against disorder and perturbations, e.g., surface roughness, that may arise during the synthesis of physical nanowires \cite{PhysRevB.97.085142,PhysRevB.97.235108,PhysRevB.96.201401}. 

To test the robustness of the surface transmission against disorder, we simulate surface roughness in our nanowire transmission calculations for TiS, ZrB$_2$ and MoN by randomly removing $1\%$ and $ 5\%$ of the surface unit cells prior to evaluating the transmission at the Fermi energy (\autoref{fig:Fig4}). 
Here we use nanowire geometries that produced maximal $T_S$ in our initial screening.
We repeat this simulation across at least 100 random configurations per vacancy density, and results are averaged to assess the statistical reliability.
Rather than comparing absolute transmission values, we focus on the  relative decrease in surface transmission. 
If the surface conduction channels were not topologically protected, surface disorder would induce localization, driving  $T_{s}\rightarrow 0$; however, as  shown in \autoref{fig:Fig4}a,c,e, this behavior is not observed, indicating that $T_S>0$ arises from  topologically protected states. 
Upon increasing the surface vacancy density from 1\% to 5\%, we find a reduction in $T_S$ of {48\%, 65\% and 75\%} for TiS, ZrB$_2$, and MoN respectively.
In contrast, similar calculations for NbAs show a reduction of $T_S$ by 51\% (see \supf5), suggesting that the extended surface states in TiS may offer greater protection to surface roughness in the ultra thin limit and 0 K limit, where no surface and bulk phonon scattering is absent.

Next, we examine the resistance-area scaling as a function of nanowire thickness \cite{kumar2024surface}, to better understand at which dimensions the surface states dominate transmission over the bulk. 
The slab resistance area, $RA_{Slab}=A/(T_{S}+W\,g_\mathrm{bulk})$, where the denominator follows \autoref{eq:G}. 
Upon substituting $A=W H$, we extract $RA_\mathrm{bulk}$ by taking the limit $W\rightarrow \infty$. 
The normalized $RA$ product then can be approximated as 
${(RA)}_{{{{\rm{slab}}}}}/{(RA)}_{{{{\rm{bulk}}}}}\approx ({1+\alpha /W})^{-1}$, where $\alpha$ describes how much of the conductance is due to surface states versus bulk states. 
For a trivial conductor like copper, the ratio of these products is approximately unity ($\alpha\rightarrow0$) \cite{kumar2024surface}, and hence independent of slab thickness, because bulk states dominate the conduction.  
In our analysis, we additionally include surface roughness into the assessment using the data from \autoref{fig:Fig4}.

\begin{figure*}
    \centering
    \includegraphics[width=0.98\textwidth]{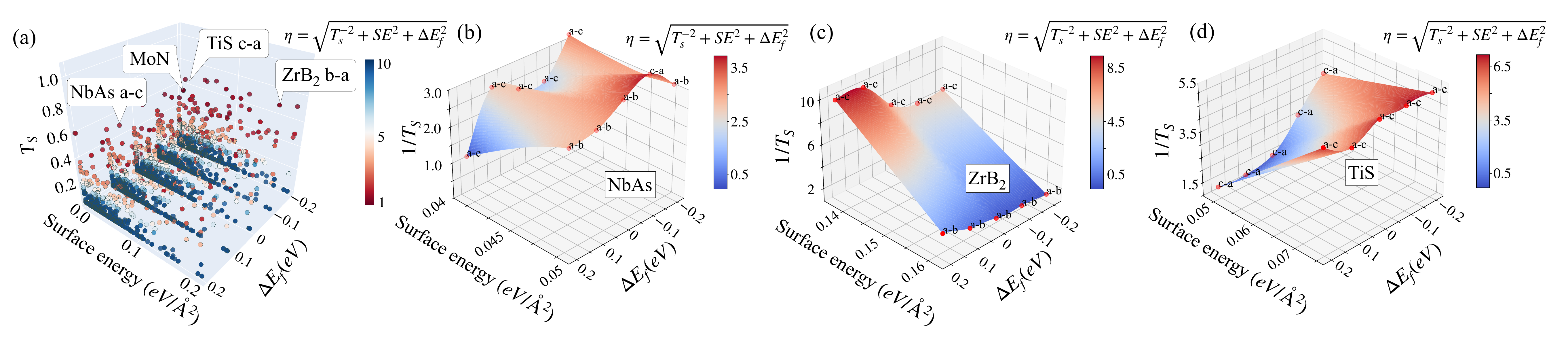}
    \caption{\sffamily \textbf{Pareto frontier candidates in the interconnect design space.}
    (a) Scatter plot of all {3000} screened compound-surface-direction combinations, with each point labeled by its maximal surface transmission ($T_S$), surface energy, and Fermi level shift ($\Delta E_F$). 
    Points are colored by $\mathbf{\eta} = (\Delta E_F^2 + \gamma^2 + R_S^{2})^{1/2}$, with $R_S=T_S^{-1}$, where lower values indicate optimal interconnect candidates. 
    Selected data points are labeled by transmission direction and surface orientation. 
    An interactive version of this figure is available online at Ref. \cite{Data}.
    Pareto subsurface for (b) NbAs, (c) TiS, and (d) ZrB$_2$ showing interpolated trends in the $\Delta E_{F}-\gamma$ plane as a function of doping.}
    \label{fig:Pareto}
\end{figure*}

\autoref{fig:Fig4}b,d,f shows the ratio of the slab-to-bulk $RA$ products for the nanowire with surface roughness, corresponding to 1\% and 5\% surface vacancy densities for TiS, ZrB$_{2}$ and MoN respectively. 
As expected, surface roughness has the effect of increasing the $RA$ product; however for TiS the rate at which it approaches the bulk limit appears reduced relative to ZrB$_{2}$ and MoN. This is in part due to the increased Fermi surface area of ZrB$_{2}$ and MoN (see \supf2). The greater number of bulk conducting channels reduces the ratio of surface to bulk conductance, $\alpha$, despite the two materials exhibiting comparable surface transmission to TiS.
Our numerical fits to the data give $\alpha=45$, $10.2$ and $8.03$ in the pristine (zero roughness limit) for TiS, Zr$B_2$ and MoN which are comparable to values simulated for pristine TaAs ($\alpha=28$) and NbAs ($\alpha=60$) \cite{kumar2024surface}. This indicates that TiS could support enhanced stability to surface roughness/defects. It also suggests that TiS may suffer less from scattering between topological surface states and bulk states. As a near perfect semimetal, surface states dominate conductance in the thin film limit for NbAs, elevating $\alpha$. This suggests methods to engineer the band structure and shift trivial bulk states away from the Fermi energy\cite{lei2021band} may be beneficial for future optimization of compounds displaying high surface transmission.

\begin{table}
\caption{\sffamily \textbf{Multiobjective design space.} Representative subset of critical interconnect property objectives considered for optimization. 
}
\begin{ruledtabular}
\begin{tabular}{lll}
Objective   & Variable & Target Range \\ 
\hline
Transmission direction & $[h\, k\, l]$        & Low   index \\
Surface orientation    & $(h\, k\, l)$        & Low   index \\
Surface transmission   & $T_S$          & maximum     \\ 
Doping level away from $E_F$  & $\Delta E_F$   & $\sim$0           \\
Surface energy         & $\gamma_{hkl}$ & Small       \\
\end{tabular}
\end{ruledtabular}
\label{tab:opt}
\end{table}

\subsection{Multi-variable optimization}
The outcome of the employed workflow is a dataset comprising zero-temperature nanowire transmission values for all surfaces orthogonal to principal lattice vectors within the conventional unit cell, as a function of Fermi energy. 
While this data is vital for identifying promising interconnect candidates, it must be evaluated along with other practical considerations.
These include crystallographic orientation and microstructure, self-limiting oxidation behavior, finite-temperature effects (e.g., bulk and surface phonon scattering and thermal expansion), and long-term reliability concerns like electromigration. 

Together, these considerations necessitate a multi-variable optimization framework, for which we formulate a simplified approach to 
assess and rank interconnect materials under realistic operating conditions by optimizing over a subset requirements \autoref{tab:opt}. 
Specifically, we consider the transmission direction/surface orientation, 
surface transmission value ($T_{S}$, 
electron chemical potential, i.e.,  doping level relative to the Fermi level $\Delta E_{F}$), and 
surface energy ($\gamma$), which serves as a proxy for the ease of synthesizability on a given $(h\, k\, l)$ surface.
Although a Utopian composition seeks to simultaneously achieve ideal target values across all objectives, such a solution is typically unattainable in practice.
Consequently, the Pareto front within this reduced parameter space enables a tractable yet physically meaningful optimization strategy, which can be combined with other parameters, for identifying viable interconnect candidates and delineating the inherent trade-offs among competing objectives.
Surface energies represent the only missing objective values  in the surface transmission workflow previously described. 
Thus, we computed the surface energies for all $(h\,k\,l)$ surfaces as $\gamma_{hkl}=({E^{hkl}_\mathrm{slab}-E^{hkl}_\mathrm{bulk}\times n_\mathrm{slab}})/({2\, A_\mathrm{slab}})$, where $E^{hkl}_\mathrm{slab}$ is the total energy of a slab exposing the $(hkl)$ surface and $E^{hkl}_\mathrm{bulk}$ is the per atom total energy of the bulk unit cell oriented along the same direction, $n_\mathrm{slab}$ is the number of atoms in the slab, and $A$ is the surface area of the slab \cite{tran2016surface}. 
For each material, we construct a slab consisting of five unit cells and insert $20$\,\AA\ of vacuum to expose the surfaces. 
All surface energy calculations are carried out using the CHGNet machine learned interatomic potential (v0.3.0, \cite{deng_2023_chgnet}) for consistency. 
\autoref{fig:Pareto}a presents 3000 evaluated objective values within a three-dimensional representation, where each coordinate $x,y,z$ is given by the surface energy, doping level, and surface transmission for a compound  ($\gamma_{hkl},\Delta E_F,T_S$). 
Within this design space, we aim to identify an optimal interconnect candidate that satisfies the following criteria: 
($i$) Requires no doping, i.e., the maximum $T_S$ occurs at the Fermi energy such that $\Delta E_F = 0$, 
($ii$) Exhibits minimal surface energy, with Cu serving as a reference exhibiting experimental surface energies of 0.08 eV/\AA$^2$\cite{mills2006review}, {which is close to our simulated value of $\sim$0.06\,eV/\AA$^2$)}, and
($iii$) maximal surface transmission, corresponding to minimal surface resistance $R_S=T_S^{-1}$.
We combine these objectives into a figure-of-merit, $\mathbf{\eta}$, which quantifies the Euclidean distance of the ideal point in the three-dimensional design space: $\mathbf{\eta}=({\Delta E_{F}^2+\gamma^2+R_{S}^{2}})^{1/2}$.
Minimizing $\mathbf{\eta}$ thus recasts the problem of identifying an optimal interconnect candidate into a tractable multi-objective optimization task.

Among the optimal compounds identified in \autoref{fig:Fig1} by examining $T_S$ alone, we find that the known Weyl semimetal NbAs and our identified compounds TiS, ZrB$_2$, and MoN appear on the Pareto front (\autoref{fig:Pareto}a). 
Other promising compounds including CoSi, NbP and TaP fall among these compounds, represented as dark red points. They can be seen by accessing the interactive version of this figure at \url{https://mtd.mccormick.northwestern.edu/interconnect-database}.
In addition to examining the Pareto front, we further examine the subspace manifold for a given compound as shown for NbAs, TiS, and ZrB$_2$ (\autoref{fig:Pareto}b--d). 
Here we directly plot $R_S$, $\gamma$, and $\Delta E_F$ and to evaluate the anisotropy with doping away from the Fermi level, we construct an interpolating function connecting data points in the $\Delta E_{F}- SE$ plane and color this surface according to the value of $\mathbf{\eta}.$ We note that the Pareto designs are not being interpolated from the constrained design space given that $\gamma_{hkl}$ is a nonlinear function of $h, k, l$.  
\par
For NbAs (\autoref{fig:Pareto}b) we find that the $(001)$ surface supports a lower surface energy, making it more likely to be experimentally accessible. In addition, nanowire transmission due to surface states on the $(001)$ surface is also maximal. Optimization of NbAs then reduces to tracking the Pareto Front produced by the interpolated surface as a function of doping. 
In contrast, the design manifold for ZrB$_2$ (\autoref{fig:Pareto}c) shows that the highest $T_S$ is along [100] or [001] for the (010) and (100) surfaces, respectively, however, the (001) has lowest surface energy, which makes it most accessible. On both surfaces we find that doping below the Fermi energy can enhance surface transmission.
\autoref{fig:Pareto}d shows that transmission due to surface states on the $(100)$ surface is maximal for TiS. Advantageously, the (100) surface also supports a minimal surface energy. We further note that surface transmission is maximal at $E_F$ and remains high when doping below it but not above. 

The tradeoff among accessible surface orientations and optimal transmission direction shows the complexity in finding an interconnect compound that satisfies all design objectives and informing subsequent manufacturing, i.e., whether growth in trenches or blanket films followed by pattering would be pursued. 

\begin{figure}
    \centering
    \includegraphics[width=8cm]{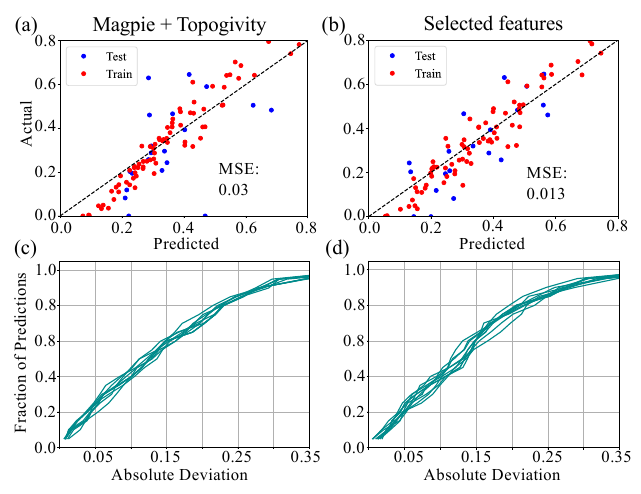}
    \caption{\sffamily \textbf{Machine learning surface transmission.} (a) Results of random forest model trained to predict surface transmission using all Magpie~\cite{ward2016general} features as well as the topogivity index of a given compound~\cite{ma2023topogivity}. (b) Refined random forest model trained to predict maximal surface transmission using only most important features, topogivity, mode Mendeleev number of the constituent elements and space group number. (c)-(d) Regression error characteristic (REC) curve for model trained (c) using all Magpie features and topogivity and (d) only most important features. Each colored line represents the median REC curve from one of 10 random seeds. For each seed, a 5-fold cross-validation is carried out.}
    \label{fig:Fig5}
\end{figure}

\subsection{Chemical features for compound discovery}

This work has significantly expanded the number of materials for which surface transmission has been screened and provided a workflow that can be automated without sacrificing accuracy. A benefit of such progress is the ability to apply data mining strategies and create a machine-learning network to rapidly explore larger areas of material space, including virtual materials databases such as Gnome \cite{merchant2023scaling} to identify optimal interconnect candidates.
Using a tree-based model with a random $80\%/20\%$ train/test split, we obtained reasonable performance of our model with a mean-squared-error (MSE) of 0.03 
using all possible Magpie features and the topogivity index of the compound (\autoref{fig:Fig5}a). 
By examining feature importance,  we find the most influential descriptors are the Mendeleev numbers of the constituent atoms, the topogivity index, space group, and the number of $d$ electrons in the valence shell. 
These features are intuitive as that provide insight into the topological nature of the compound. %
For example, the space group symmetry is relevant to the presence of non-trivial electronic band topology as non-magnetic Weyl semimetals require lifting of inversion symmetry. 
Retraining the model using only these key features further improves the performance as shown in \autoref{fig:Fig5}b.
Despite the relatively small dataset, the model demonstrates reasonable accuracy, suggesting that our approach can be extended to efficiently down select broad compositions spaces for high-throughput screening in future interconnect searches. 

\section{Summary and outlook}
In this work we have proposed and implemented a workflow in an effort to identify optimal topological conductors for use as interconnects in next-generation integrated circuits. Our proposed workflow relies on utilization of Wannier tight-binding models which replicate the electronic structure of the bulk material precisely, however by working in a Wannier basis it is possible to exploit sparse matrix methods to access large system sizes in computation of transmission via the NEGF formalism. We have identified TiS, ZrB$_{2}$ and $A$N ($A=\mathrm{Mo},\, \mathrm{Ta},\, \mathrm{W}$) to be added to the list of potentially optimal interconnect candidates which already includes CoSi, and $M$Pn, $M=(\mathrm{Nb},\,\mathrm{Ta}$) and $\mathrm{Pn}=(\mathrm{P},\, \mathrm{As})$. We have further investigated the degree to which the computed surface transmission is robust to surface roughness. Our findings suggest that a maximal value of surface transmission is a high-priority metric and can be correlated with a large number of chiral Fermi arcs as found in prior works. However, our screening also brings to light the impact of additional bulk bands in the vicinity of the Fermi energy. No existing candidate is an ideal TSM with a point-like Fermi surface so the impact of additional bulk bands which intersect the Fermi energy is important to consider. While these bands do not impact the surface transmission, they decrease two other key metrics, $\alpha$ and $\rho_{0} \lambda$ through an increase in the bulk conductance.

While a decreased value of $\rho_{0} \lambda$ is beneficial for a trivial interconnect candidate due to lower bulk resistance, a decreased value of $\alpha$ is viewed negatively for a topological interconnect candidate as it suggests surface states may not remain the dominant conducting channel for finite thickness films. This situation is particularly relevant for topological metals like MoN. While topological metals are identified as interconnect candidates by the surface transmission values computed in this work, the computed value of $\rho_{0} \lambda$ indicates it is the most important metric as the bulk will dominate nanowire transmission. This situation underscores the need to study the multivariable Pareto frontier and leverage the protection against localization offered to bulk states as well as surface states by non-trivial topology.
In addition, crystallographic orientation is critical for maximizing $T_S$, given its strong anisotropy. 
Our simulations suggest robustness to surface roughness and disorder, supporting feasibility under realistic fabrication conditions; however, experimental validation is essential to assess long-term reliability.

Future work to extend this protocol to further ternary compounds will benefit from refinement of the nanowire transmission computations, primarily through efficient parallelization of the non-equilibrium Greens function formalism. Additionally the potential to utilize foundational machine learning networks for generation of the Hamiltonian, such as that presented in Refs.\  \cite{zhong2023transferable,gong2023general}, may provide an avenue to efficiently screening a large number of materials. 
Magnetic semimetals also present a promising class of material which has been reserved for future study. While a number of magnetic TSMs have been experimentally synthesized \cite{belopolski2019discovery,yang2017topological,PhysRevB.83.205101,liu2018giant,wang2018large,xu2020high,bernevig2022progress}, the impact of thermal fluctuations and defects on the magnetic ordering and Fermi arc states must be carefully studied.

Beyond intrinsic properties evaluated in this work, practical integration of the proposed compounds into interconnect technologies requires consideration of process compatibility (with back-end-of-line (BEOL) temperatures), orientation control for maximizing transmission, electromigration reliability, diffusion barrier properties, liner/wetting layer and adhesion layer considerations. 
State-of-the-art interconnects use Cu  embedded in low-$K$ dielectrics like SICOH, with multilayer stacks including TaN barriers and Co adhesion layers. Emerging materials such as MoN, TiS and ZrB$_2$ may reduce or eliminate the need for these layers due to superior bonding and diffusion resistance (see \supd VII). 
For example, binary MoP exhibits enhanced electromigration resistance \cite{han2023topological}, potentially simplifying BEOL integration.
Future first-principles studies on the compounds identified in this work should prioritize evaluating atomic diffusion barriers, as these are critical for assessing electromigration reliability under operational conditions. Achieving the correct phase and desired crystallographic orientation is essential for optimal performance and must be considered alongside the availability of reactive ion etching, wet etch, and chemical mechanical polishing chemistries. Area-selective deposition of topological metals offers potential for contact formation beyond the damascene process, where oxidation effects should be considered.

\section*{Methods}
We perform density functional theory (DFT) computations with spin-orbit coupling (SOC) using the Quantum Espresso software package \cite{QE-2020,Perdew1996}, and 
fully relativistic norm-conserving pseudopotentials from the PseudoDojo library~\cite{van2018pseudodojo}  along with a maximal plane-wave energy cutoff of 60\,Ry. 
The Brillouin zone was sampled using a 9 $\times$ 9 $\times$ 9 Monkhorst-Pack $k$-point grid \cite{monkhorst1976special}. This grid is sufficient for convergence when considering the smallest unit cell considered in this work. The choice of a uniform grid is made to prioritize the creation of high-quality Wannier tight-binding models rather than maximize computationally efficiency.   

For the initial screening, the Wannier tight-binding models are constructed using the selected columns of the density matrix (SCDM) protocol put forth in Ref. ~\cite{vitale2020automated}, allowing this process to be fully automated. 
Once constructed, the Wannier tight-binding models are inspected for accuracy, determined by comparison of the band structure for the tight-binding model and \emph{ab initio} data within $2$\, eV of the Fermi energy. If the band structure for the tight-binding model faithfully reproduces the bulk band structure in this window, the second stage of the workflow is initiated, estimation of surface transmission. 

\begin{acknowledgments}
Work by W.R.\ and J.M.R.\ was supported by the SUPeRior Energy-efficient Materials and dEvices (SUPREME) Center SUPREME, one of seven centers in the JUMP 2.0, a Semiconductor Research Corporation (SRC) program sponsored by DARPA.
NORDITA is supported in part by NordForsk.
H. Lin acknowledges the support by the National Science and Technology Council (NSTC) in Taiwan under grant number MOST 114-2112-M-001-055-MY3. Y.-H. Tu and G.C. Liang acknowledges the support from NSTC under grant number NSTC 112-2112-M-A49-047-MY3, and the Co-creation Platform of the Industry-Academia Innovation School, NYCU, under the framework of the National Key Fields Industry-University Cooperation and Skilled Personnel Training Act.
The authors thank D.\ Gall, C.\ Hinkle, E.\ Pop, J.\ Cha, and G.\ Parsons for useful discussions.
\end{acknowledgments}

\bibliography{Ref.bib}

\end{document}


\preprint{APS/123-QED}

\title{{\sc supplementary information}\\[0.5em]%
Accelerated Discovery of Topological Conductors for Nanoscale Interconnects}

\author{Alexander C.\ Tyner}
\affiliation{NORDITA, KTH Royal Institute of Technology and Stockholm University, Stockholm, Sweden}
\affiliation{Department of Physics, University of Connecticut, Storrs, CT, USA}

\author{William Rogers}
\affiliation{Graduate Program in Applied Physics, Northwestern University, Evanston, IL, 60208, USA}

\author{Po-Hsin Shih}
\affiliation{IBM Thomas J. Watson Research Center, Yorktown Heights, NY, USA}

\author{Yi-Hsin Tu}
\affiliation{Industry Academia Innovation School, National Yang Ming Chiao Tung University, Hsinchu, Taiwan}

\author{Gengchiau Liang}
\affiliation{Industry Academia Innovation School, National Yang Ming Chiao Tung University, Hsinchu, Taiwan}

\author{Hsin Lin}
\affiliation{Institute of Physics, Academia Sinica, Taipei, Taiwan}

\author{Ching-Tzu Chen}
\affiliation{IBM Thomas J. Watson Research Center, Yorktown Heights, NY, USA}

\author{James M.\ Rondinelli}
\affiliation{Department of Materials Science and Engineering, Northwestern University, Evanston, IL, USA}

\date{\today}
\maketitle

\tableofcontents

\clearpage
\newpage

\section{Dataset Details}

We restrict this work to mainly binary and elemental topological conductors with selected ternaries for purposes of computational efficiency, primarily due to the high memory demands of the nanowire transmission computation via non-equilibrium Greens function which remains a computational bottleneck. We reiterate that compounds that have either been experimentally synthesized or exhibit evidence of non-trivial bulk topology in prior computational studies are prioritized.

\begin{figure}[h]
    \centering
    \includegraphics[width=16cm]{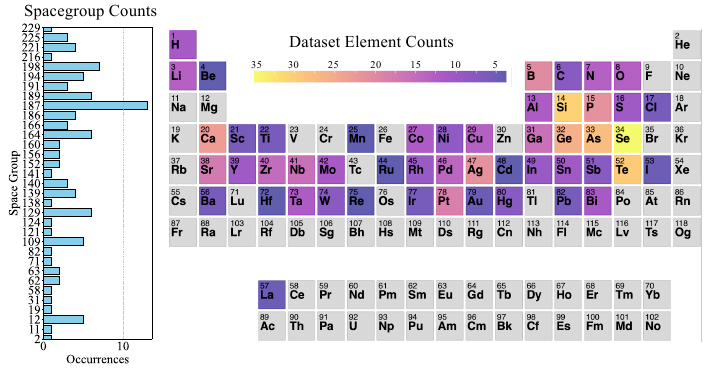}
    \caption{\sffamily \textbf{Dataset chemistry and space group diversity.} Occurrence of each space-group and element among the dataset of candidate interconnects explored in this work. }
    \label{fig:Dataset}
\end{figure}

\autoref{fig:Dataset} provides information regarding the breakdown of the dataset in terms of elemental composition and space group. 
A full list of the candidate materials is available in \autoref{sec:ResultSummary} and online at \jmr{URL}.

\newpage
\section{Surface and bulk electronic structure of TiS, Z\lowercase{r}B$_{2}$, N\lowercase{b}A\lowercase{s}, M\lowercase{o}N and M\lowercase{o}P}
\begin{figure}[h]
    \centering
    \includegraphics[width=16cm]{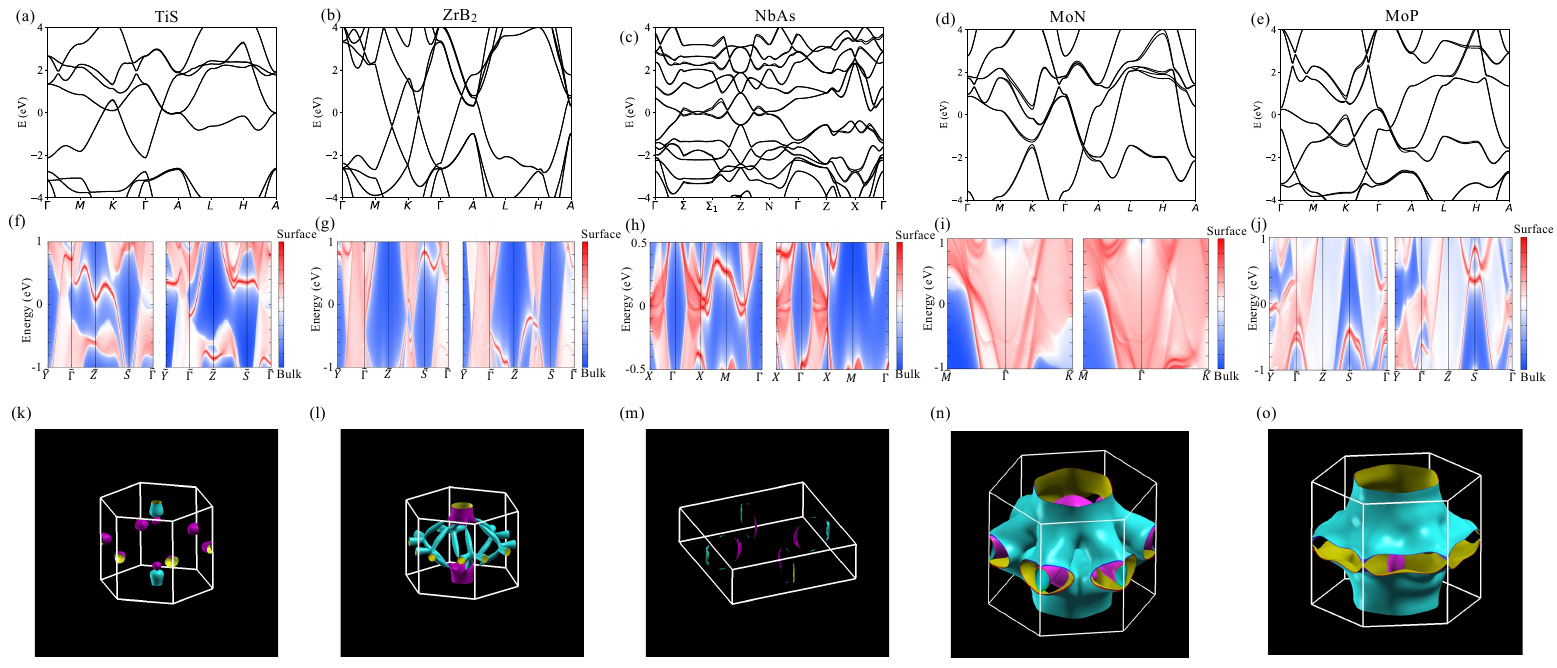}
    \caption{\sffamily \textbf{Electronic properties of select interconnect compounds.} (a)-(e) Bulk band structure of TiS, ZrB$_{2}$, NbAs, MoN and MoP respectively along high-symmetry paths in the Brillouin zone. (f)-(g) (100) surface spectral density along a high-symmetry path on the surface for TiS and ZrB$_{2}$ respectively. The spectral density for the top and bottom surfaces of the slab are shown on the left  and right sides respectively. (h)-(i) (001) surface spectral density along a high-symmetry path on the surface for NbAs and MoN respectively. The spectral density for the top and bottom surfaces of the slab are shown on the left and right sides respectively. (j) (100) surface spectral density along a high-symmetry path on the surface for MoP. (k)-(o) Bulk Fermi surface for TiS, ZrB$_{2}$, NbAs, MoN and MoP respectively. Different colors are assigned to contributions from distinct bands.}
    \label{fig:ElectronicDetails}
\end{figure}

TiS, ZrB$_{2}$ and MoN are identified as promising interconnect candidates and compared to MoP and NbAs topological materials which have been proposed as interconnect candidates in prior works. Here we provide the bulk band structure (\autoref{fig:ElectronicDetails}a-e), the surface spectral density (\autoref{fig:ElectronicDetails}f-j), and the bulk Fermi surface (\autoref{fig:ElectronicDetails}k-o) allowing for a direct comparison among the three compounds.   
\par 
We emphasize that although all compounds support Fermi arcs, as detailed in Fig.\ (3) of the main body, both TiS, ZrB$_{2}$, MoN and MoP exhibit a greater number of bulk bands that intersect with the Fermi energy when compared to NbAs. This leads to a larger Fermi surface as seen in \autoref{fig:ElectronicDetails}k-o, and contributes to the decreased value of $\alpha$ relative to NbAs. 

\newpage
\section{Dependence of surface transmission on Wannier basis choice}
\par
The nanowire transmission computations employed in this work require the electronic structure to be transformed into a Wannier basis from the plane-wave basis utilized by Quantum Espresso via the Wannier90 software package \cite{Pizzi2020}. The Wannierization process requires selecting a set of orbitals for each atom which constitute to the basis for the Wannier tight-binding model. One option for creation of high-quality tight-binding models is to consider all possible orbitals. While effective for production of high-quality Wannier tight-binding models, this approach increases the computational expense of the subsequent transmission computations by incorporating occupied bands far below the Fermi energy; increasing the degrees of freedom for the sparse matrix representing the electronic structure of a single unit cell. A computationally advantageous approach is to consider only a subset of orbitals which dominate the density of states near the Fermi energy. This leads to Wannier tight-binding models with considerably less degrees of freedom, limiting the computational expense of subsequent transmission computations. 

\begin{figure} [h]
    \centering
    \includegraphics[width=16cm]{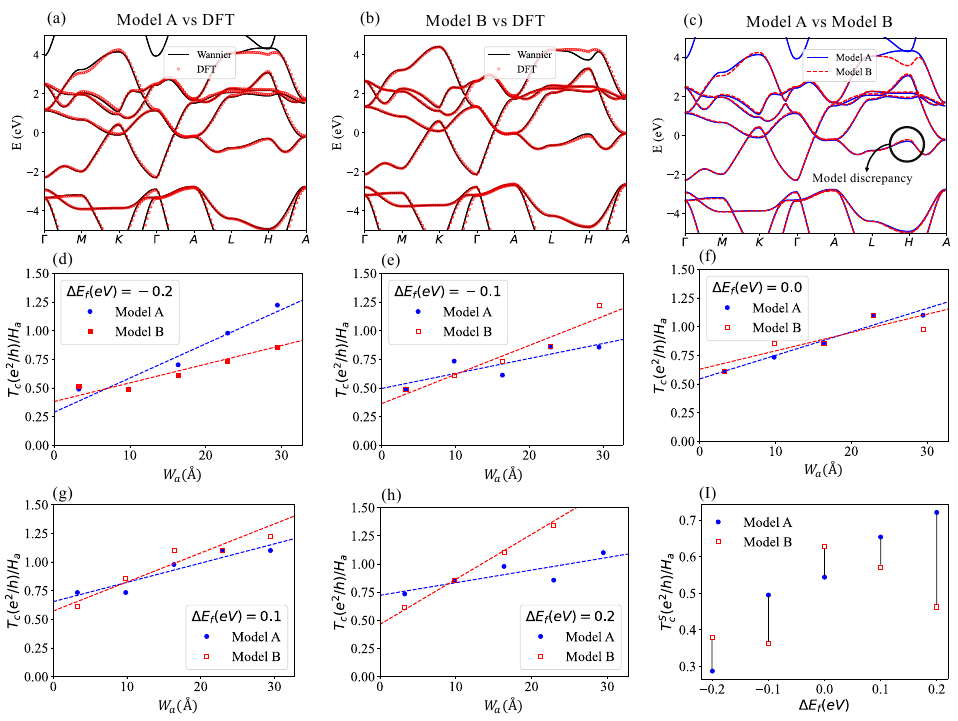}
    \caption{\sffamily \textbf{Wannier basis sensitivity analysis: TiS.} (a)-(b) Bulk band structure of Wannier tight-binding model using (a) only d-orbitals for Ti and $p$-orbitals for S and (b) $s,p,d$-orbitals for Ti and $s,p$-orbitals for S, compared to density functional theory band structure along a high-symmetry path in the Brillouin zone. (c) Comparison of the bulk band structure between the two Wannier tight-binding models. (d)-(h) Nanowire transmission as a function of thickness for the two tight-binding models varying the doping between $\Delta E_{f}=-0.2$ and $\Delta E_{f}=0.2$. (I) Extracted value of surface transmission for both Wannier tight-binding models as a function of doping.}
    \label{fig:ModelComparison}
\end{figure}

\begin{figure} [h]
    \centering
    \includegraphics[width=16cm]{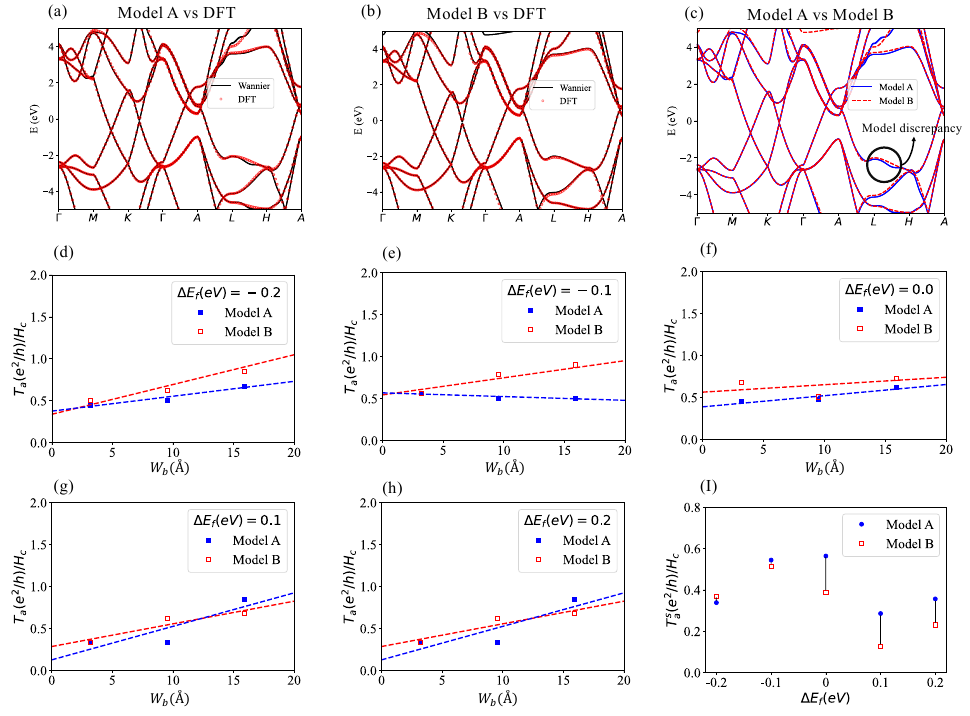}
    \caption{\sffamily \textbf{Wannier basis sensitivity analysis: ZrB$_{2}$.} (a)-(b) Bulk band structure of Wannier tight-binding model using (a) only d-orbitals for Zr and $p$-orbitals for B and (b) $s,p,d$-orbitals for Zr and $s,p$-orbitals for B, compared to density functional theory band structure along a high-symmetry path in the Brillouin zone. (c) Comparison of the bulk band structure between the two Wannier tight-binding models. (d)-(h) Nanowire transmission as a function of thickness for the two tight-binding models varying the doping between $\Delta E_{f}=-0.2$ and $\Delta E_{f}=0.2$. (I) Extracted value of surface transmission for both Wannier tight-binding models as a function of doping.}
    \label{fig:ModelComparisonZrB2}
\end{figure}

In this work, we first prioritize accuracy of the tight-binding model, as measured by faithful reproduction of the DFT band structure within 1\,eV of the Fermi energy. The secondary goal is to use as few orbitals as possible to accomplish this, limiting the computational expense of the nanowire transmission calculations. To understand how this approach can influence the results of computing nanowire transmission we consider first the case of TiS, constructing two Wannier tight-binding models. The first, denoted model A, incorporates $s$, $p$, and $d$ orbitals for Ti and $s$ and $p$ orbitals for S, the second, denoted model B, incorporates only $d$ orbitals for Ti and $p$ orbitals for S. The resulting band structure of the two models compared to the DFT band structure and each other is shown in \autoref{fig:ModelComparison}a-c. Both structures reproduce the DFT band structure with high accuracy, the main discrepancy is marked by a black circle in \autoref{fig:ModelComparison}c.

It is now important to understand how the basis choice impacts the resulting values of nanowire transmission. To do so, we compute transmission along the [001] direction due to states on the (010) surface. We select this transmissions direction for comparison as it was found to support maximal surface transmission in the main body. The results for Model A and Model B as a function of doping are shown in \autoref{fig:ModelComparison}d-h with the extracted values of the surface transmission as a function of doping compared in \autoref{fig:ModelComparison}I. We note that the extracted values of the surface transmission are within 15\% for doping of 0.1\,eV. However, when considering a doping of 0.2\,eV above the Fermi energy, the discrepancy between the two models increases. This is not surprising as utilizing a smaller subset of orbitals requires the Wannier90 software package to fit conduction bands which may support a partial orbital density of states due to orbitals not included in the provided basis. Nevertheless, it is important to observe that the maximum value of transmission measured within 0.1\,eV of the Fermi energy is nearly identical between Model A and Model B, occurring at a doping of 0.1\,eV and at the Fermi energy, respectively.  

\par 

This analysis is repeated for ZrB$_{2}$ where Model A considers contribution of Zr $d$-orbitals and B $p$-orbitals while Model B considers $s,p,d$-orbitals of Zr and $s,p$-orbitals of B. The resulting band structures of the tight-binding model are compared to the DFT band structure and each other in \autoref{fig:ModelComparisonZrB2}(a)-(c). Slight discrepancies in the band structure of the two models is visible both above and below the Fermi energy between the $L$ and $H$ point. The resulting nanowire transmission is compared between the models in  \autoref{fig:ModelComparisonZrB2}(d)-(i). Once again the discrepancy remains smallest below the Fermi energy however the extracted surface transmission remains comparable throughout the range of doping values sampled.

\newpage
\section{Robustness of topological surface states to impurities and disorder}
In the main body the utility of topological surface states as conducting channels for next-generation interconnects is explored. Such topological surface states are often referred to as ``protected" and considered immune from back-scattering in the context of idealized low-energy models\cite{prodan2011disordered,Kobayashi2013}. This is not true in general. In particular, for symmetry protected topological materials, the surface states are not protected from disorder which breaks the underlying symmetry\cite{ASPT,Araki2019,Li2020,Chaou2025}. Furthermore, realistic materials can support multiple states which cross the Fermi energy at the same transverse momenta. Such states can hybridize, opening scattering channels\cite{Hsu2014,PhysRevB.97.235108}.
\par 
However, even in the case of symmetry protected topological systems, topological surface states have been shown to be less-susceptible to localization\cite{ASPT,Araki2019,Li2020,Chaou2025}. Their existence as conducting surface channels in the clean system is also guaranteed by the bulk topological invariant. This combination continues to make them strong candidates for next-generation interconnects despite the realistic considerations.

\newpage
\section{Effect of surface vacancies on N\lowercase{b}A\lowercase{s}}

In Fig.\ 4 of the main body the effect of disorder, in the form or random surface vacancies, on the surface transmission is explored. In particular we explore the relative change in surface transmission as a function of surface vacancy density for TiS and ZrB$_{2}$. The relative decrease in the surface transmission as a function of the vacancy density is compared to NbAs, a promising candidate interconnect TSM which has been previously studied in Ref.\ \cite{cheon2025surface}. 

\begin{figure}[h]
    \centering
    \includegraphics[width=16cm]{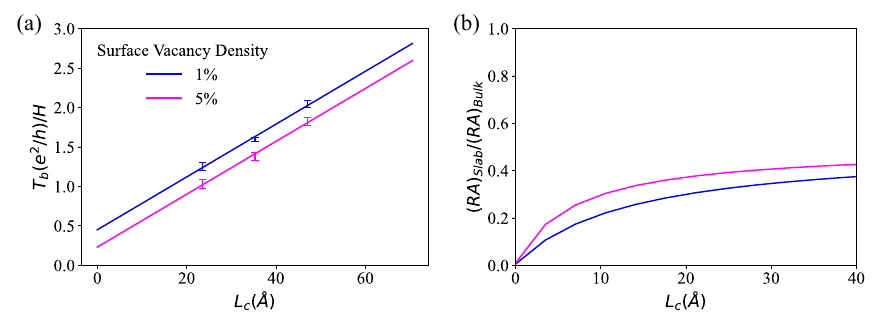}
    \caption{\sffamily \textbf{Disorder effects on transport proeprties of NbAs.}  (a) Nanowire surface transmission is extracted for $1\%$ and $5\%$ surface vacancy density in NbAs at $\Delta E_{F}=-0.1$ where surface transmission is maximal. (b) Utilizing the extracted value of surface and bulk transmission, corresponding to the intercept and slope of the linear fit to nanowire transmission as a function of system size respectively, the ratio of slab to bulk resistance area is plotted as a function of the slab surface roughness. }
    \label{fig:NbAsDisoder}
\end{figure}

In this section, we detail computation of nanowire transmission as a function of thickness in the presence of surface vacancies for NbAs at $\Delta E_{f}=-0.1eV$ where surface transmission is maximal. The results are show in Fig.\ \autoref{fig:NbAsDisoder}a, detailing that as the surface vacancy density is increased from one to five percent, the extracted surface vacancy density is reduced by approximately $51\%$. We additionally compute the ratio of the slab and bulk resistance-area product as a function of the surface vacancy density and display the results in \autoref{fig:NbAsDisoder}b. The result for one percent disorder has been displayed in the main body along side this ratio for the optimal nanowire geometry of TiS, ZrB$_{2}$ and MoN for comparison. This ratio increases at a reduced rate as a function of system size relative to TiS, ZrB$_{2}$ and MoN, due in part to the increased size of the NbAs bulk Fermi surface, allowing the Fermi arc states to serve as the dominate conducting channels at the Fermi energy. 

\newpage
\section{Computation of $\rho\, \lambda$ product}
The product of the bulk resistivity and mean-free path, $\rho \lambda$, is a value commonly used to identify optimal trivial conductors \cite{Gall:2022}. This quantity is computed for TiS ZrB$_{2}$ and MoN as, 
\begin{equation}
    \frac{1}{\rho \lambda}=\frac{e^2}{4\pi^{3}\hbar}\sum_{n}\iint_{S^{n}_{F}}\frac{v^{2}_{t,n}(\mathbf{k})}{\mathbf{v}^{2}_{n}(\mathbf{k})}dS,
\end{equation}
where $\mathbf{v}_{n}(\mathbf{k})=\frac{1}{\hbar}\nabla_{k}E_{n}(\mathbf{k})$ is the electron velocity of band $n$ and $S^{n}_{F}$ is the Fermi surface of band $n$. For cubic systems, this relation simplifies to, 
$\rho \lambda=12 \pi^3 \hbar/(e^2 A_{f})$ where $A_{f}$ is the total Fermi surface area, indicating that an increased Fermi surface area lowers  $\rho \lambda$.

We compute this quantity using a $\Gamma$ centered $400 \times 400 \times 400$ grid of $\mathbf{k}$-points. Visualizations of the Fermi surface for TiS, ZrB$_{2}$ and MoN are provided in ~\autoref{fig:ElectronicDetails}g-h. Immediately, we find an enhanced Fermi surface area of MoN and ZrB$_{2}$ relative to that of TiS, indicating a decreased value of the $\rho \lambda$ product. 

We further note that spin-orbit coupling (SOC) is included in our computation to maintain a consistent level of theory across all computations. SOC is not included in prior works computing $\rho \lambda$, such as Refs.\ \cite{Gall:2016,gall2021materials,Gall:2022}, and has the effect of generally increasing the Fermi surface area and lowering the value of $\rho \lambda$.

\newpage
\section{Computation of electromigration barrier}
The electromigration barrier is computed using the Nudged Elastic Band (NEB) method. We consider a $4 \times 4 \times 4$ supercell with a nearest neighbor vacancy of the same species. As TiS, ZrB$_{2}$, MoN and MoP all support hexagonal unit cells, we consider vacancies along the $c$ axis and $a$ axis separately. The NEB path contains 11 points as the atom diffuses from the starting to ending vacancy site. We use $\Gamma$-point sampling due to the large size of the supercells and employ a plane-wave cutoff of 60 Ry. 
\par
\begin{figure}[h]
    \centering
    \includegraphics[width=\linewidth]{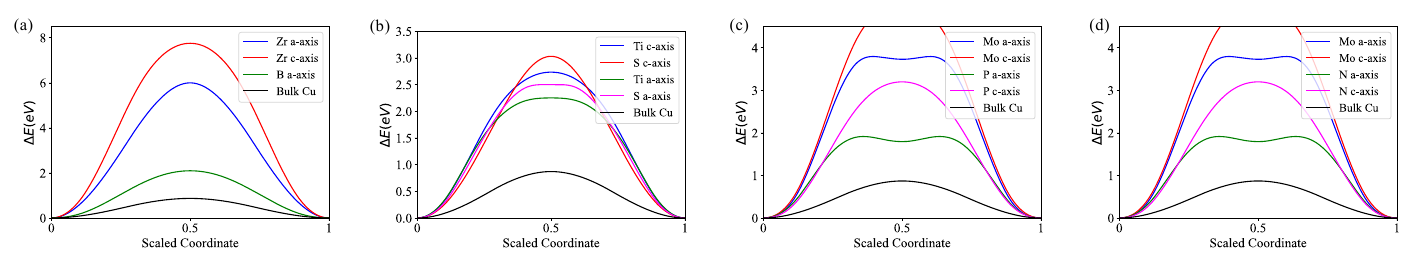}
    \caption{\sffamily \textbf{Electromigration barriers for select compounds.} Computed electromigration barriers for diffusion between nearest neighbor vacancies of the same atomic species along the $a$ and $c$ axis for hexagonal systems ZrB$_{2}$, TiS, MoP, and MoN, respectively. In each system, we further compare to the electromigration barrier computed an in identical manner for bulk copper.}
    \label{fig:Electromigration}
\end{figure}

 
The results are shown in \autoref{fig:Electromigration}(a)-(d) along with the NEB computation for a $4 \times 4 \times 4$ supercell of Cu. In all cases, the electromigration barrier is higher than that for Cu vacancy diffusion.

\clearpage
\section{Summary of results}\label{sec:ResultSummary}


\section*{Supplementary material}

# VIII.   SUMMARY OF RESULTS

Compound: AgN$_3$
Materials Project ID: 571297

### Lattice (conventional cell):

| Parameter | Value | Unit |
|-----------|-------|------|
| a | 5.8273 | Å |
| b | 5.8273 | Å |
| c | 6.0257 | Å |
| α (alpha) | 90.0000 | ° |
| β (beta) | 90.0000 | ° |
| γ (gamma) | 90.0000 | ° |

### Crystal structure (conventional cell):

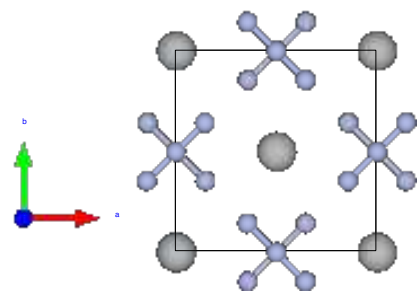

### Nanowire Transmission:

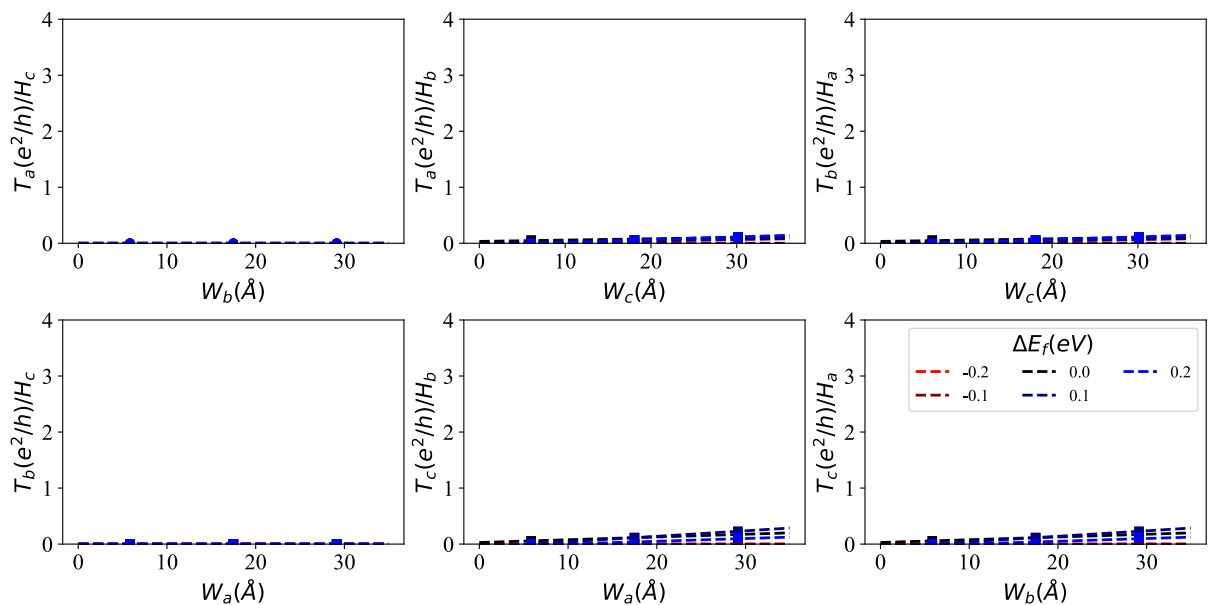

Compound: Ag₂Se

Materials Project ID: 568936

Lattice (conventional cell):

| Parameter | Value | Unit |
|-----------|-------|------|
| a | 4.4975 | Å |
| b | 7.1133 | Å |
| c | 7.6707 | Å |
| α (alpha) | 90.0000 | ° |
| β (beta) | 90.0000 | ° |
| γ (gamma) | 90.0000 | ° |

Crystal structure (conventional cell):

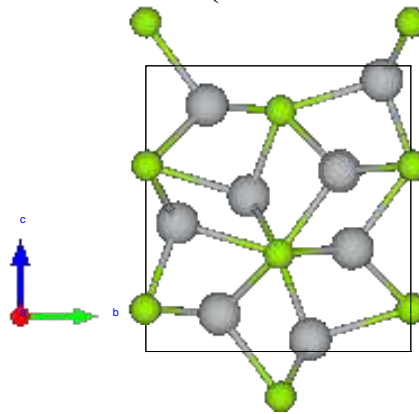

Nanowire Transmission:

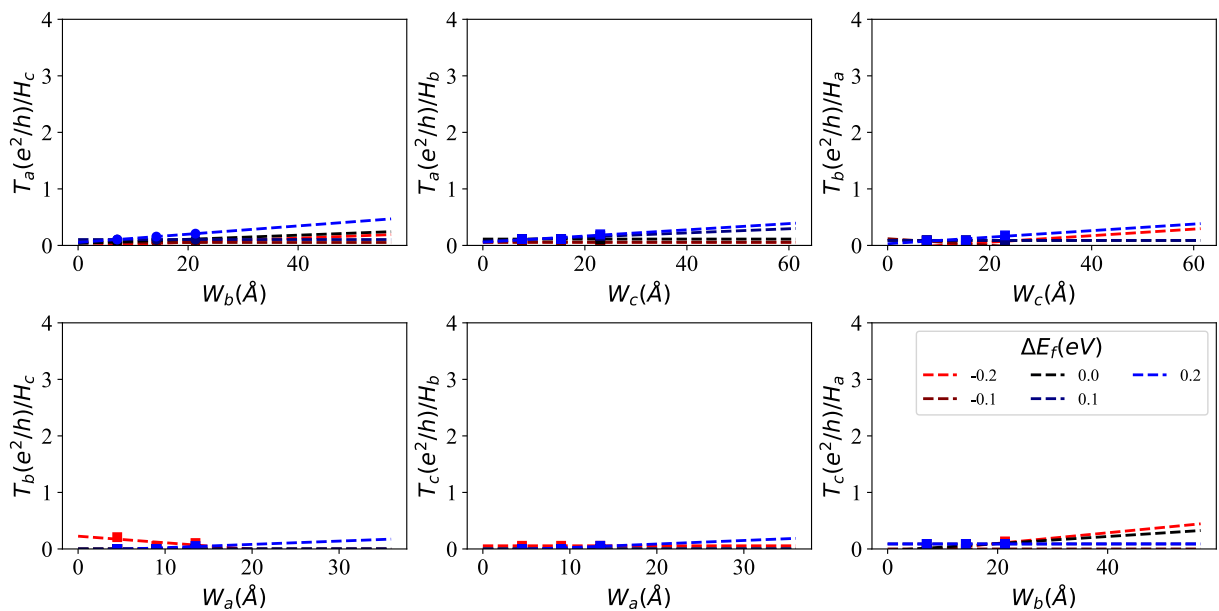

Compound: AlPd$_5$I$_2$
Materials Project ID: 27393

Lattice (conventional cell):

| Parameter | Value | Unit |
|-----------|-------|------|
| a | 4.0783 | Å |
| b | 4.0783 | Å |
| c | 20.2662 | Å |
| α (alpha) | 90.0000 | ° |
| β (beta) | 90.0000 | ° |
| γ (gamma) | 90.0000 | ° |

Crystal structure (conventional cell):

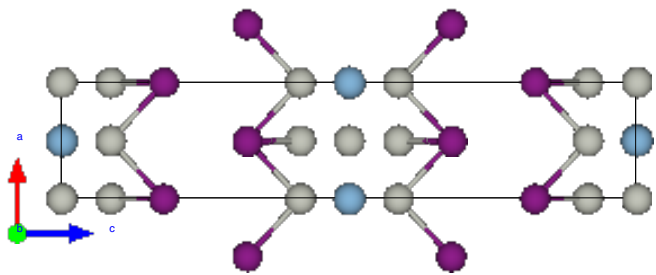

Nanowire Transmission:

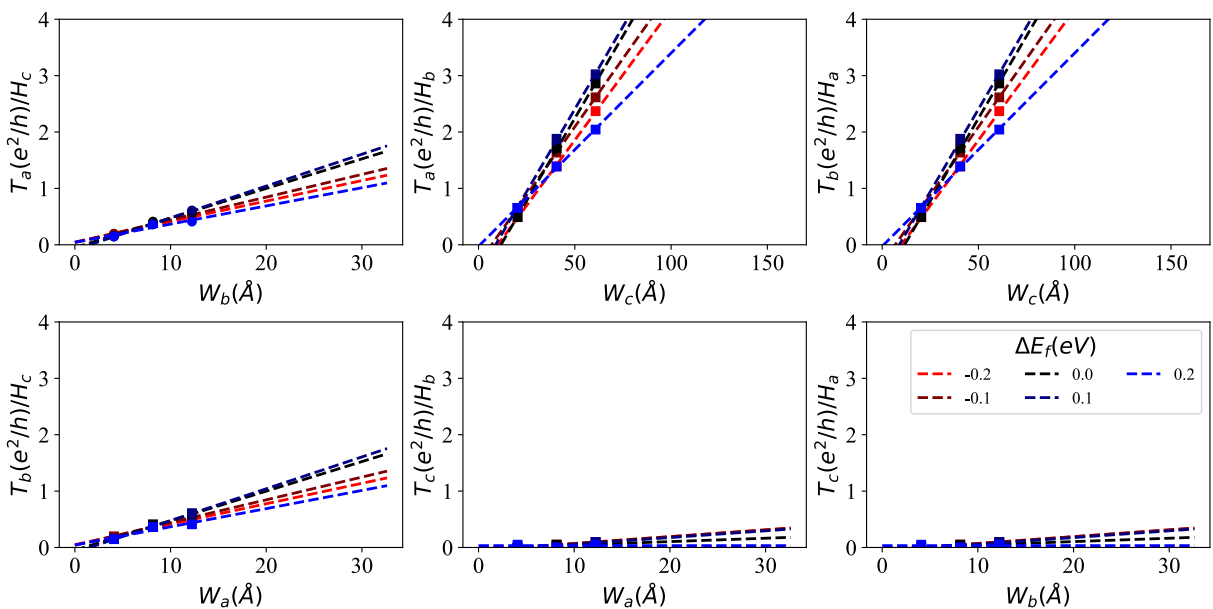

Compound: AlPt
Materials Project ID: 10904

Lattice (conventional cell):

| Parameter | Value | Unit |
|---|---|---|
| a | 4.9174 | Å |
| b | 4.9174 | Å |
| c | 4.9174 | Å |
| α (alpha) | 90.0000 | ° |
| β (beta) | 90.0000 | ° |
| γ (gamma) | 90.0000 | ° |

Crystal structure (conventional cell):

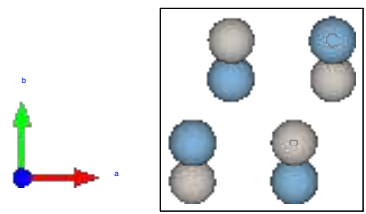

Nanowire Transmission:

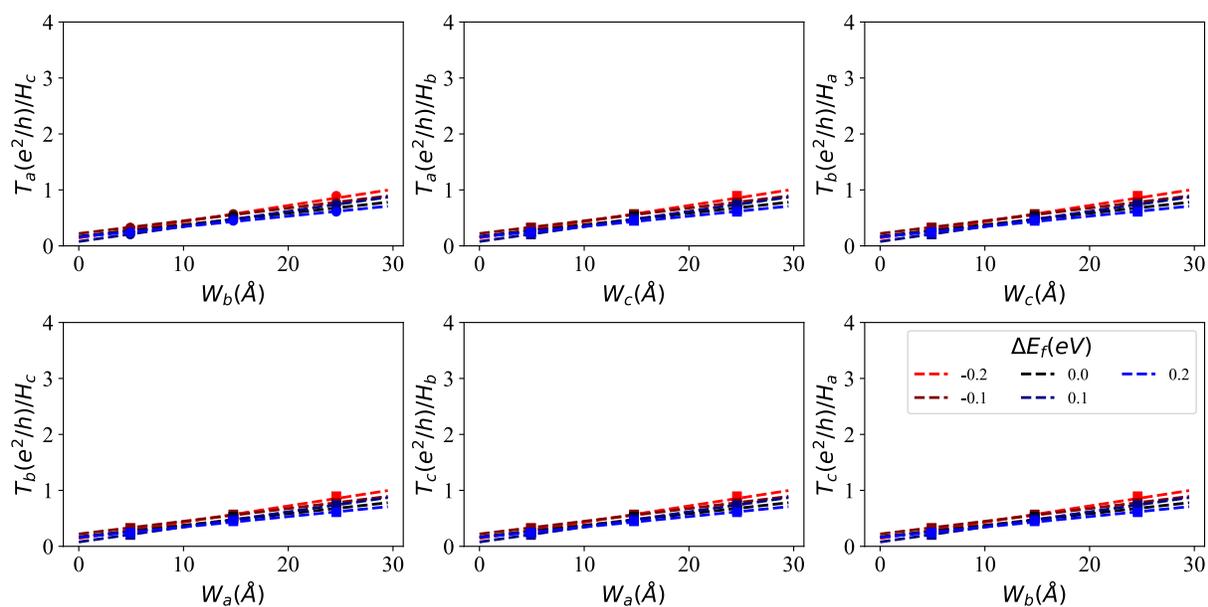

Compound: Au
Materials Project ID: 81

Lattice (conventional cell):

| Parameter | Value | Unit |
|-----------|-------|------|
| a | 4.1713 | Å |
| b | 4.1713 | Å |
| c | 4.1713 | Å |
| α (alpha) | 90.0000 | ° |
| β (beta) | 90.0000 | ° |
| γ (gamma) | 90.0000 | ° |

Crystal structure (conventional cell):

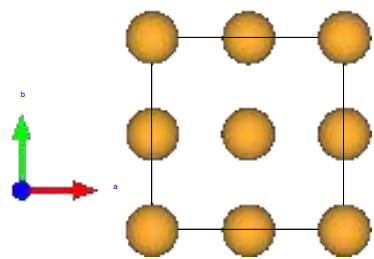

Nanowire Transmission:

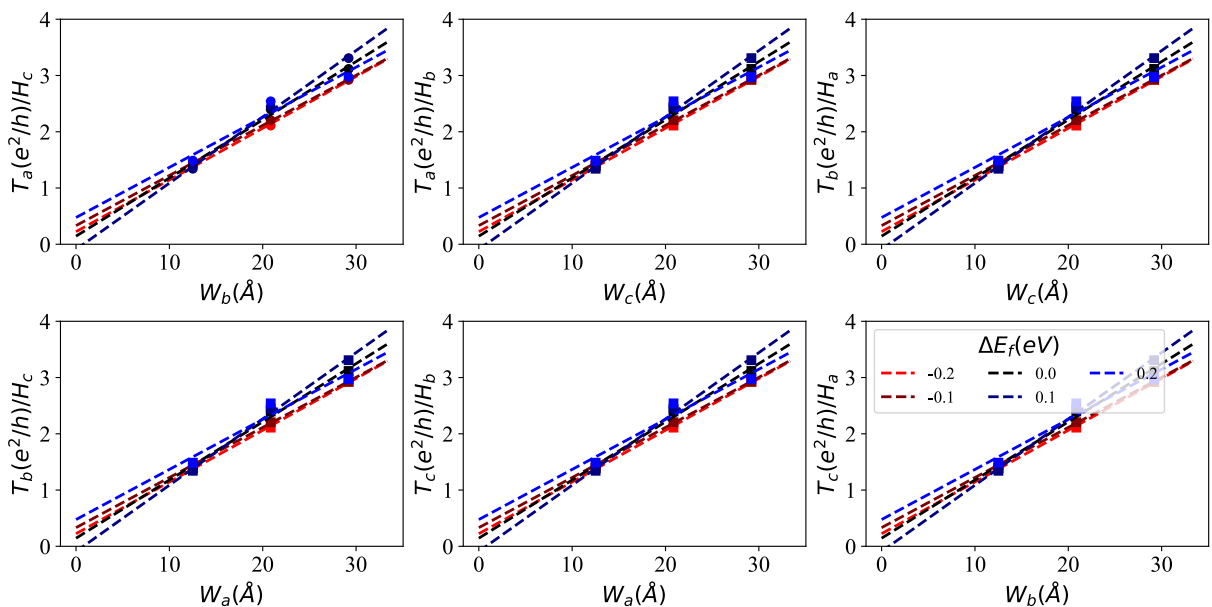

Compound: BaAl$_4$
Materials Project ID: 1903

Lattice (conventional cell):

| Parameter | Value | Unit |
|-----------|-------|------|
| a | 4.5765 | Å |
| b | 4.5765 | Å |
| c | 11.3531 | Å |
| α (alpha) | 90.0000 | ° |
| β (beta) | 90.0000 | ° |
| γ (gamma) | 90.0000 | ° |

Crystal structure (conventional cell):

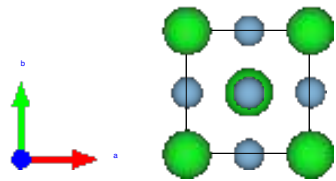

Nanowire Transmission:

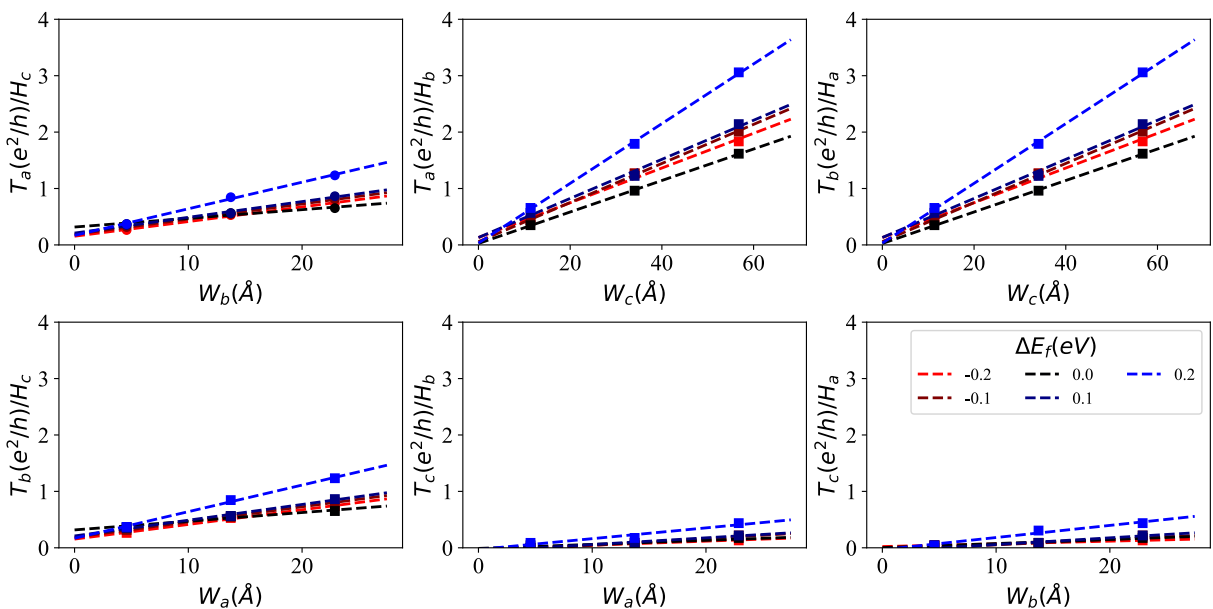

Compound: BaB$_6$
Materials Project ID: 954

Lattice (conventional cell):

| Parameter | Value | Unit |
|---|---|---|
| a | 4.2801 | Å |
| b | 4.2801 | Å |
| c | 4.2801 | Å |
| α (alpha) | 90.0000 | ° |
| β (beta) | 90.0000 | ° |
| γ (gamma) | 90.0000 | ° |

Crystal structure (conventional cell):

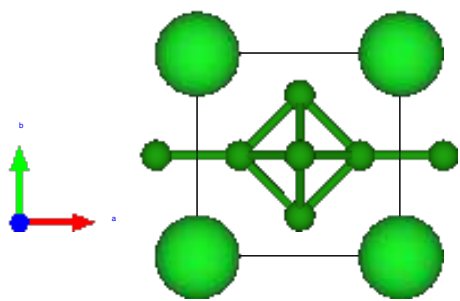

Nanowire Transmission:

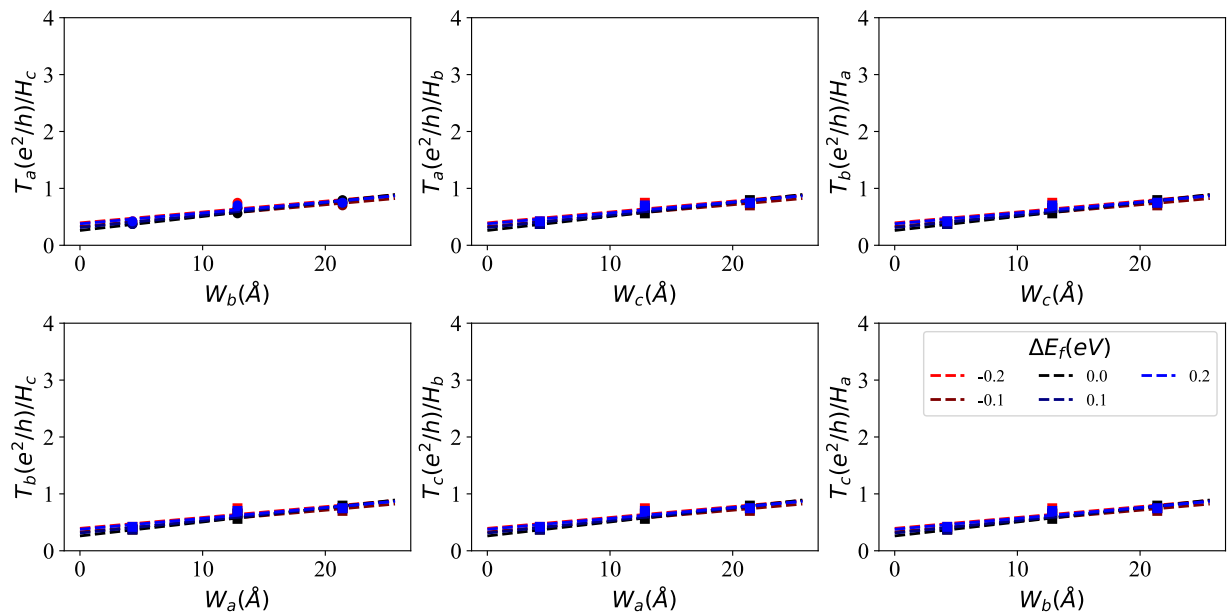

Compound: BaSi$_2$
Materials Project ID: 7655

Lattice (conventional cell):

| Parameter | Value | Unit |
|---|---|---|
| a | 4.0928 | Å |
| b | 4.0928 | Å |
| c | 5.3629 | Å |
| α (alpha) | 90.0000 | ° |
| β (beta) | 90.0000 | ° |
| γ (gamma) | 120.0000 | ° |

Crystal structure (conventional cell):

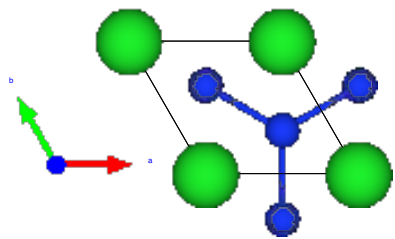

Nanowire Transmission:

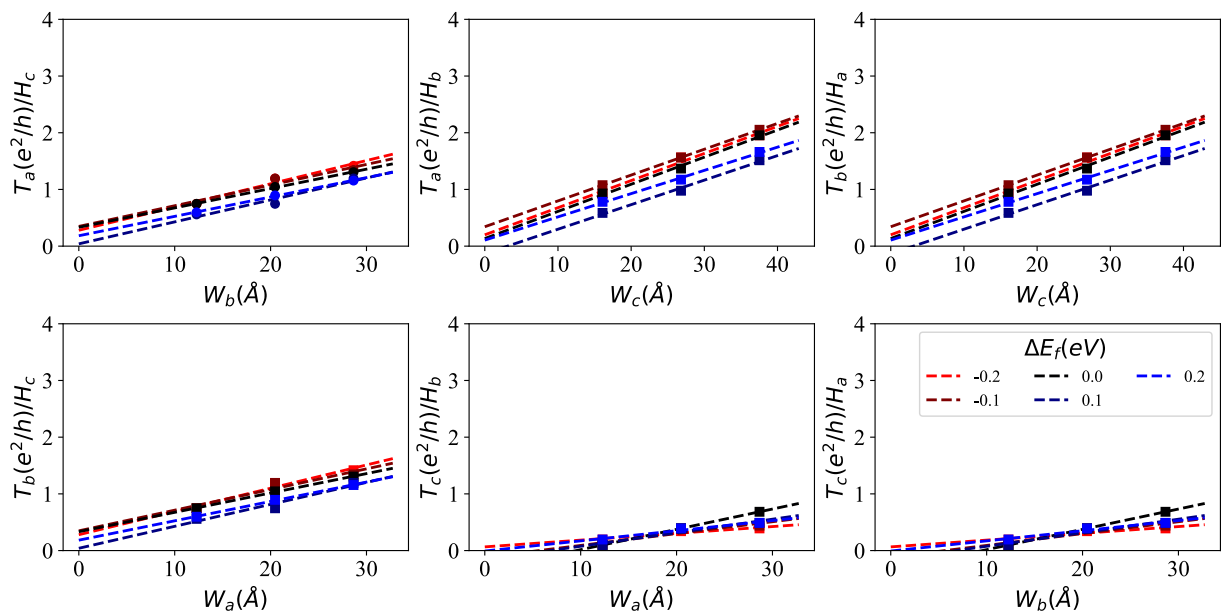

Compound: BaSn$_2$
Materials Project ID: 567510

Lattice (conventional cell):

| Parameter | Value | Unit |
|---|---|---|
| a | 4.7650 | Å |
| b | 4.7650 | Å |
| c | 5.6084 | Å |
| α (alpha) | 90.0000 | ° |
| β (beta) | 90.0000 | ° |
| γ (gamma) | 120.0000 | ° |

Crystal structure (conventional cell):

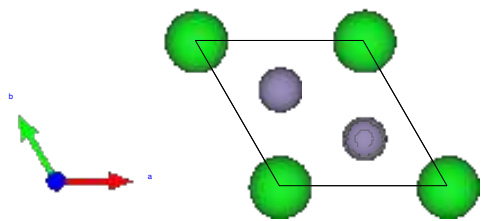

Nanowire Transmission:

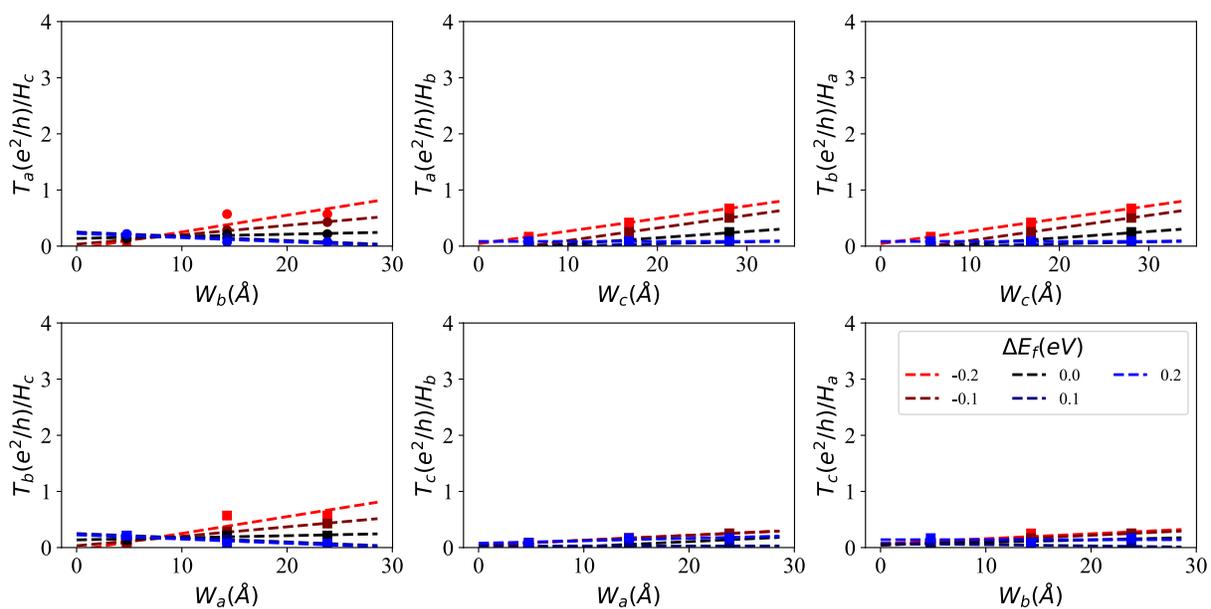

Compound: Be
Materials Project ID: 87

## Lattice (conventional cell):

| Parameter | Value | Unit |
|---|---|---|
| a | 2.2598 | Å |
| b | 2.2598 | Å |
| c | 3.5699 | Å |
| $\alpha$ (alpha) | 90.0000 | ° |
| $\beta$ (beta) | 90.0000 | ° |
| $\gamma$ (gamma) | 120.0000 | ° |

## Crystal structure (conventional cell):

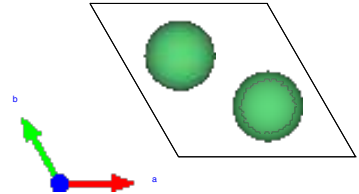

## Nanowire Transmission:

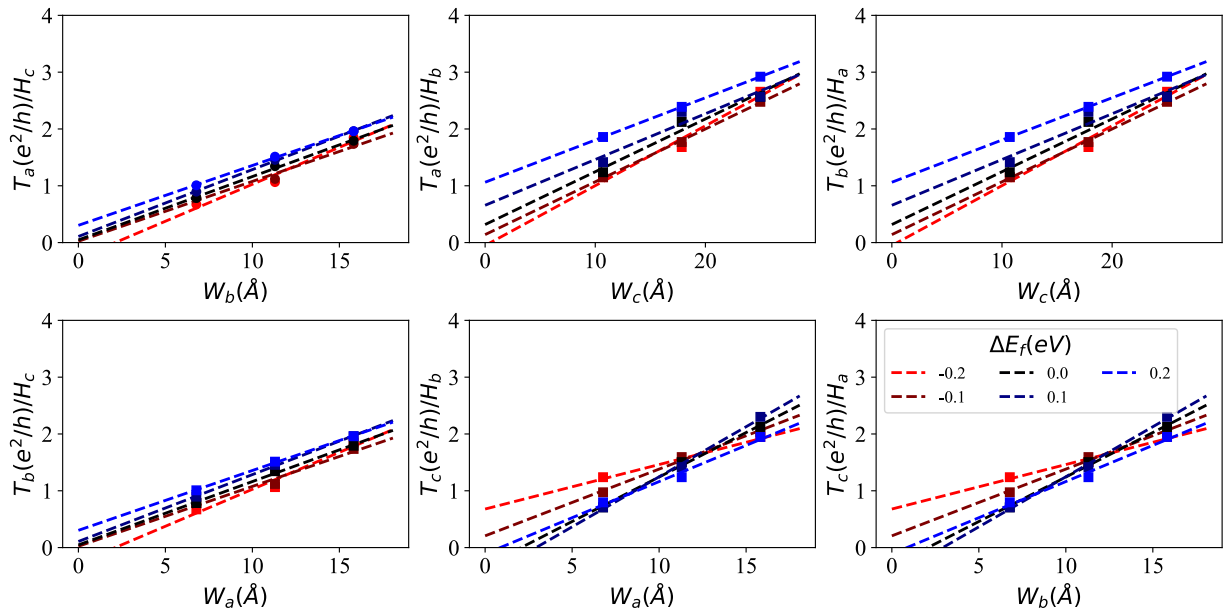

Compound: Bi
Materials Project ID: 23152

Lattice (conventional cell):

| Parameter | Value | Unit |
|-----------|-------|------|
| a | 4.6096 | Å |
| b | 4.6096 | Å |
| c | 11.9755 | Å |
| α (alpha) | 90.0000 | ° |
| β (beta) | 90.0000 | ° |
| γ (gamma) | 120.0000 | ° |

Crystal structure (conventional cell):

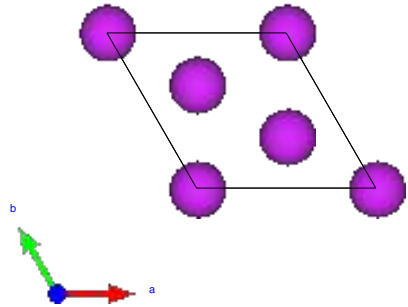

Nanowire Transmission:

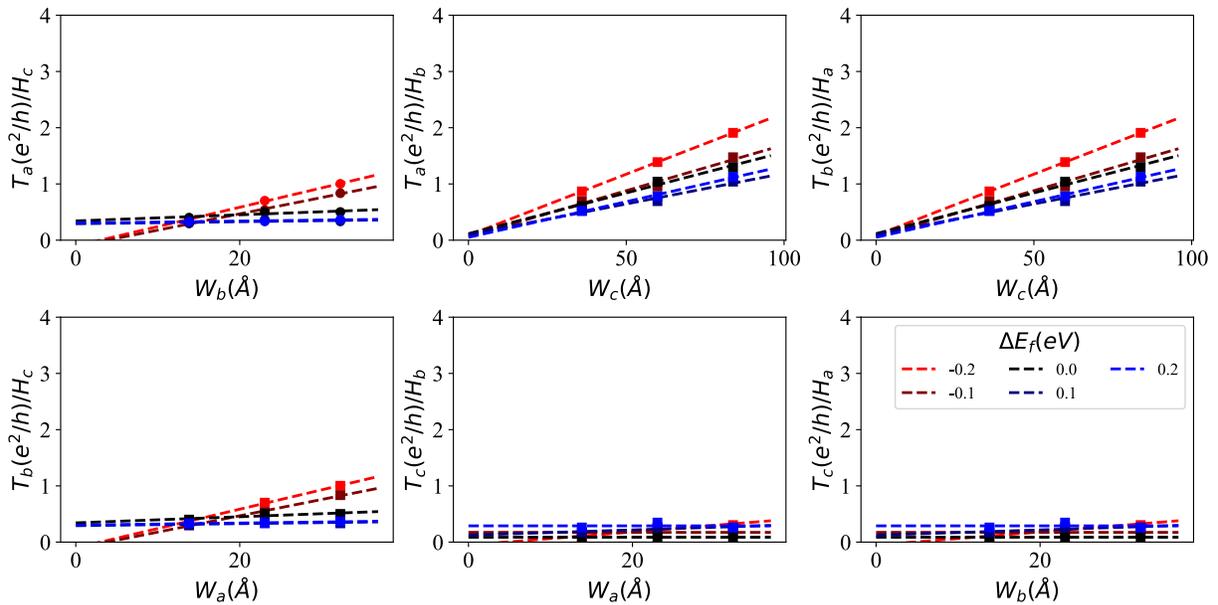

Compound: BiTeI
Materials Project ID: 22965

Lattice (conventional cell):

| Parameter | Value | Unit |
|---|---|---|
| a | 4.4252 | Å |
| b | 4.4252 | Å |
| c | 7.3781 | Å |
| $\alpha$ (alpha) | 90.0000 | ° |
| $\beta$ (beta) | 90.0000 | ° |
| $\gamma$ (gamma) | 120.0000 | ° |

Crystal structure (conventional cell):

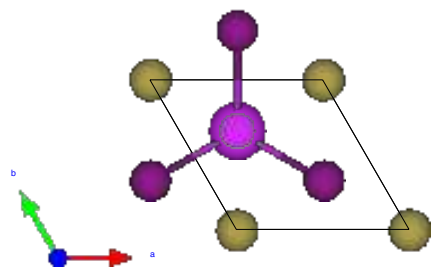

Nanowire Transmission:

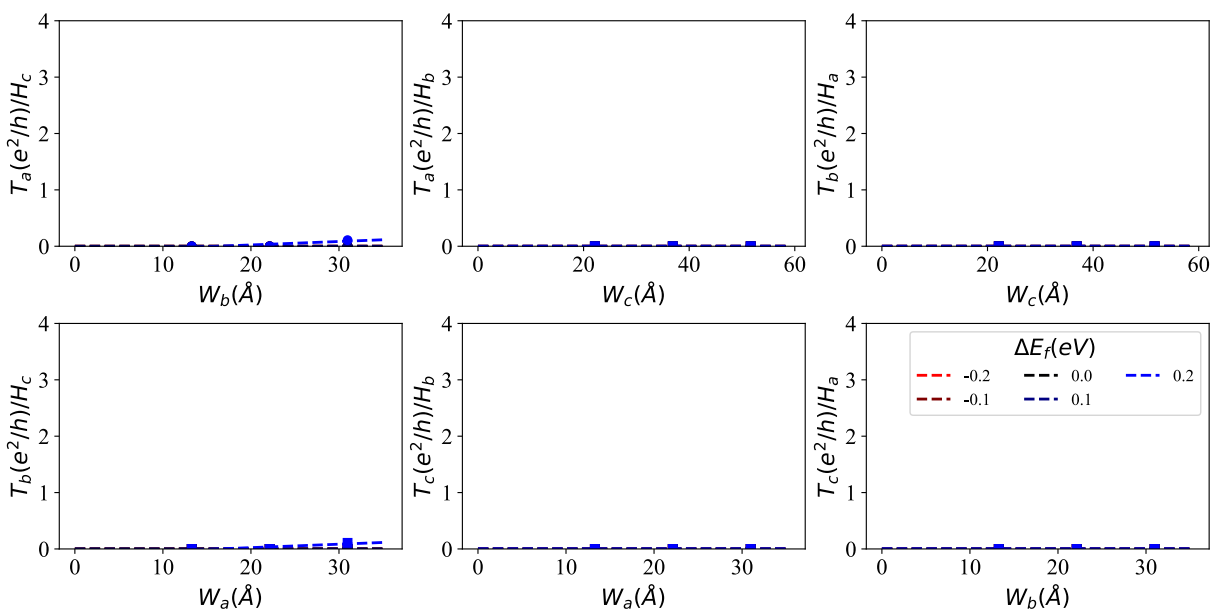

Compound: Ca$_2$As
Materials Project ID: 10106

Lattice (conventional cell):

| Parameter | Value | Unit |
|---|---|---|
| a | 4.5724 | Å |
| b | 4.5724 | Å |
| c | 15.8762 | Å |
| α (alpha) | 90.0000 | ° |
| β (beta) | 90.0000 | ° |
| γ (gamma) | 90.0000 | ° |

Crystal structure (conventional cell):

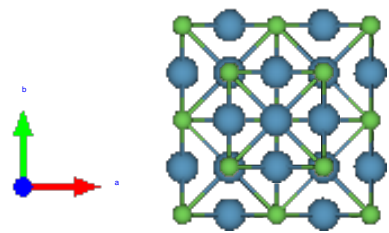

Nanowire Transmission:

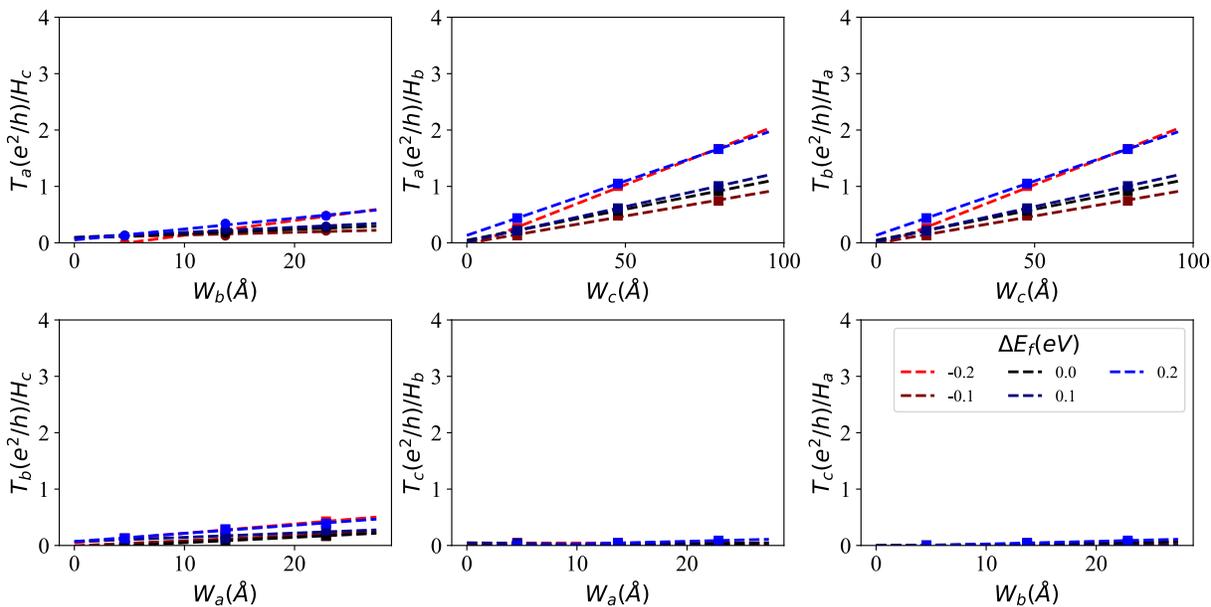

Compound: CaAgAs
Materials Project ID: 5615

Lattice (conventional cell):

| Parameter | Value | Unit |
|---|---|---|
| a | 7.2663 | Å |
| b | 7.2663 | Å |
| c | 4.3127 | Å |
| $\alpha$ (alpha) | 90.0000 | ° |
| $\beta$ (beta) | 90.0000 | ° |
| $\gamma$ (gamma) | 120.0000 | ° |

Crystal structure (conventional cell):

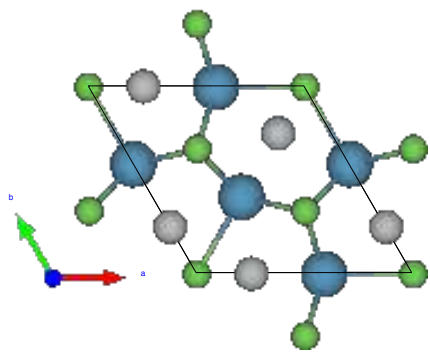

Nanowire Transmission:

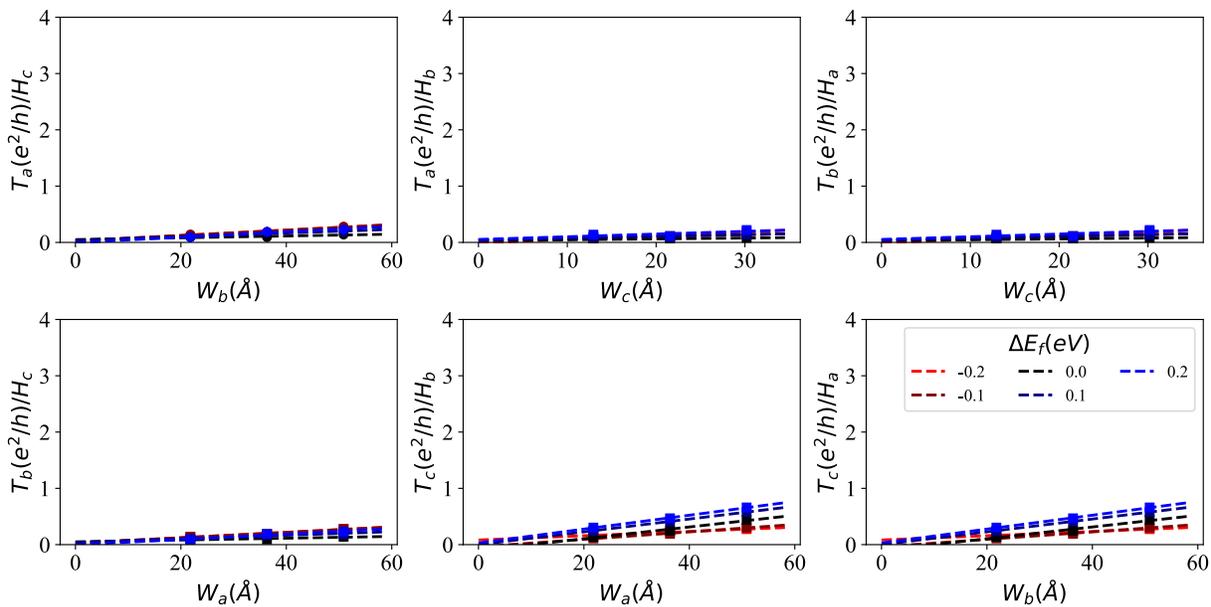

Compound: CaAgBi

Materials Project ID: 568664

Lattice (conventional cell):

| Parameter | Value | Unit |
|-----------|-------|------|
| a | 4.9065 | Å |
| b | 4.9065 | Å |
| c | 7.8429 | Å |
| $\alpha$ (alpha) | 90.0000 | ° |
| $\beta$ (beta) | 90.0000 | ° |
| $\gamma$ (gamma) | 120.0000 | ° |

Crystal structure (conventional cell):

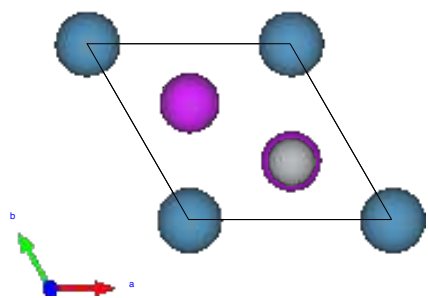

Nanowire Transmission:

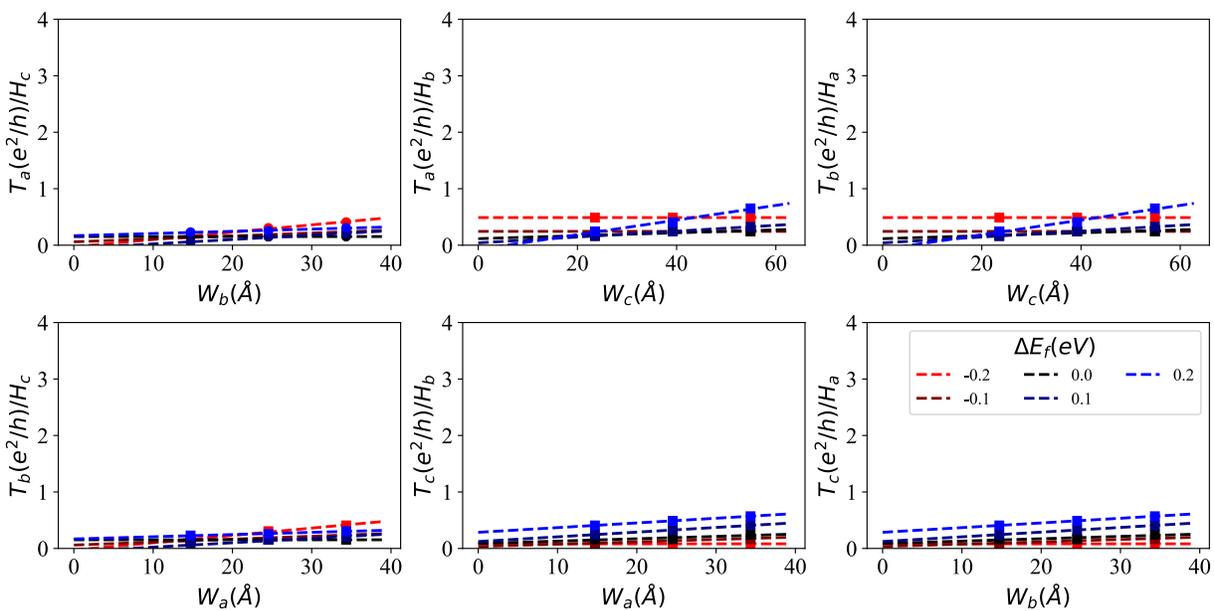

Compound: CaAgP
Materials Project ID: 12277

Lattice (conventional cell):

| Parameter | Value | Unit |
|-----------|---------|------|
| a | 7.0768 | Å |
| b | 7.0768 | Å |
| c | 4.2030 | Å |
| $\alpha$ (alpha) | 90.0000 | ° |
| $\beta$ (beta) | 90.0000 | ° |
| $\gamma$ (gamma) | 120.0000 | ° |

Crystal structure (conventional cell):

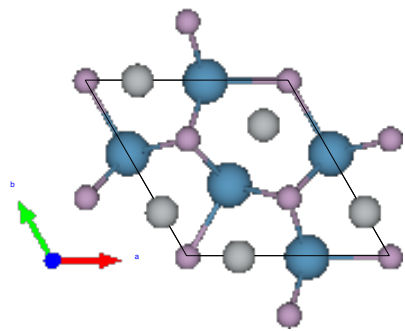

Nanowire Transmission:

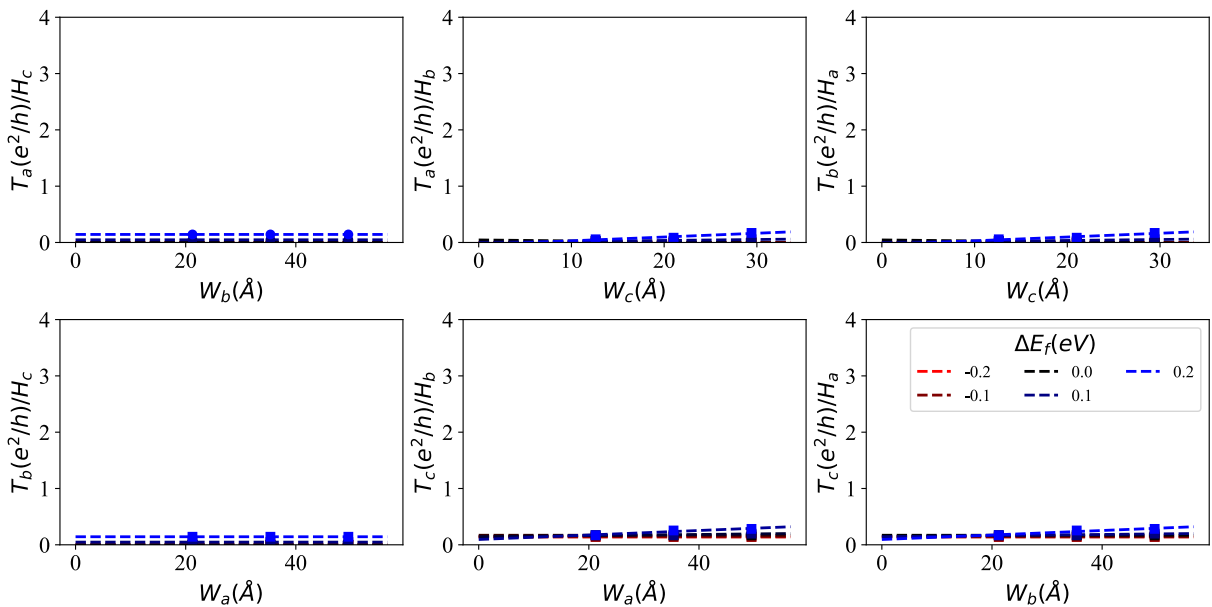

Compound: CaB$_6$
Materials Project ID: 865

Lattice (conventional cell):

| Parameter | Value | Unit |
|---|---|---|
| a | 4.1501 | Å |
| b | 4.1501 | Å |
| c | 4.1501 | Å |
| $\alpha$ (alpha) | 90.0000 | ° |
| $\beta$ (beta) | 90.0000 | ° |
| $\gamma$ (gamma) | 90.0000 | ° |

Crystal structure (conventional cell):

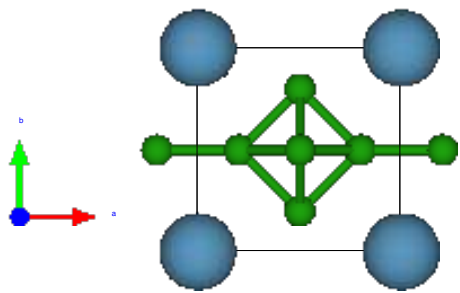

Nanowire Transmission:

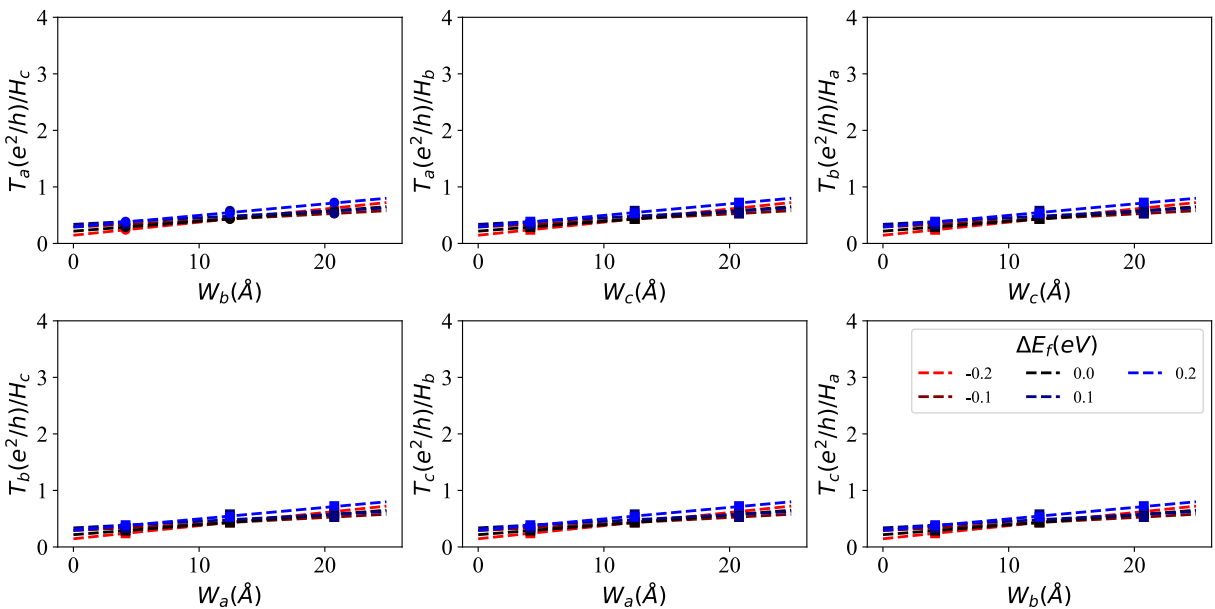

Compound: CaBiAu
Materials Project ID: 1018657

Lattice (conventional cell):

| Parameter | Value | Unit |
|---|---|---|
| a | 4.8612 | Å |
| b | 4.8612 | Å |
| c | 7.9224 | Å |
| $\alpha$ (alpha) | 90.0000 | ° |
| $\beta$ (beta) | 90.0000 | ° |
| $\gamma$ (gamma) | 120.0000 | ° |

Crystal structure (conventional cell):

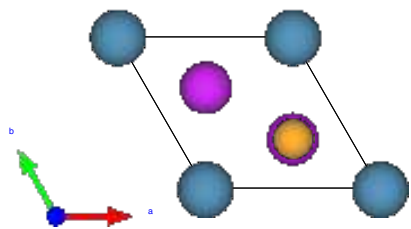

Nanowire Transmission:

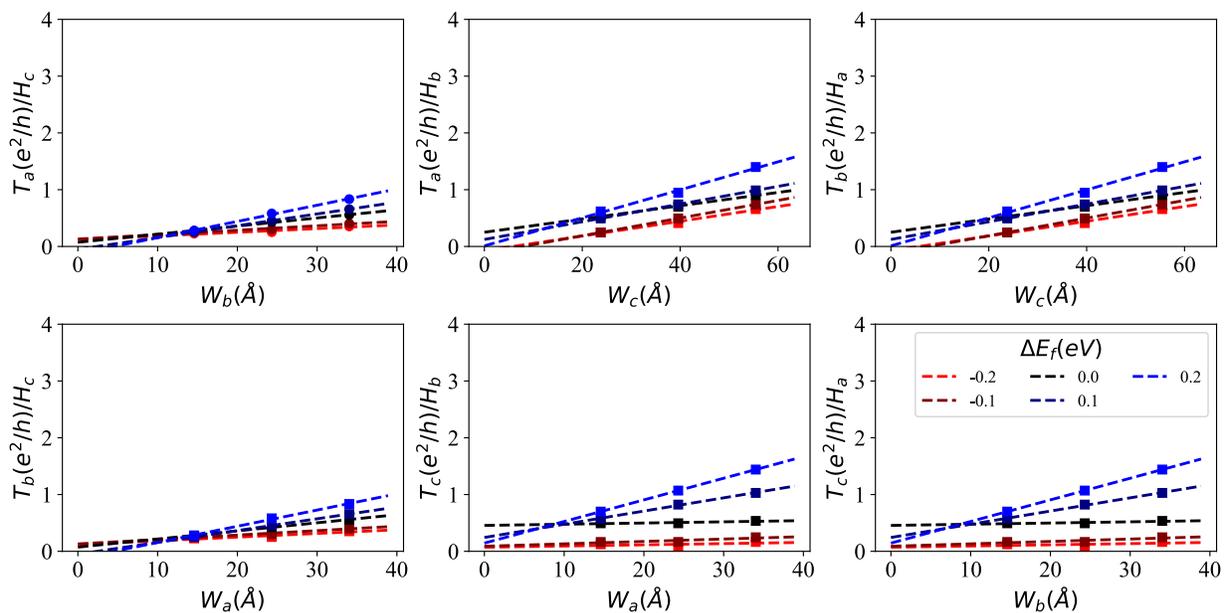

Compound: CaC$_2$
Materials Project ID: 1077545

Lattice (conventional cell):

| Parameter | Value | Unit |
|---|---|---|
| a | 3.7480 | Å |
| b | 8.7387 | Å |
| c | 4.7766 | Å |
| $\alpha$ (alpha) | 90.0000 | ° |
| $\beta$ (beta) | 90.0000 | ° |
| $\gamma$ (gamma) | 90.0000 | ° |

Crystal structure (conventional cell):

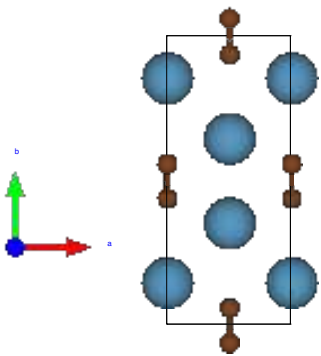

Nanowire Transmission:

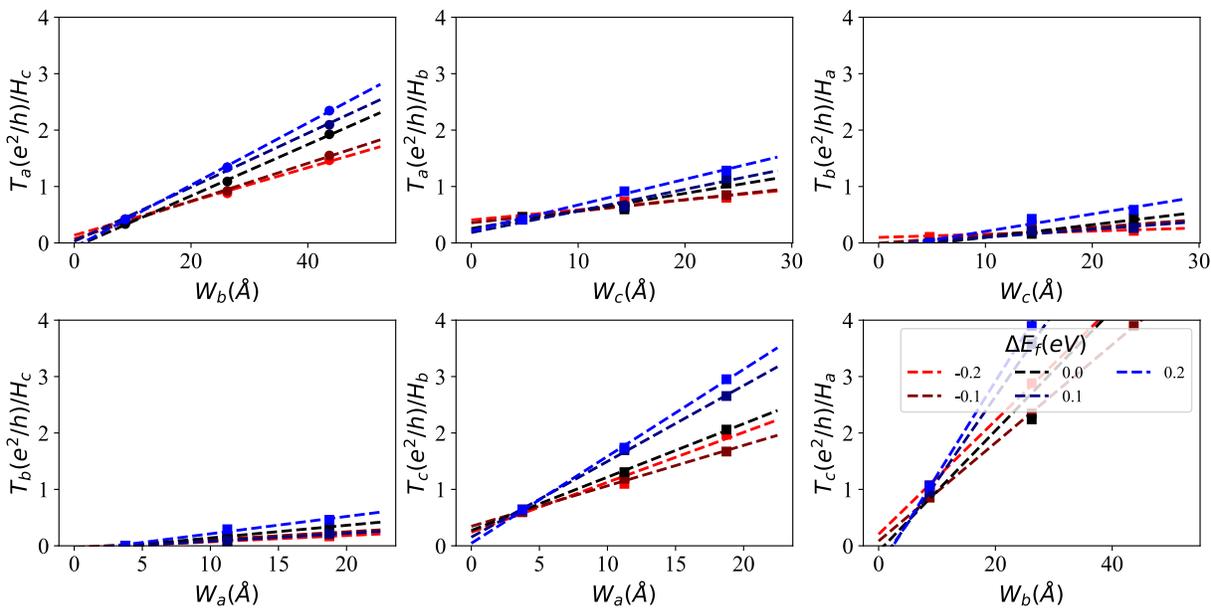

Compound: CaSi$_2$
Materials Project ID: 8372

Lattice (conventional cell):

| Parameter | Value | Unit |
|-----------|---------|------|
| a | 3.8878 | Å |
| b | 3.8878 | Å |
| c | 4.9460 | Å |
| α (alpha) | 90.0000 | ° |
| β (beta) | 90.0000 | ° |
| γ (gamma) | 120.0000 | ° |

Crystal structure (conventional cell):

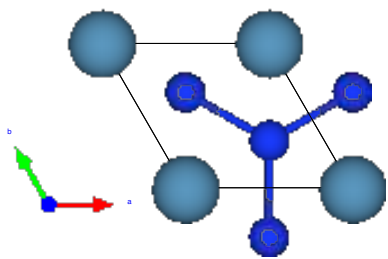

Nanowire Transmission:

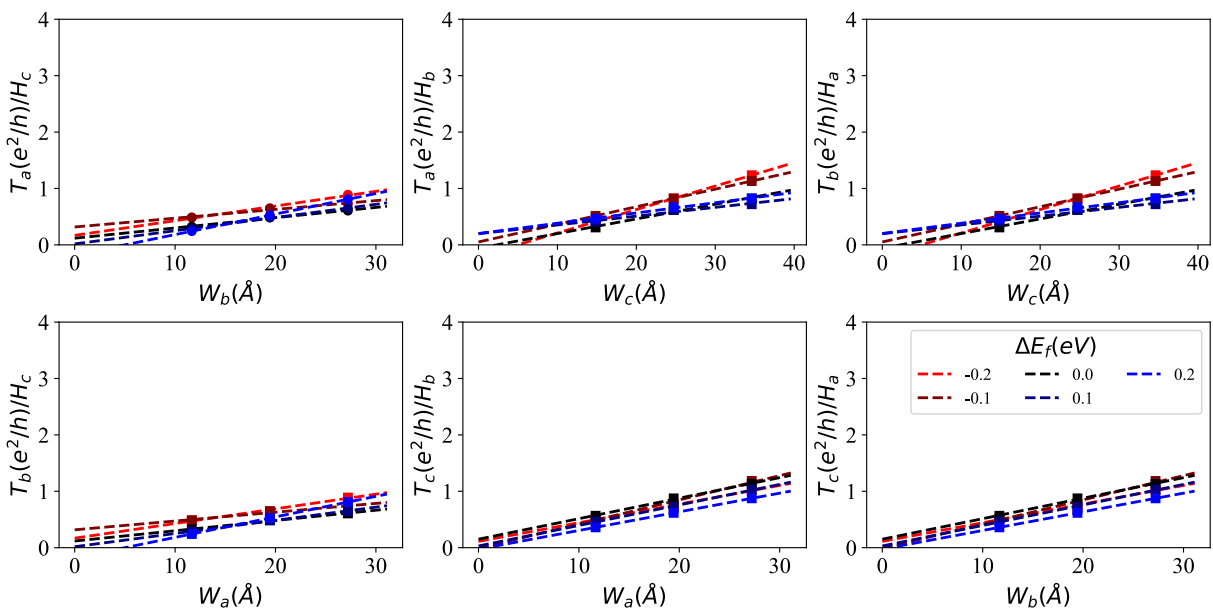

Compound: CaSnHg
Materials Project ID: 1019104

Lattice (conventional cell):

| Parameter | Value | Unit |
|-----------|-------|------|
| a | 4.9042 | Å |
| b | 4.9042 | Å |
| c | 7.8216 | Å |
| $\alpha$ (alpha) | 90.0000 | ° |
| $\beta$ (beta) | 90.0000 | ° |
| $\gamma$ (gamma) | 120.0000 | ° |

Crystal structure (conventional cell):

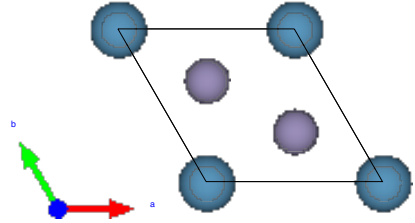

Nanowire Transmission:

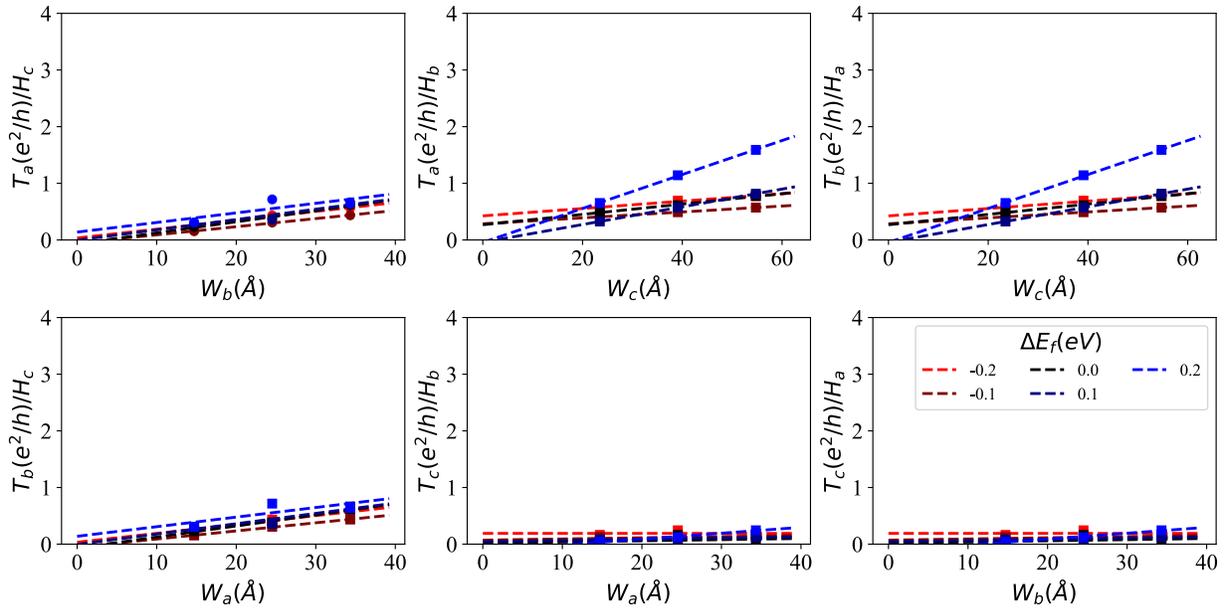

Compound: CaTe
Materials Project ID: 10684

Lattice (conventional cell):

| Parameter | Value | Unit |
|-----------|-------|------|
| a | 3.9261 | Å |
| b | 3.9261 | Å |
| c | 3.9261 | Å |
| α (alpha) | 90.0000 | ° |
| β (beta) | 90.0000 | ° |
| γ (gamma) | 90.0000 | ° |

Crystal structure (conventional cell):

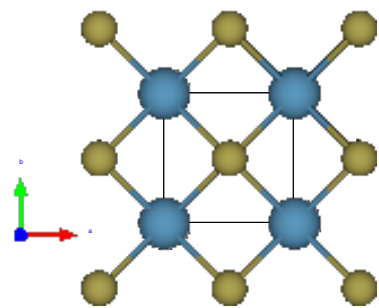

Nanowire Transmission:

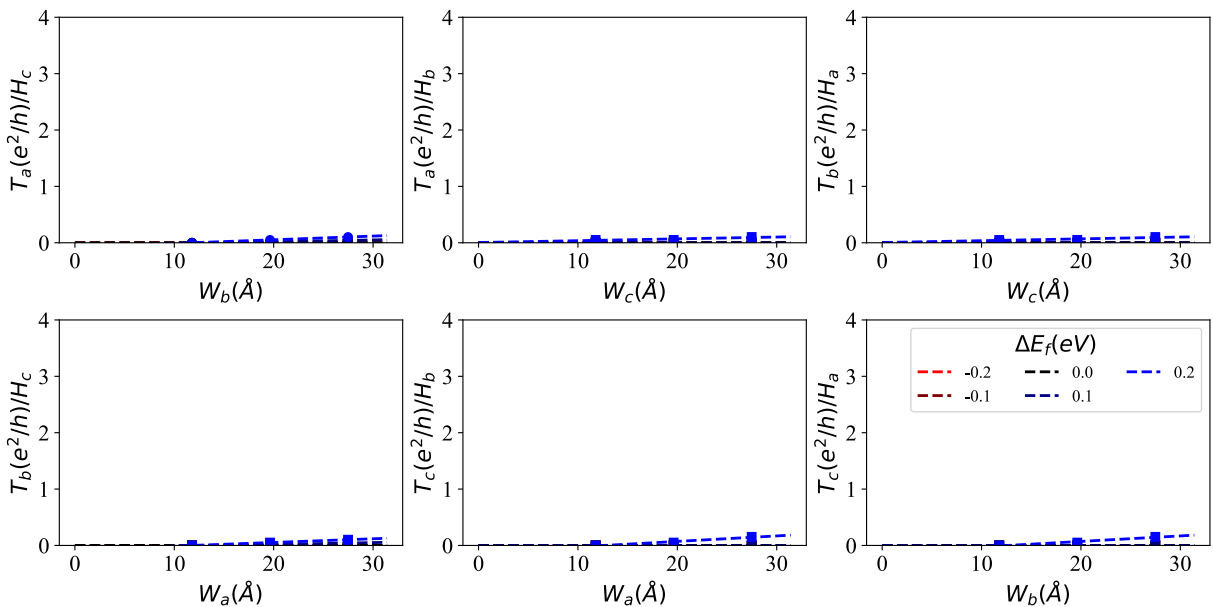

Compound: CaP$_3$
Materials Project ID: 9122

Lattice (conventional cell):

| Parameter | Value | Unit |
|-----------|-------|------|
| a | 5.6095 | Å |
| b | 5.6390 | Å |
| c | 5.6946 | Å |
| α (alpha) | 70.0761 | ° |
| β (beta) | 79.6272 | ° |
| γ (gamma) | 74.6843 | ° |

Crystal structure (conventional cell):

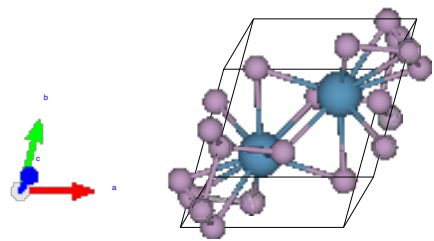

Nanowire Transmission:

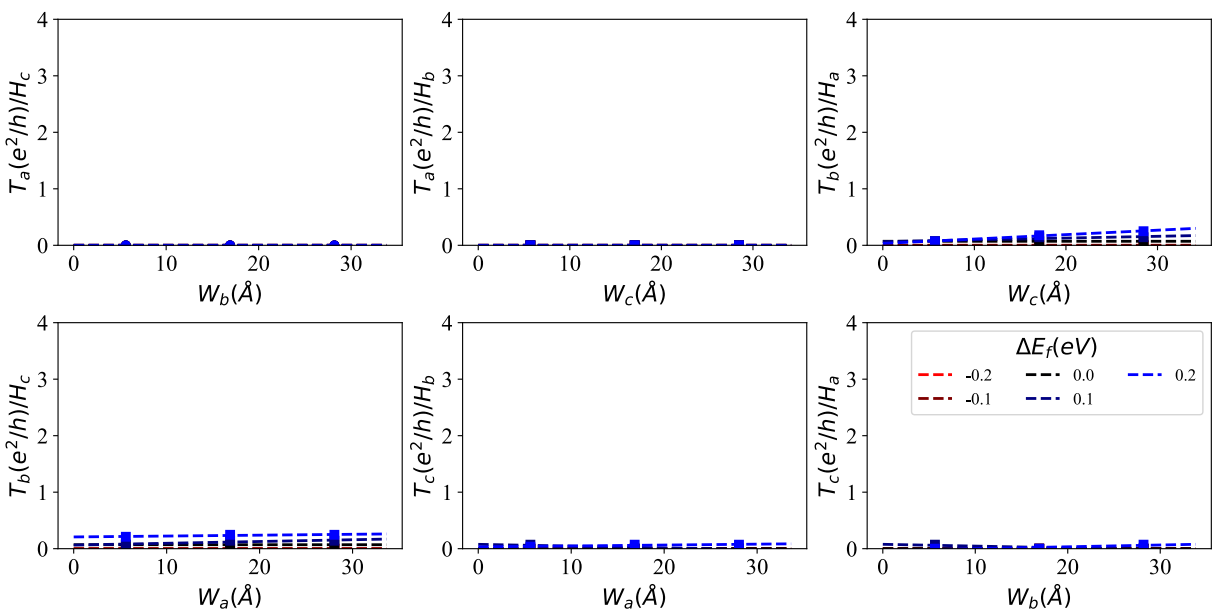

Compound: Co
Materials Project ID: 54

**Lattice (conventional cell):**

| Parameter | Value | Unit |
|-----------|-------|------|
| a | 2.5008 | Å |
| b | 2.5008 | Å |
| c | 4.0333 | Å |
| α (alpha) | 90.0000 | ° |
| β (beta) | 90.0000 | ° |
| γ (gamma) | 120.0000 | ° |

**Crystal structure (conventional cell):**

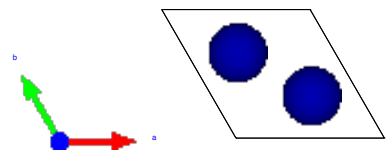

**Nanowire Transmission:**

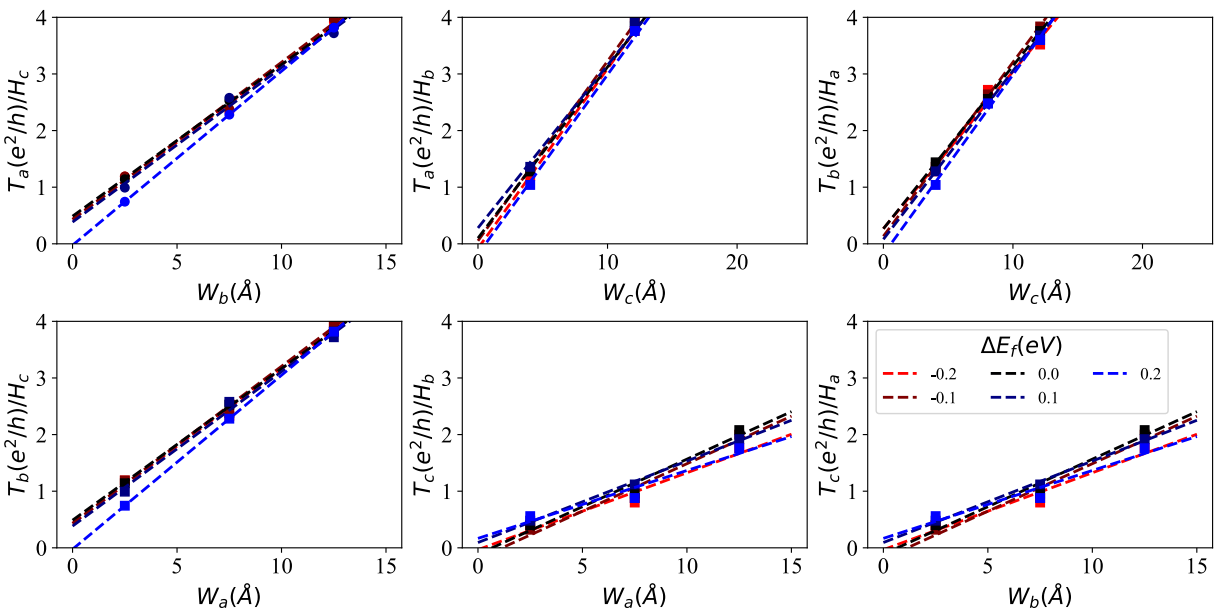

Compound: CoGe
Materials Project ID: 10692

Lattice (conventional cell):

| Parameter | Value | Unit |
|-----------|-------|------|
| a | 4.6370 | Å |
| b | 4.6370 | Å |
| c | 4.6370 | Å |
| α (alpha) | 90.0000 | ° |
| β (beta) | 90.0000 | ° |
| γ (gamma) | 90.0000 | ° |

Crystal structure (conventional cell):

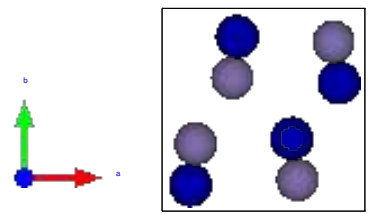

Nanowire Transmission:

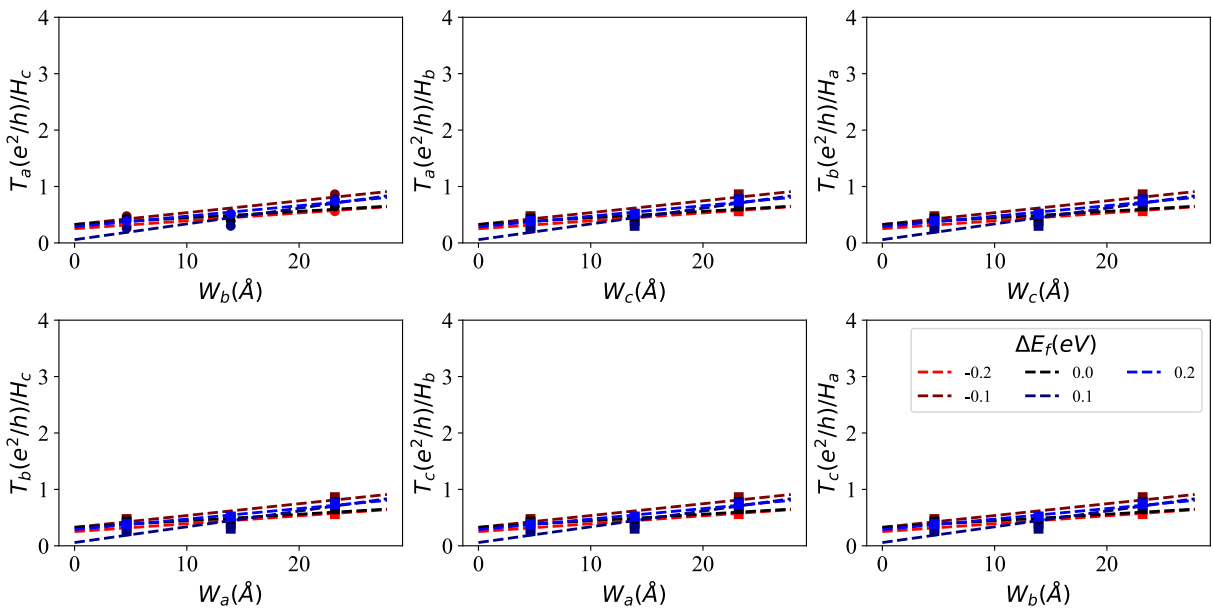

Compound: CoSi
Materials Project ID: 7577

## Lattice (conventional cell):

| Parameter | Value | Unit |
|-----------|-------|------|
| a | 4.4332 | Å |
| b | 4.4332 | Å |
| c | 4.4332 | Å |
| α (alpha) | 90.0000 | ° |
| β (beta) | 90.0000 | ° |
| γ (gamma) | 90.0000 | ° |

## Crystal structure (conventional cell):

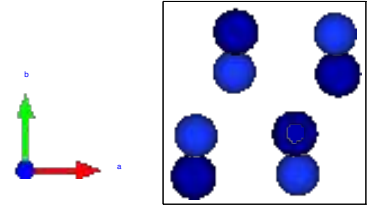

## Nanowire Transmission:

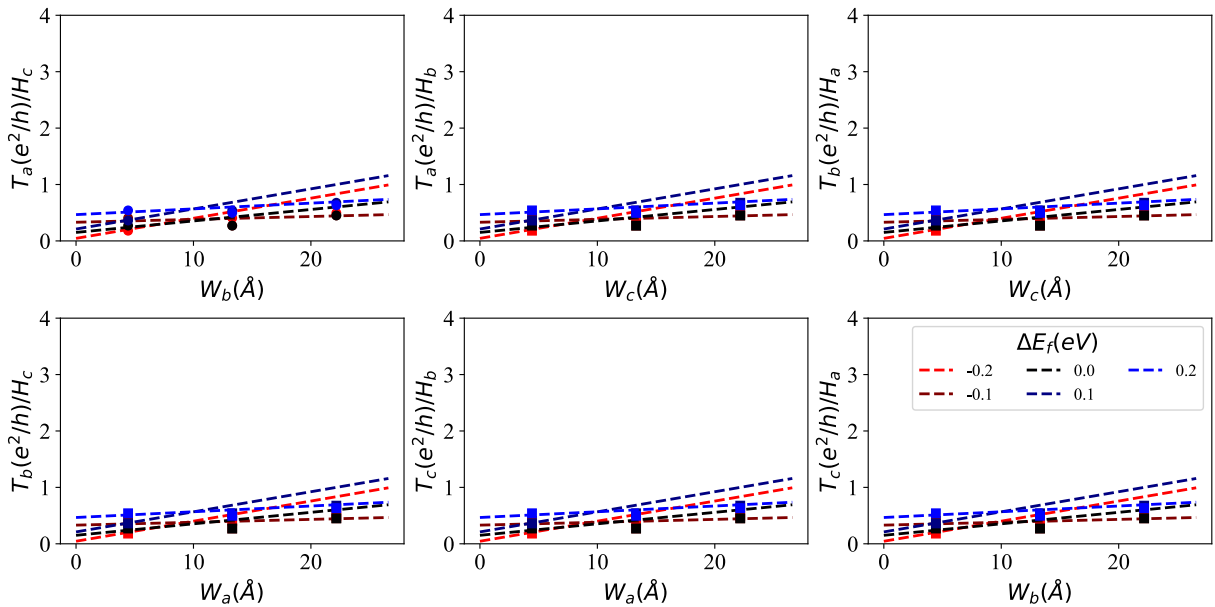

Compound: CoPtO$_2$
Materials Project ID: 19210

Lattice (conventional cell):

| Parameter | Value | Unit |
|---|---|---|
| a | 2.8847 | Å |
| b | 2.8847 | Å |
| c | 18.2033 | Å |
| α (alpha) | 90.0000 | ° |
| β (beta) | 90.0000 | ° |
| γ (gamma) | 120.0000 | ° |

Crystal structure (conventional cell):

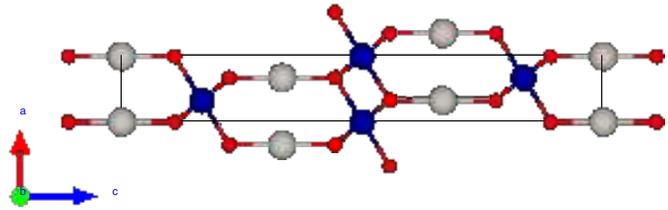

Nanowire Transmission:

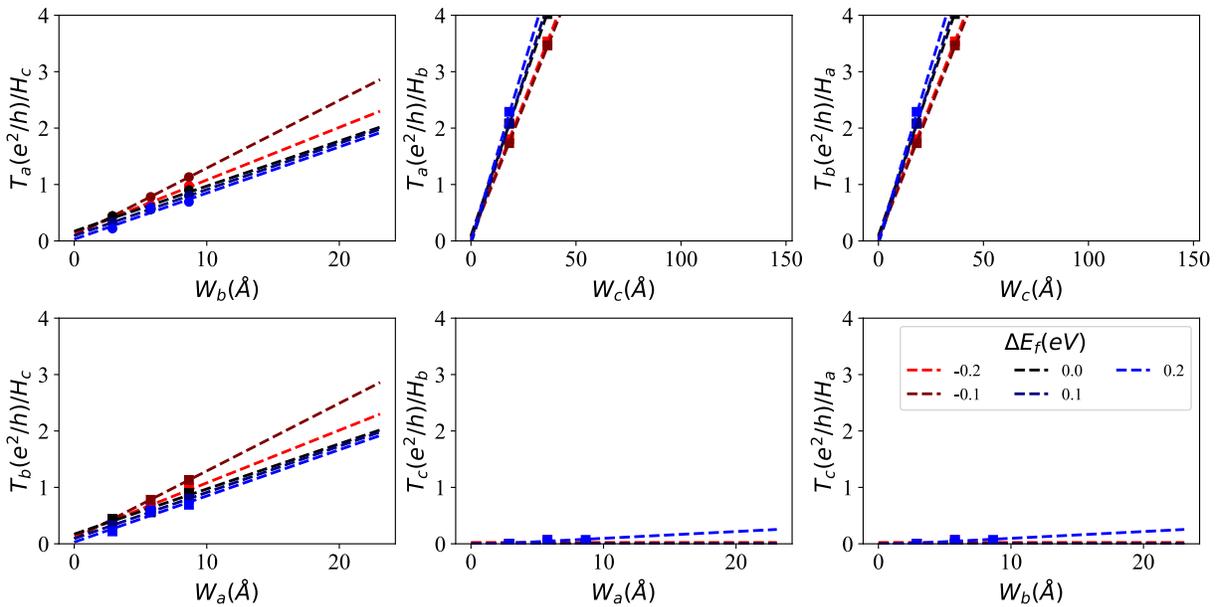

Compound: Cu
Materials Project ID: 30

**Lattice (conventional cell):**

| Parameter | Value | Unit |
|-----------|-------|------|
| a | 3.6213 | Å |
| b | 3.6213 | Å |
| c | 3.6213 | Å |
| α (alpha) | 90.0000 | ° |
| β (beta) | 90.0000 | ° |
| γ (gamma) | 90.0000 | ° |

**Crystal structure (conventional cell):**

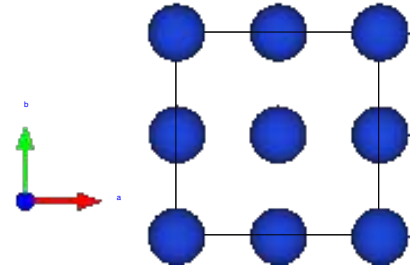

**Nanowire Transmission:**

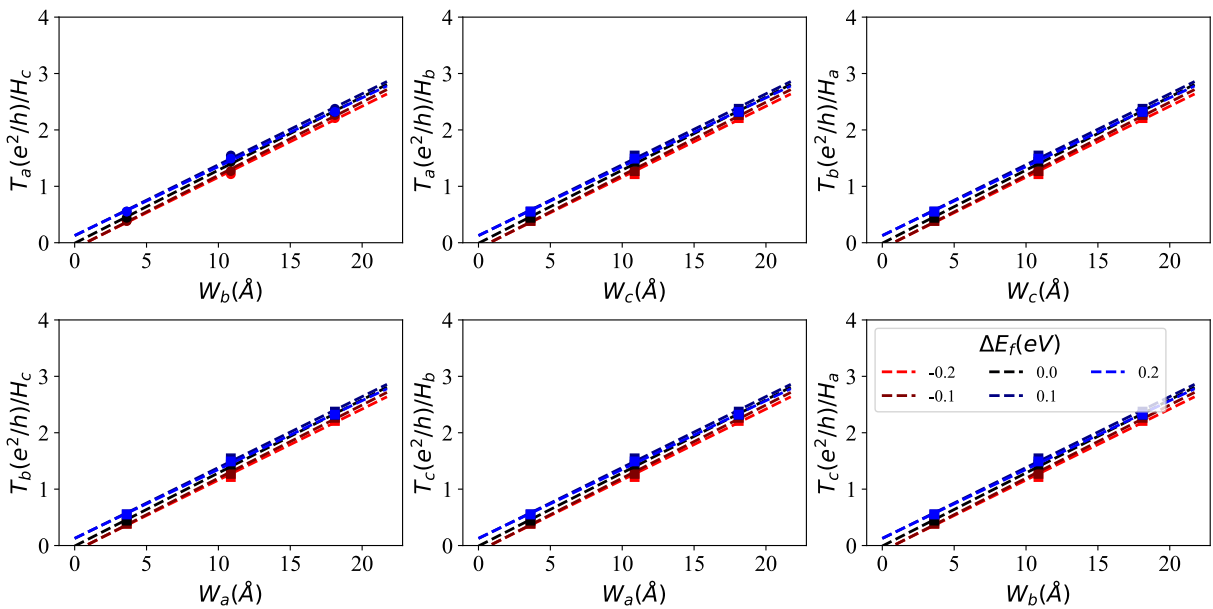

Compound: Cu₃AsSe₄
Materials Project ID: 675626

Lattice (conventional cell):

Crystal structure (conventional cell):

| Parameter | Value | Unit |
|-----------|-------|------|
| a | 5.6212 | Å |
| b | 5.6212 | Å |
| c | 11.1400 | Å |
| α (alpha) | 90.0000 | ° |
| β (beta) | 90.0000 | ° |
| γ (gamma) | 90.0000 | ° |

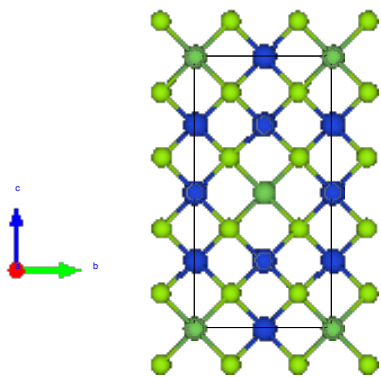

Nanowire Transmission:

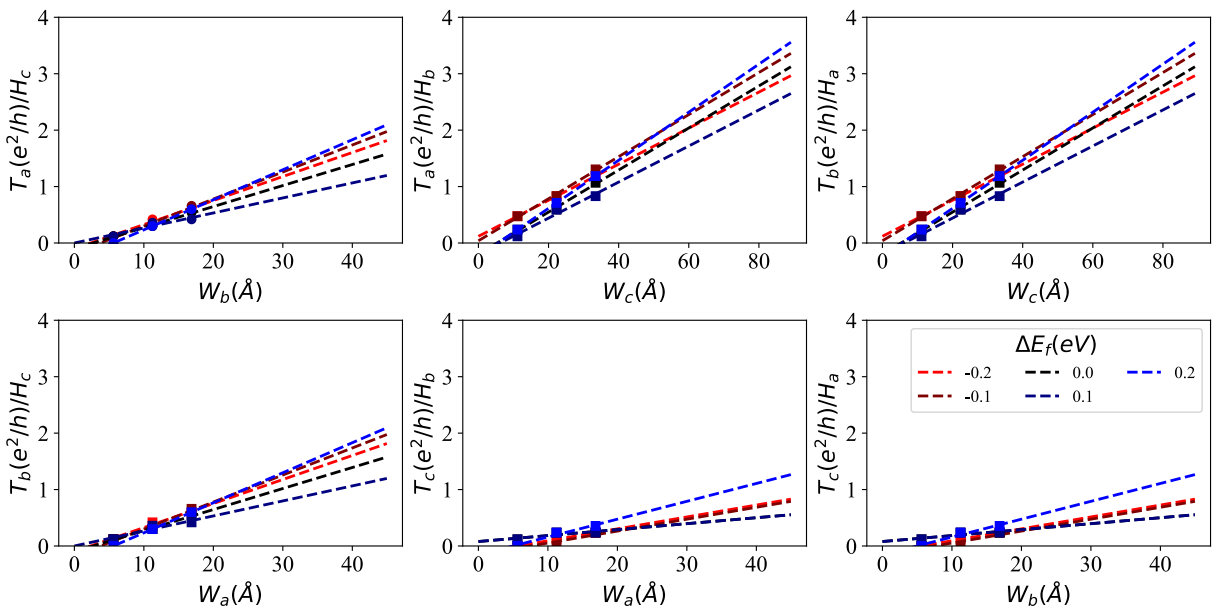

Compound: GaPd
Materials Project ID: 1078526

Lattice (conventional cell):

| Parameter | Value | Unit |
|-----------|-------|------|
| a | 4.9634 | Å |
| b | 4.9634 | Å |
| c | 4.9634 | Å |
| α (alpha) | 90.0000 | ° |
| β (beta) | 90.0000 | ° |
| γ (gamma) | 90.0000 | ° |

Crystal structure (conventional cell):

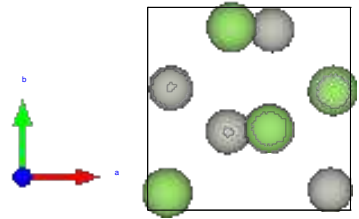

Nanowire Transmission:

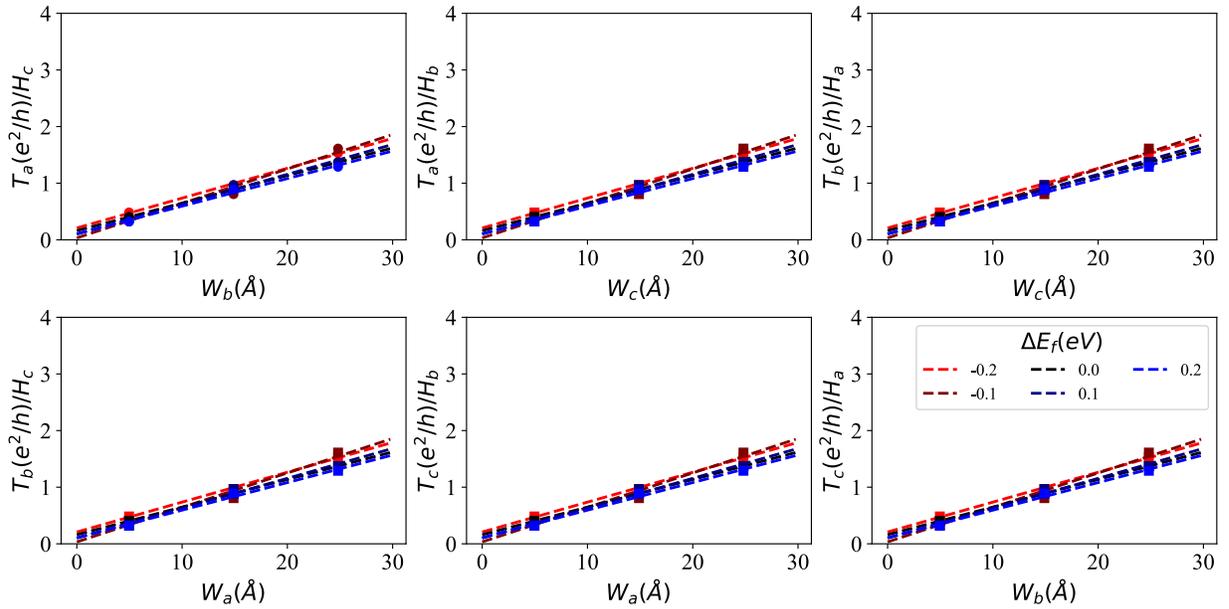

Compound: GaPt
Materials Project ID: 1025551

Lattice (conventional cell):

| Parameter | Value | Unit |
|---|---|---|
| a | 4.9729 | Å |
| b | 4.9729 | Å |
| c | 4.9729 | Å |
| $\alpha$ (alpha) | 90.0000 | ° |
| $\beta$ (beta) | 90.0000 | ° |
| $\gamma$ (gamma) | 90.0000 | ° |

Crystal structure (conventional cell):

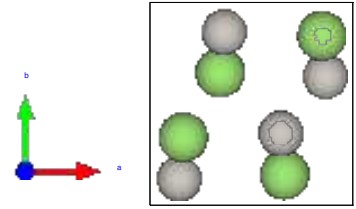

Nanowire Transmission:

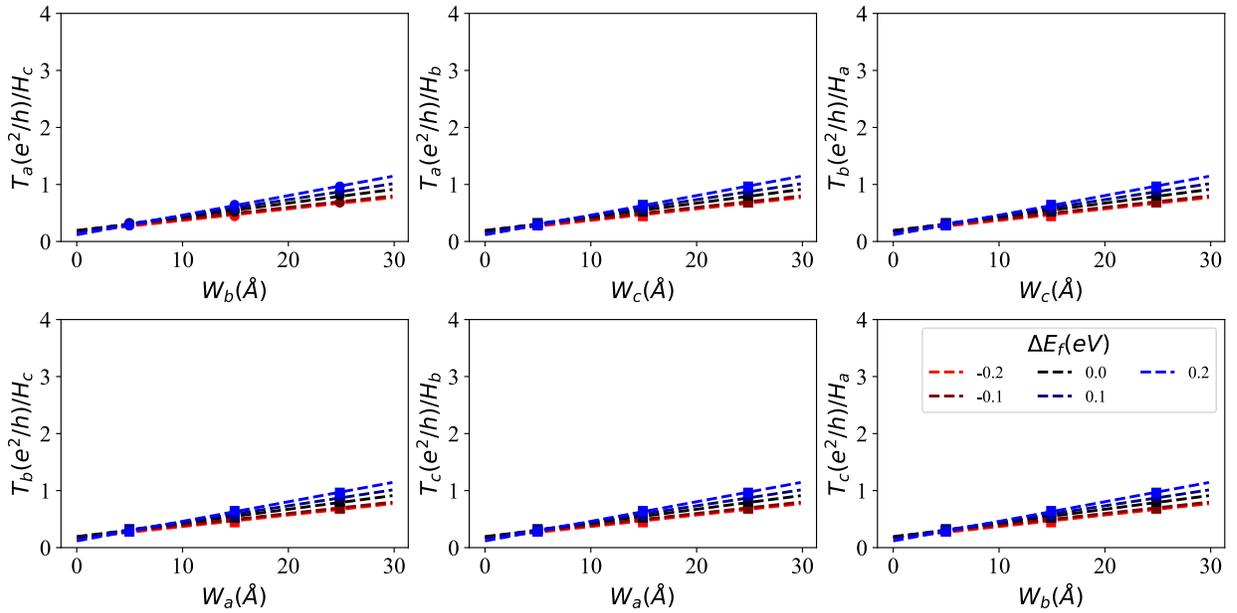

Compound: GaGeTe
Materials Project ID: 8211

Lattice (conventional cell):

| Parameter | Value | Unit |
|-----------|---------|------|
| a | 4.1264 | Å |
| b | 4.1264 | Å |
| c | 36.5601 | Å |
| $\alpha$ (alpha) | 90.0000 | ° |
| $\beta$ (beta) | 90.0000 | ° |
| $\gamma$ (gamma) | 120.0000 | ° |

Crystal structure (conventional cell):

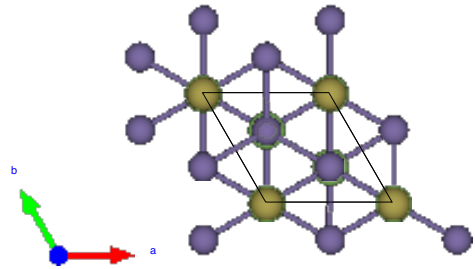

Nanowire Transmission:

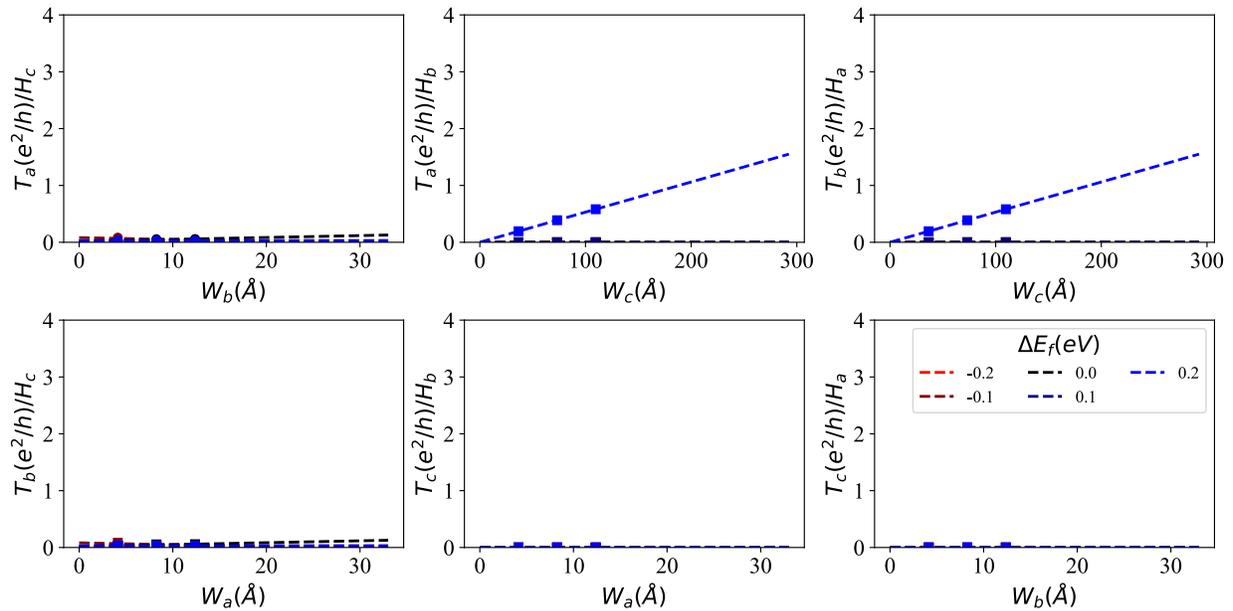

Compound: GeRh
Materials Project ID: 20866

Lattice (conventional cell):

| Parameter | Value | Unit |
|-----------|-------|------|
| a | 4.8620 | Å |
| b | 4.8620 | Å |
| c | 4.8620 | Å |
| α (alpha) | 90.0000 | ° |
| β (beta) | 90.0000 | ° |
| γ (gamma) | 90.0000 | ° |

Crystal structure (conventional cell):

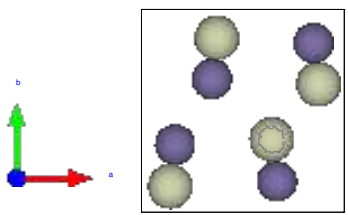

Nanowire Transmission:

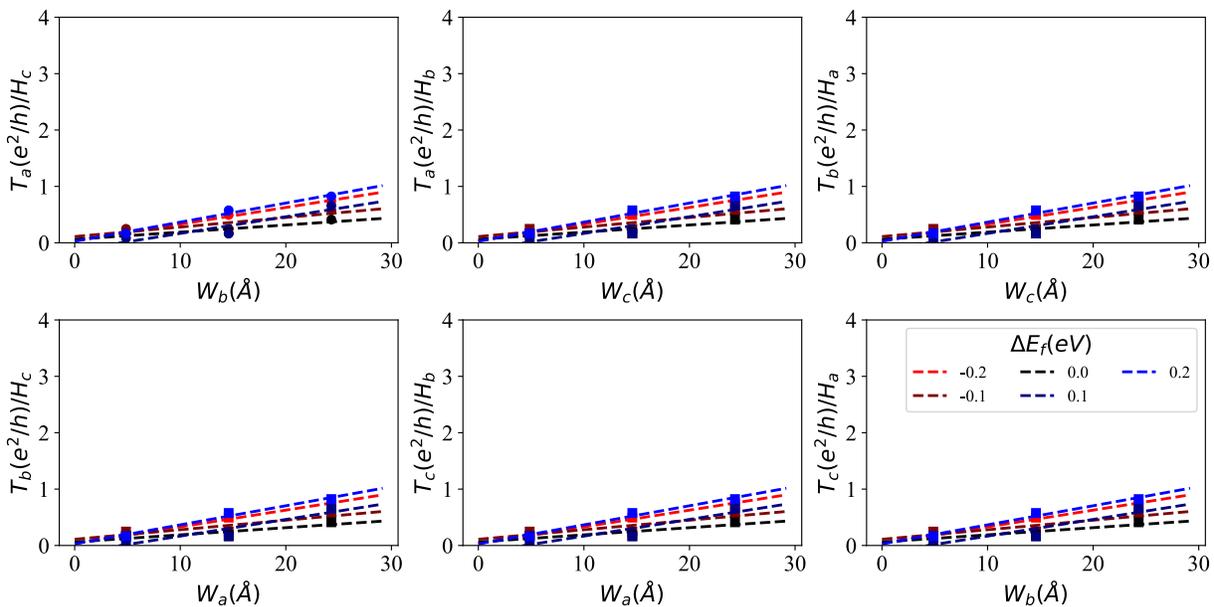

Compound: HfB$_2$
Materials Project ID: 1994

Lattice (conventional cell):

| Parameter | Value | Unit |
|---|---|---|
| a | 3.1487 | Å |
| b | 3.1487 | Å |
| c | 3.4799 | Å |
| α (alpha) | 90.0000 | ° |
| β (beta) | 90.0000 | ° |
| γ (gamma) | 120.0000 | ° |

Crystal structure (conventional cell):

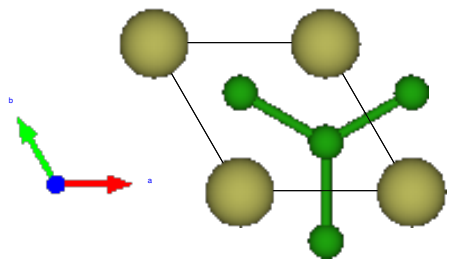

Nanowire Transmission:

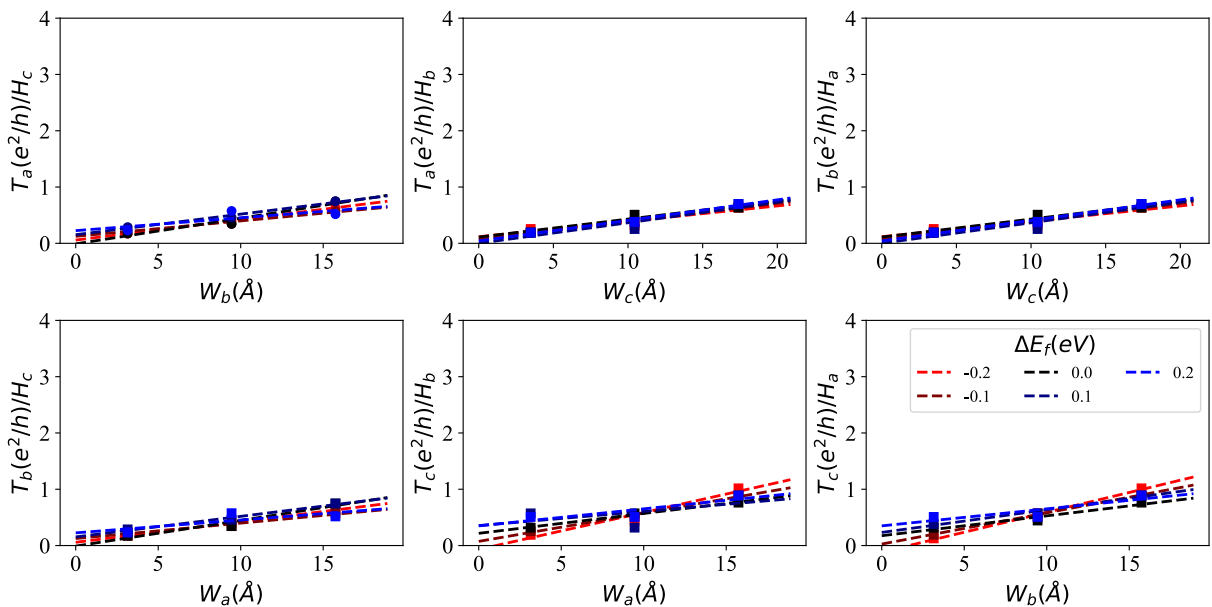

Compound: HfCuP
Materials Project ID: 569933

Lattice (conventional cell):

| Parameter | Value | Unit |
|-----------|---------|------|
| a | 3.5872 | Å |
| b | 3.5872 | Å |
| c | 5.5801 | Å |
| α (alpha) | 90.0000 | ° |
| β (beta) | 90.0000 | ° |
| γ (gamma) | 120.0000 | ° |

Crystal structure (conventional cell):

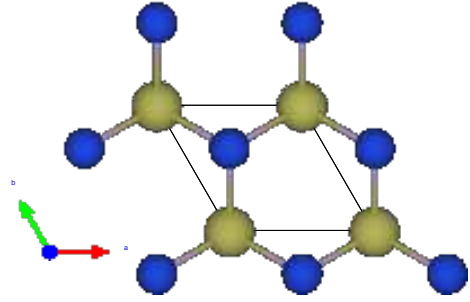

Nanowire Transmission:

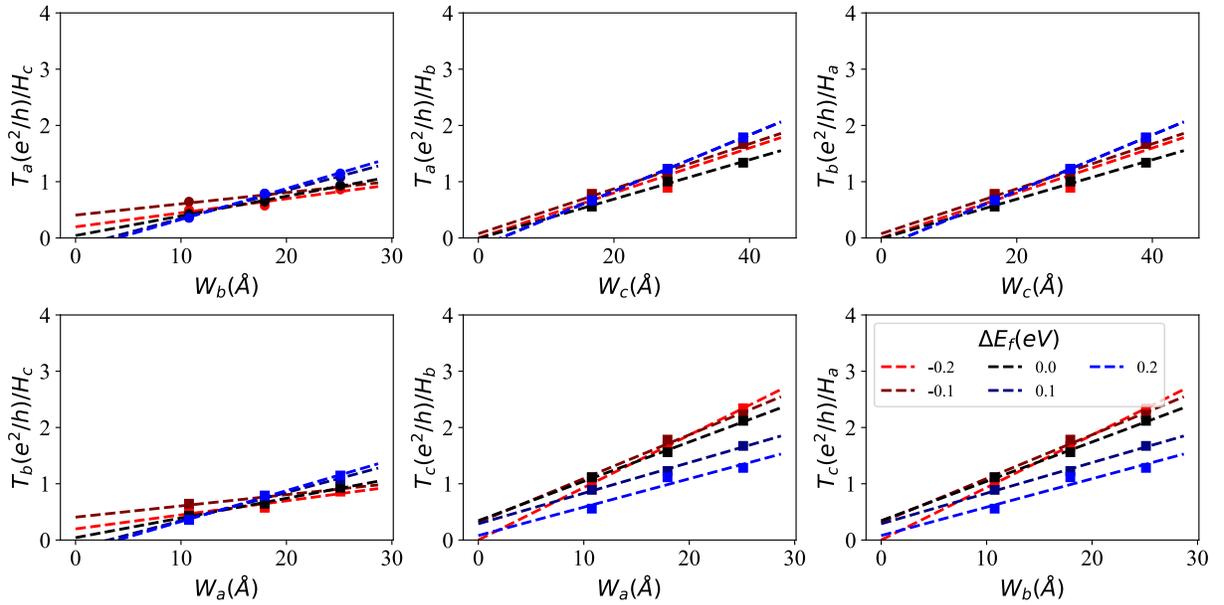

Compound: InGaSe$_2$
Materials Project ID: 1095059

Lattice (conventional cell):

| Parameter | Value | Unit |
|---|---|---|
| a | 8.1411 | Å |
| b | 8.1411 | Å |
| c | 6.4253 | Å |
| α (alpha) | 90.0000 | ° |
| β (beta) | 90.0000 | ° |
| γ (gamma) | 90.0000 | ° |

Crystal structure (conventional cell):

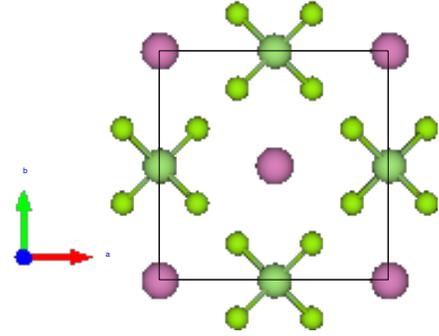

Nanowire Transmission:

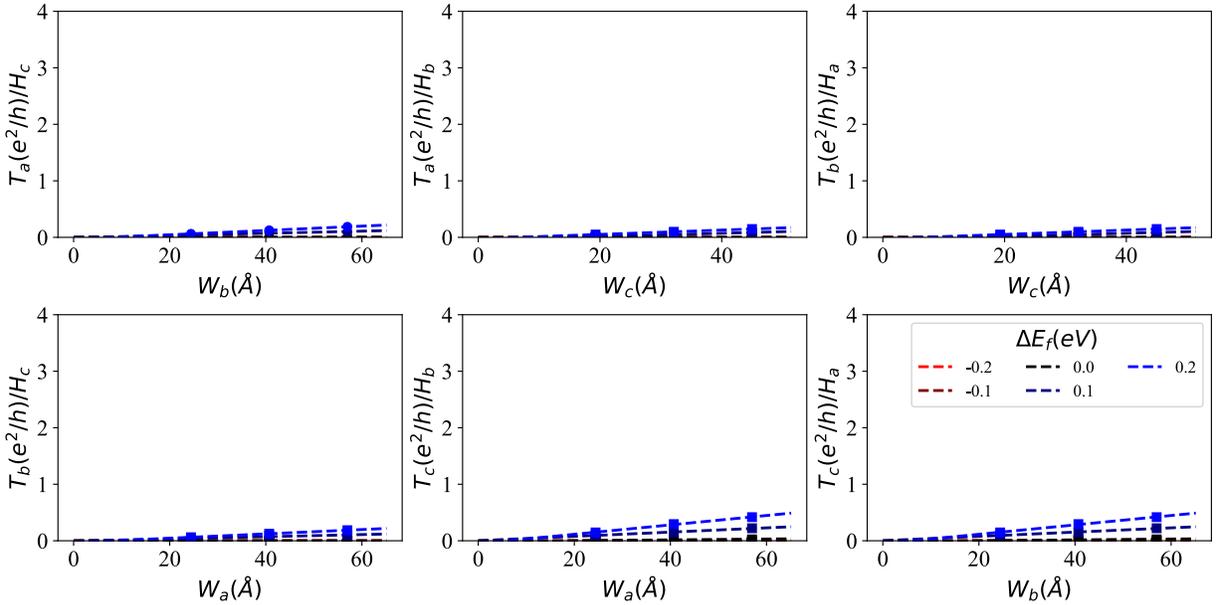

Compound: InTe
Materials Project ID: 20320

Lattice (conventional cell):

| Parameter | Value | Unit |
|-----------|-------|------|
| a | 8.6274 | Å |
| b | 8.6274 | Å |
| c | 7.2267 | Å |
| α (alpha) | 90.0000 | ° |
| β (beta) | 90.0000 | ° |
| γ (gamma) | 90.0000 | ° |

Crystal structure (conventional cell):

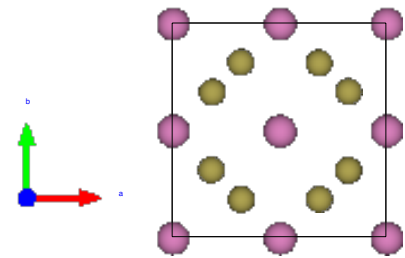

Nanowire Transmission:

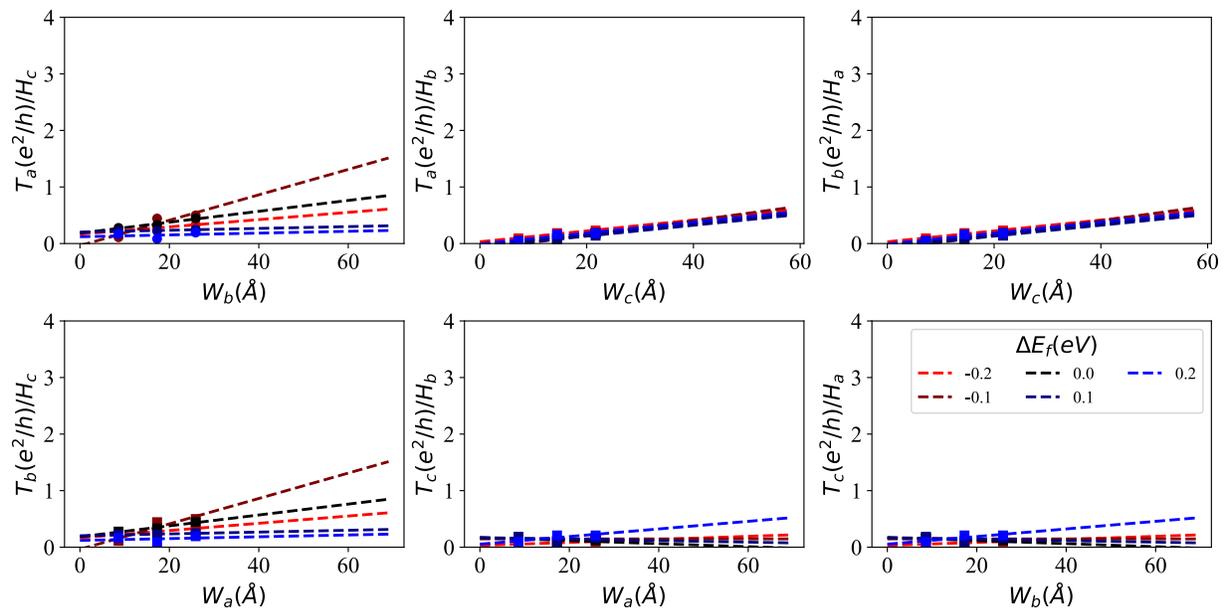

Compound: Ir
Materials Project ID: 101

## Lattice (conventional cell):

| Parameter | Value | Unit |
|-----------|---------|------|
| a | 3.8757 | Å |
| b | 3.8757 | Å |
| c | 3.8757 | Å |
| α (alpha) | 90.0000 | ° |
| β (beta) | 90.0000 | ° |
| γ (gamma) | 90.0000 | ° |

## Crystal structure (conventional cell):

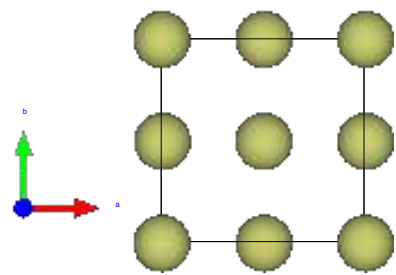

## Nanowire Transmission:

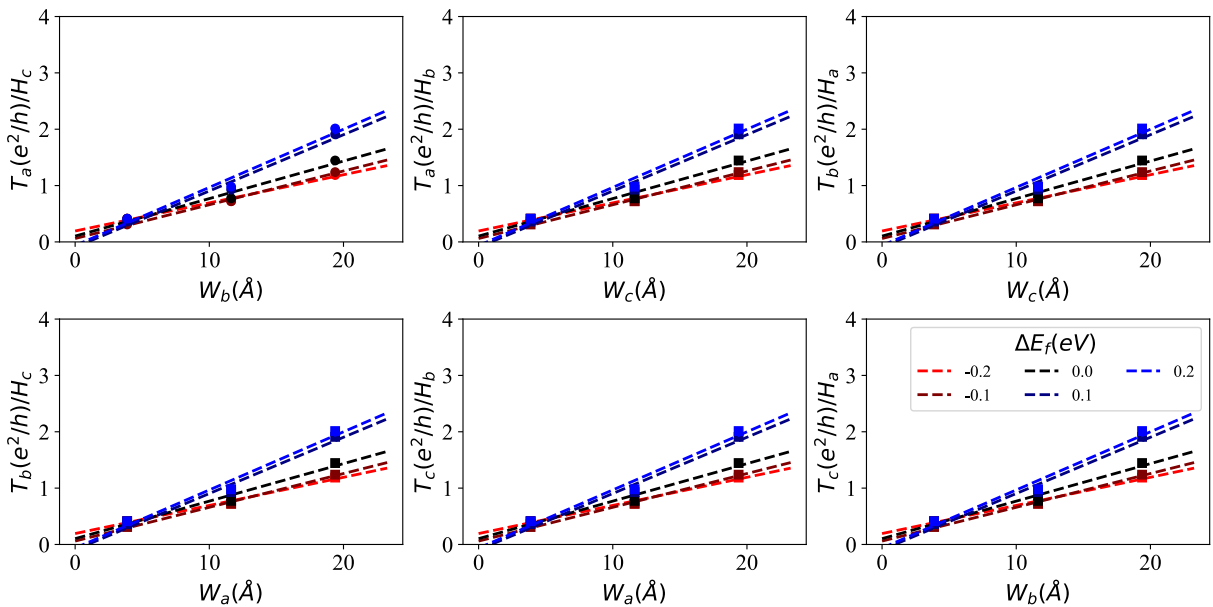

Compound: LaBiPt
Materials Project ID: 1018136

## Lattice (conventional cell):

| Parameter | Value | Unit |
|-----------|-------|------|
| a | 6.9689 | Å |
| b | 6.9689 | Å |
| c | 6.9689 | Å |
| α (alpha) | 90.0000 | ° |
| β (beta) | 90.0000 | ° |
| γ (gamma) | 90.0000 | ° |

## Crystal structure (conventional cell):

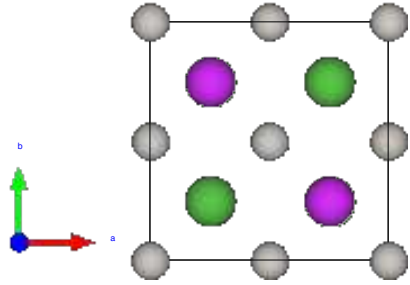

## Nanowire Transmission:

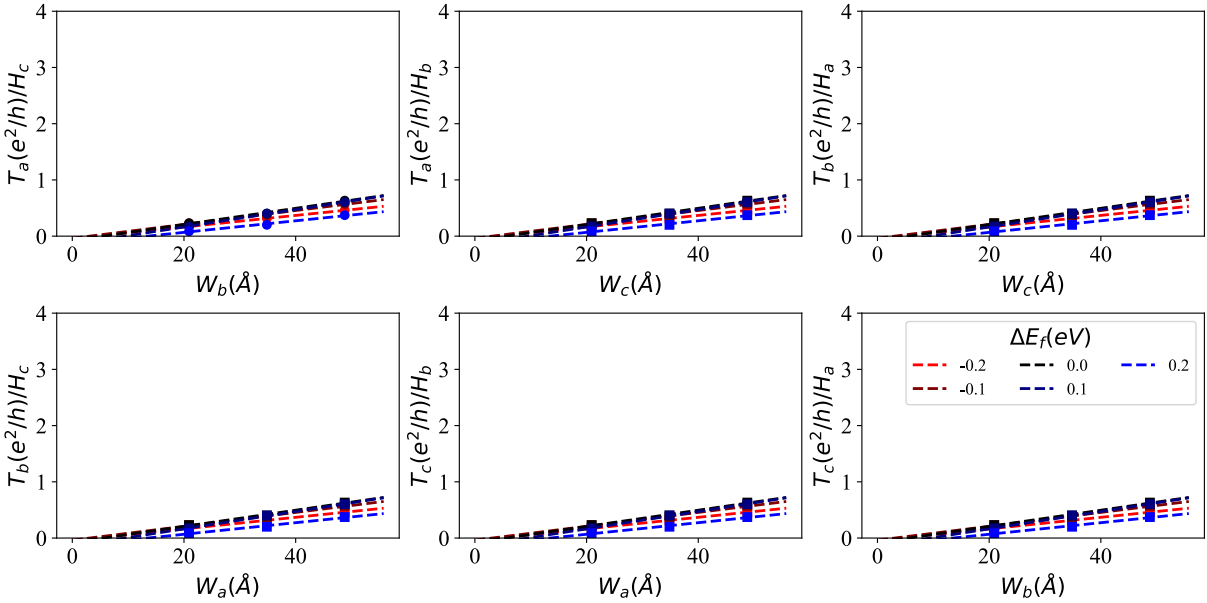

Compound: LaPIr
Materials Project ID: 12919

Lattice (conventional cell):

| Parameter | Value | Unit |
|-----------|---------|------|
| a | 4.1709 | Å |
| b | 4.1709 | Å |
| c | 14.3855 | Å |
| α (alpha) | 90.0000 | ° |
| β (beta) | 90.0000 | ° |
| γ (gamma) | 90.0000 | ° |

Crystal structure (conventional cell):

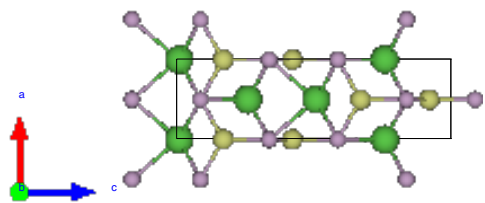

Nanowire Transmission:

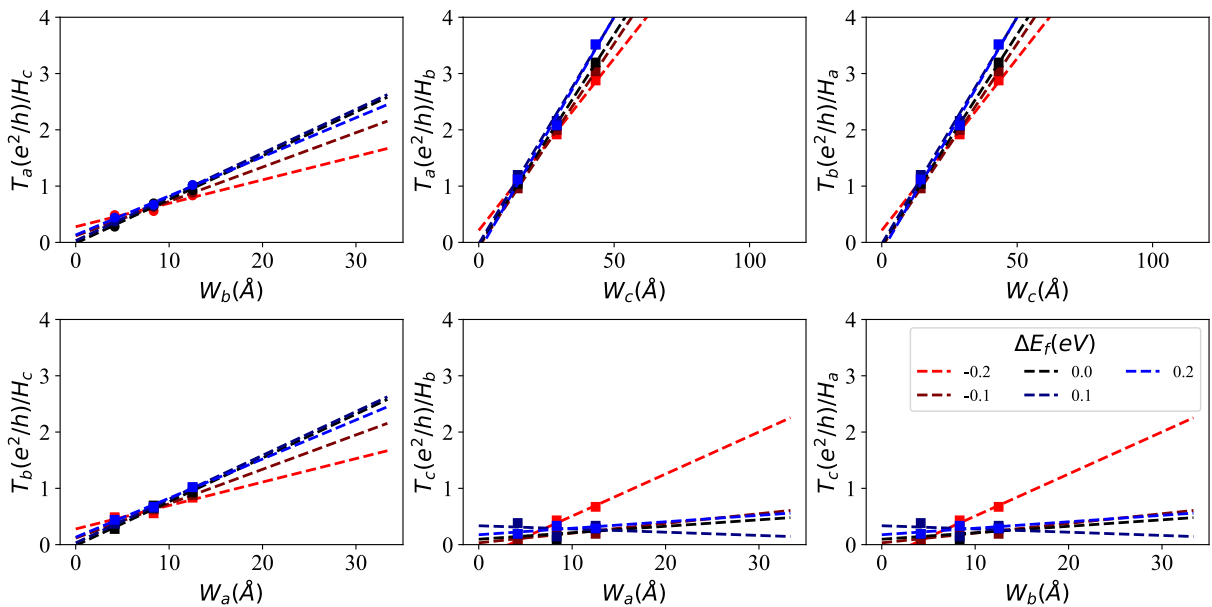

Compound: LiClO$_2$
Materials Project ID: 31367

## Lattice (conventional cell):

| Parameter | Value | Unit |
|---|---|---|
| a | 4.8313 | Å |
| b | 4.8313 | Å |
| c | 10.6884 | Å |
| α (alpha) | 90.0000 | ° |
| β (beta) | 90.0000 | ° |
| γ (gamma) | 90.0000 | ° |

## Crystal structure (conventional cell):

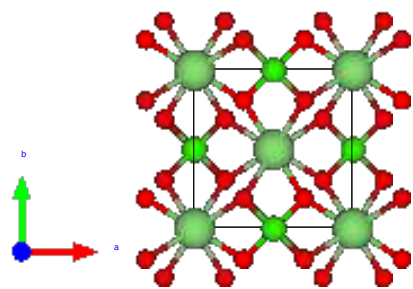

## Nanowire Transmission:

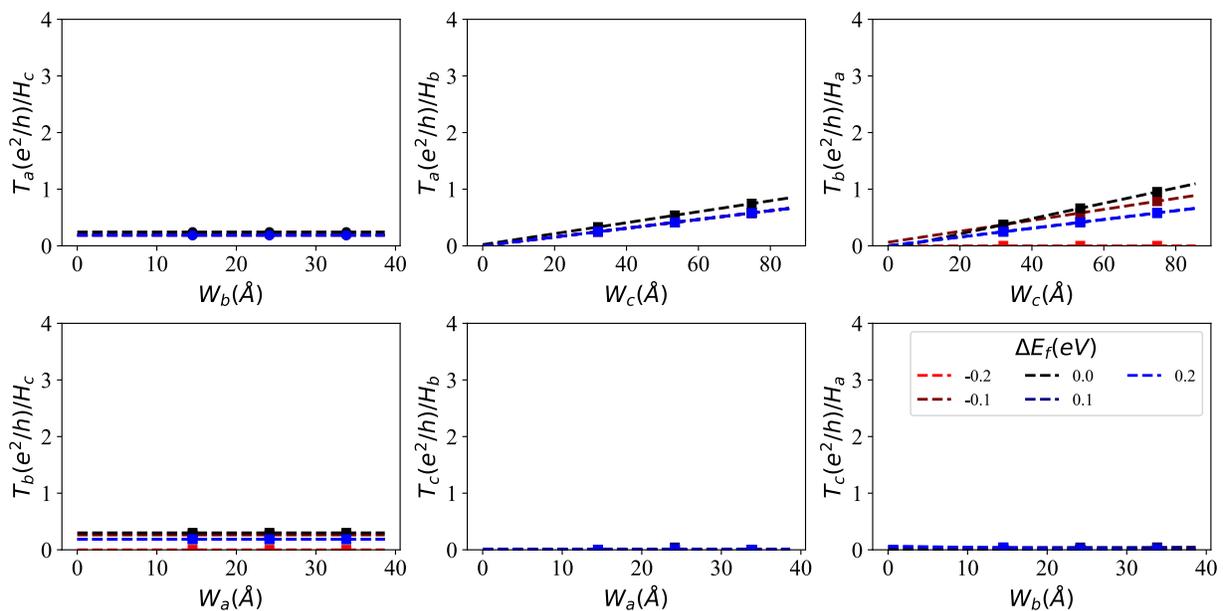

Compound: LiHO
Materials Project ID: 23856

Lattice (conventional cell):

| Parameter | Value | Unit |
|-----------|-------|------|
| a | 3.5914 | Å |
| b | 3.5914 | Å |
| c | 4.4105 | Å |
| α (alpha) | 90.0000 | ° |
| β (beta) | 90.0000 | ° |
| γ (gamma) | 90.0000 | ° |

Crystal structure (conventional cell):

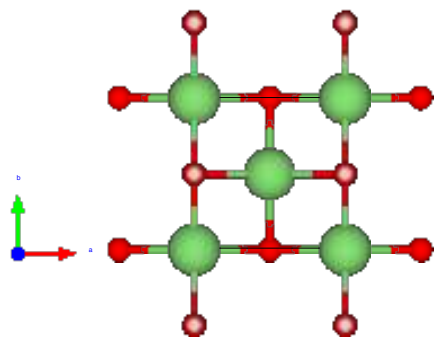

Nanowire Transmission:

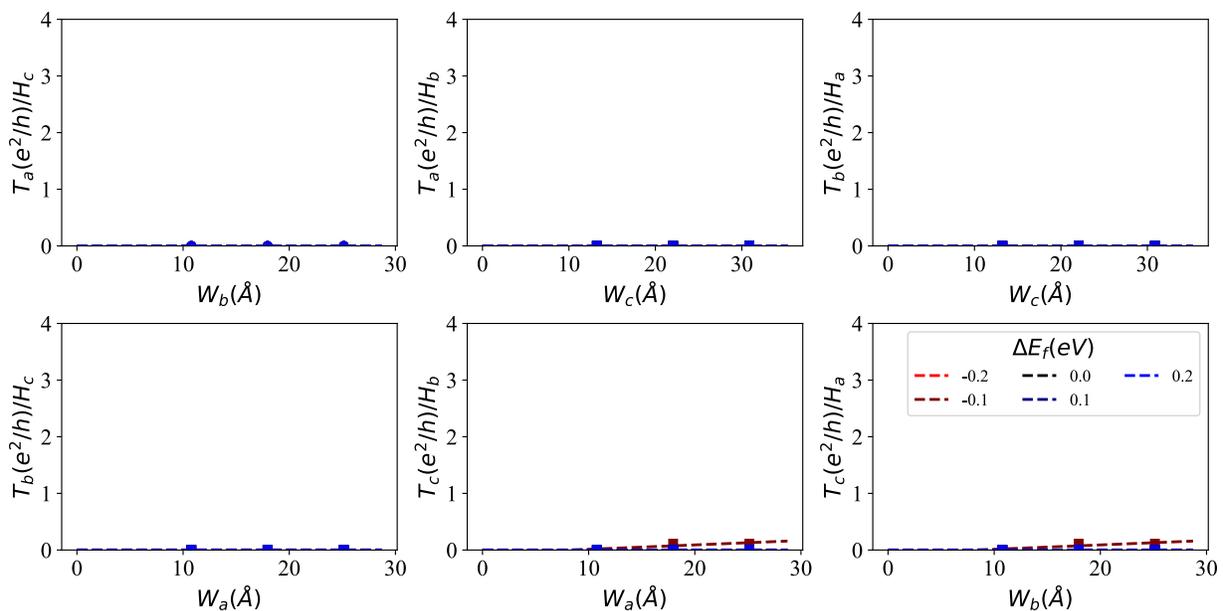

Compound: LiYGe
Materials Project ID: 14209

Lattice (conventional cell):

| Parameter | Value | Unit |
|-----------|---------|------|
| a | 7.0924 | Å |
| b | 7.0924 | Å |
| c | 4.2591 | Å |
| $\alpha$ (alpha) | 90.0000 | ° |
| $\beta$ (beta) | 90.0000 | ° |
| $\gamma$ (gamma) | 120.0000 | ° |

Crystal structure (conventional cell):

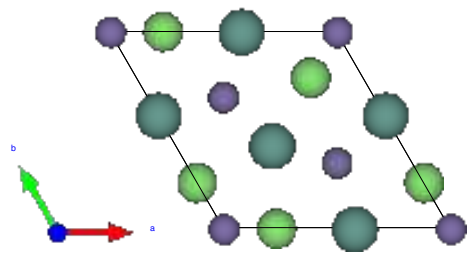

Nanowire Transmission:

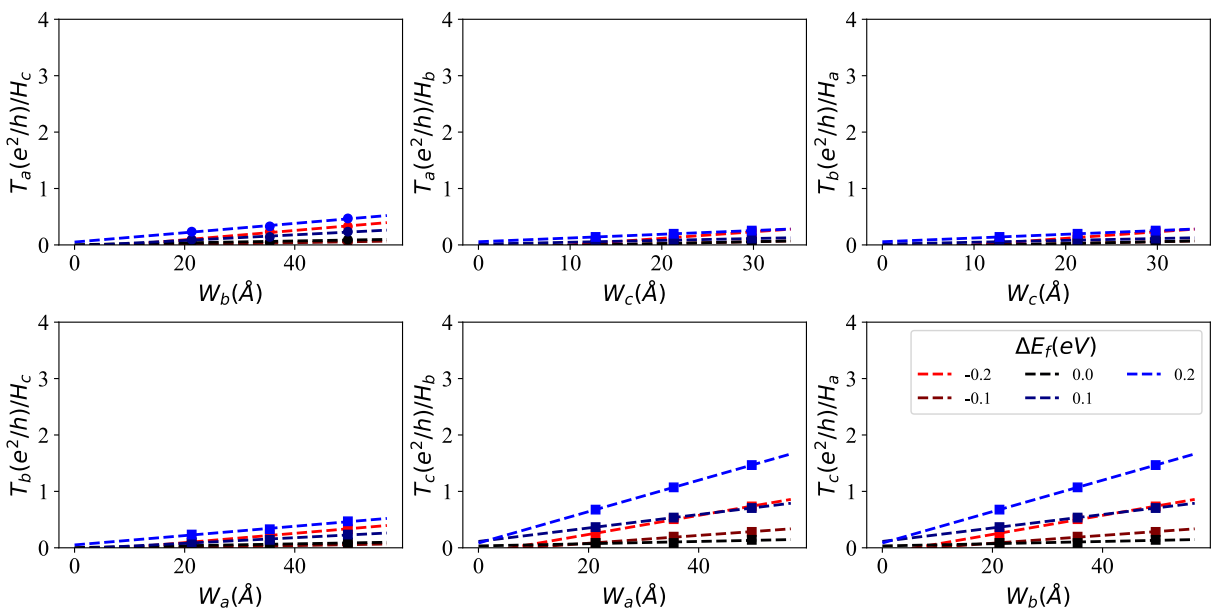

Compound: LiYSi

Materials Project ID: 14208

Lattice (conventional cell):

| Parameter | Value | Unit |
|---|---|---|
| a | 7.0152 | Å |
| b | 7.0152 | Å |
| c | 4.2400 | Å |
| $\alpha$ (alpha) | 90.0000 | ° |
| $\beta$ (beta) | 90.0000 | ° |
| $\gamma$ (gamma) | 120.0000 | ° |

Crystal structure (conventional cell):

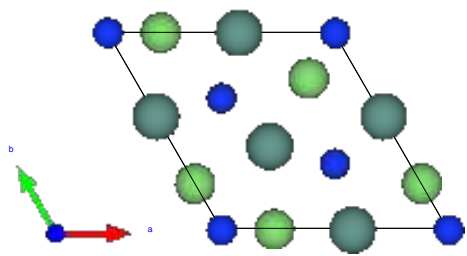

**Nanowire Transmission:**

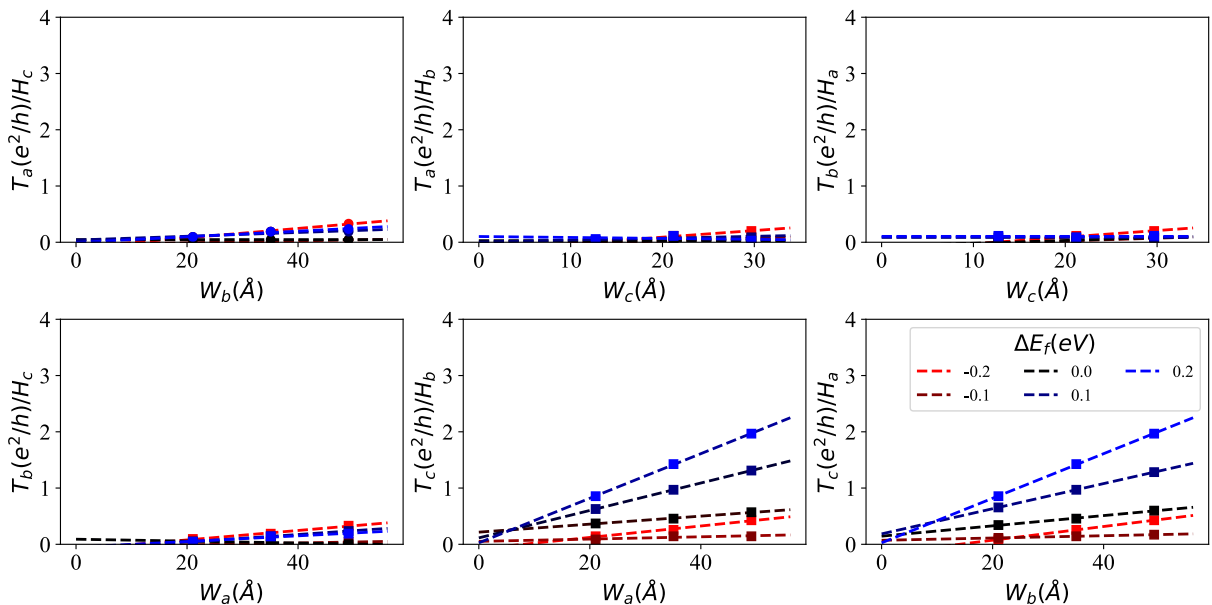

Compound: Mo
Materials Project ID: 129

Lattice (conventional cell):

| Parameter | Value | Unit |
|-----------|--------|------|
| a | 3.1676 | Å |
| b | 3.1676 | Å |
| c | 3.1676 | Å |
| α (alpha) | 90.0000 | ° |
| β (beta) | 90.0000 | ° |
| γ (gamma) | 90.0000 | ° |

Crystal structure (conventional cell):

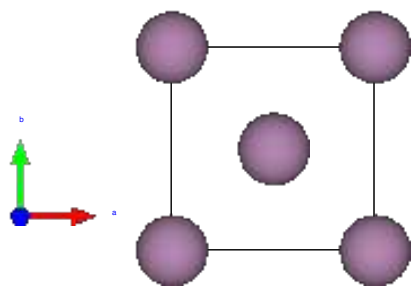

Nanowire Transmission:

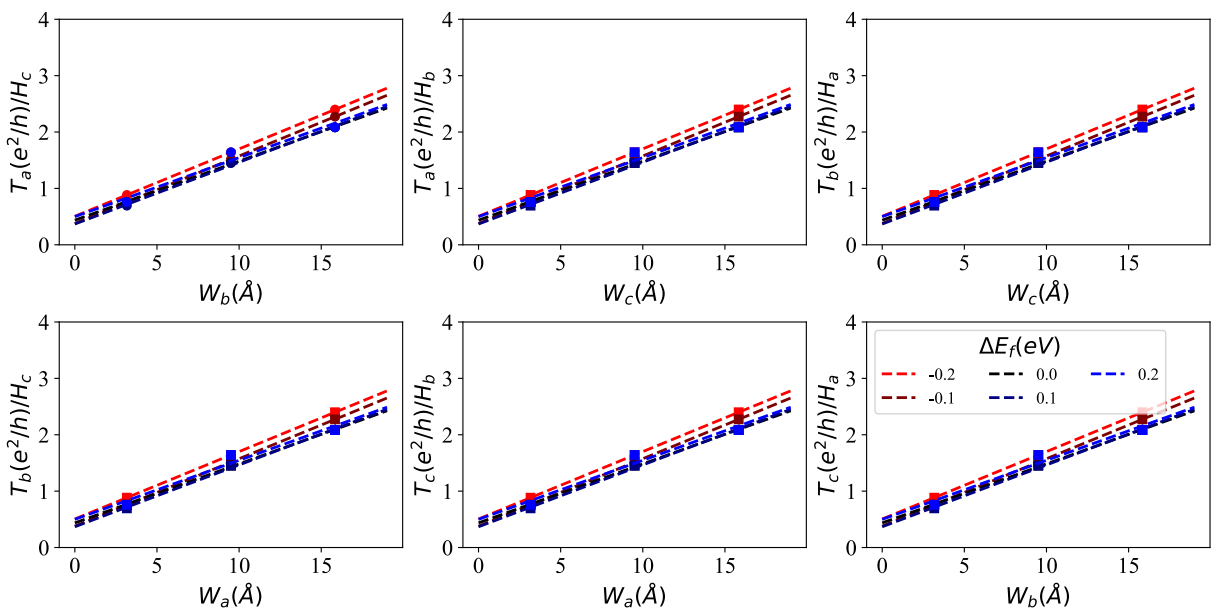

Compound: MoC
Materials Project ID: 2305

### Lattice (conventional cell):

| Parameter | Value | Unit |
|-----------|---------|------|
| a | 2.9241 | Å |
| b | 2.9241 | Å |
| c | 2.8351 | Å |
| α (alpha) | 90.0000 | ° |
| β (beta) | 90.0000 | ° |
| γ (gamma) | 120.0000 | ° |

### Crystal structure (conventional cell):

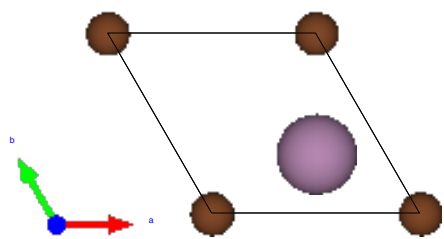

### Nanowire Transmission:

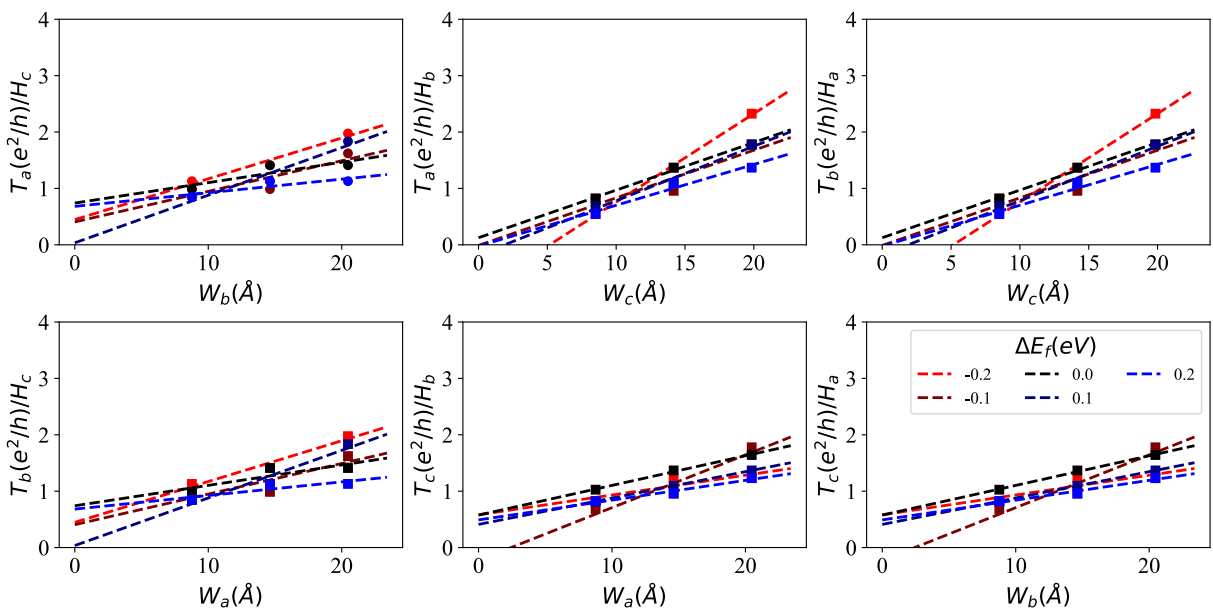

Compound: MoN

Materials Project ID: 13036

Lattice (conventional cell):

| Parameter | Value | Unit |
|-----------|-------|------|
| a | 2.8858 | Å |
| b | 2.8858 | Å |
| c | 2.8561 | Å |
| α (alpha) | 90.0000 | ° |
| β (beta) | 90.0000 | ° |
| γ (gamma) | 120.0000 | ° |

Crystal structure (conventional cell):

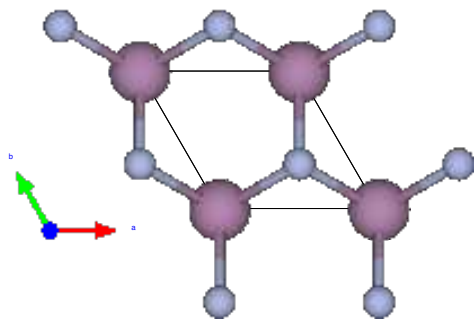

Nanowire Transmission:

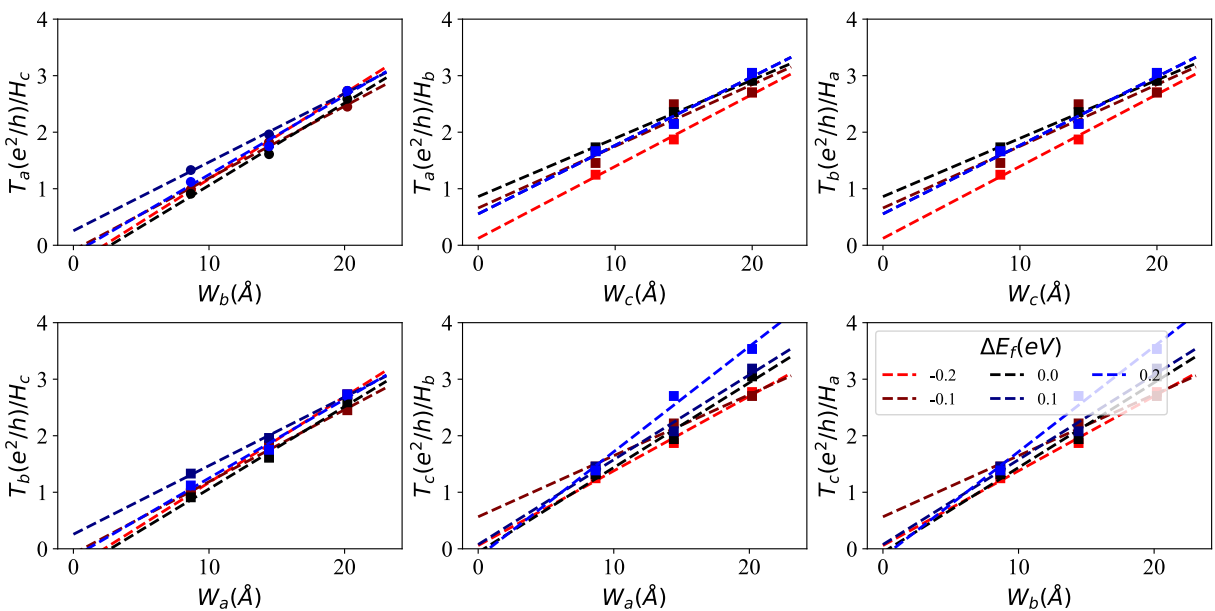

Compound: MoP
Materials Project ID: 219

Lattice (conventional cell):

| Parameter | Value | Unit |
|-----------|---------|------|
| a | 3.2446 | Å |
| b | 3.2446 | Å |
| c | 3.2005 | Å |
| α (alpha) | 90.0000 | ° |
| β (beta) | 90.0000 | ° |
| γ (gamma) | 120.0000 | ° |

Crystal structure (conventional cell):

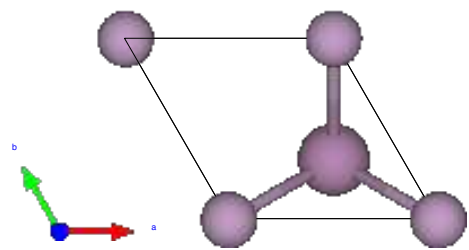

Nanowire Transmission:

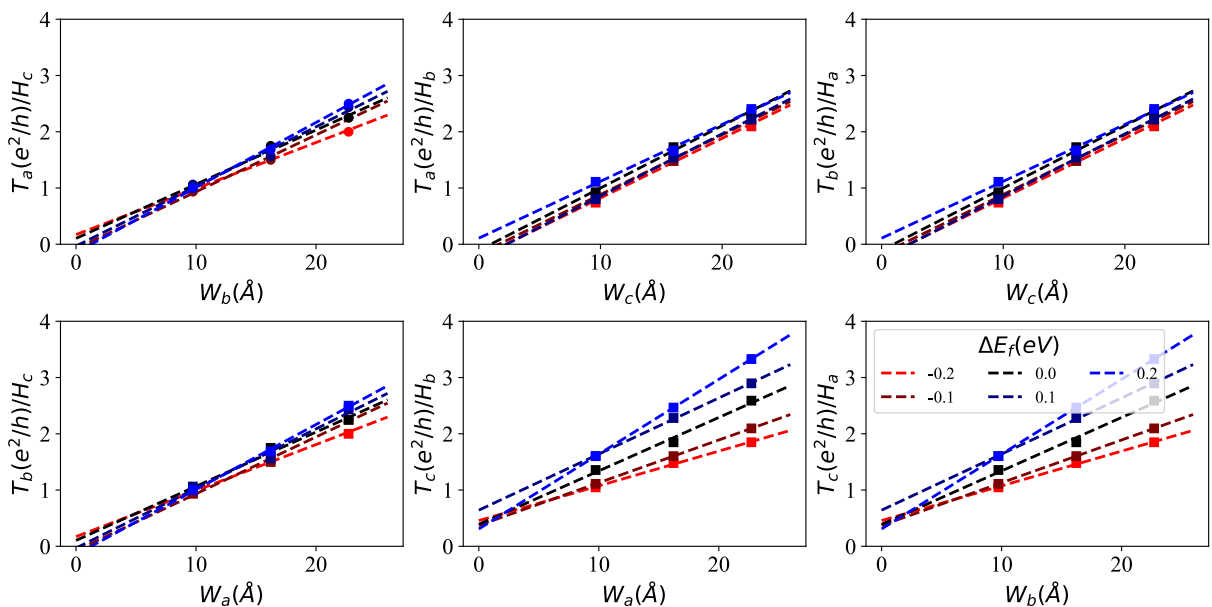

Compound: NbAs
Materials Project ID: 2059

Lattice (conventional cell):

| Parameter | Value | Unit |
|---|---|---|
| a | 3.4847 | Å |
| b | 3.4847 | Å |
| c | 11.7615 | Å |
| α (alpha) | 90.0000 | ° |
| β (beta) | 90.0000 | ° |
| γ (gamma) | 90.0000 | ° |

Crystal structure (conventional cell):

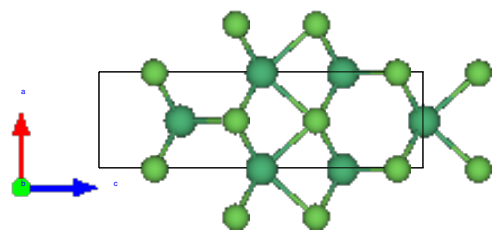

Nanowire Transmission:

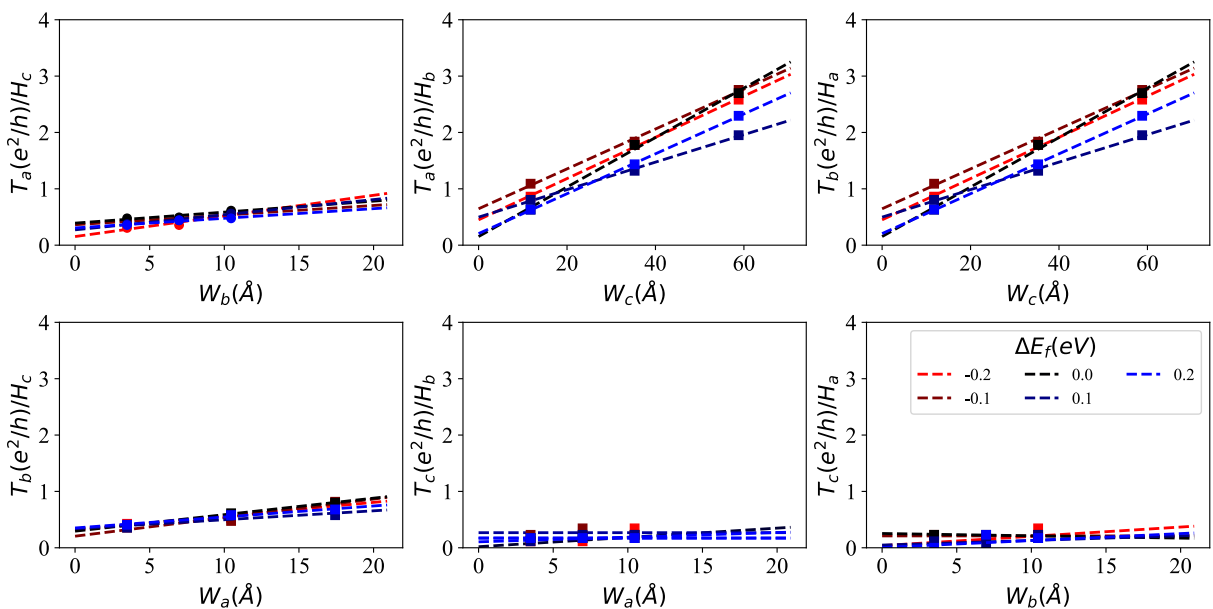

Compound: NbAs$_2$
Materials Project ID: 7598

Lattice (conventional cell):

| Parameter | Value | Unit |
|-----------|-------|------|
| a | 9.4518 | Å |
| b | 3.4178 | Å |
| c | 7.8850 | Å |
| α (alpha) | 90.0000 | ° |
| β (beta) | 119.3863 | ° |
| γ (gamma) | 90.0000 | ° |

Crystal structure (conventional cell):

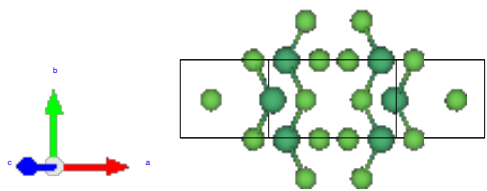

Nanowire Transmission:

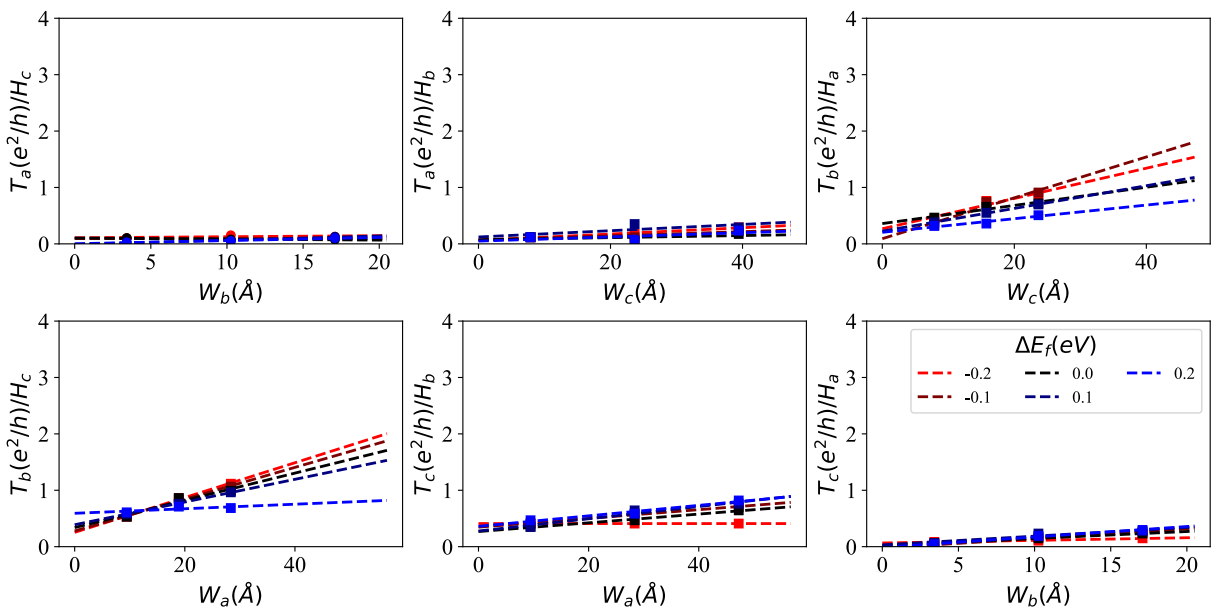

Compound: NbInSe$_2$
Materials Project ID: 20279

Lattice (conventional cell):

| Parameter | Value | Unit |
|---|---|---|
| a | 3.4646 | Å |
| b | 3.4646 | Å |
| c | 9.3914 | Å |
| α (alpha) | 90.0000 | ° |
| β (beta) | 90.0000 | ° |
| γ (gamma) | 120.0000 | ° |

Crystal structure (conventional cell):

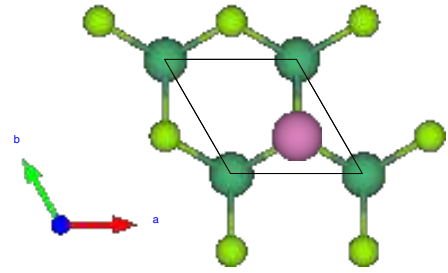

Nanowire Transmission:

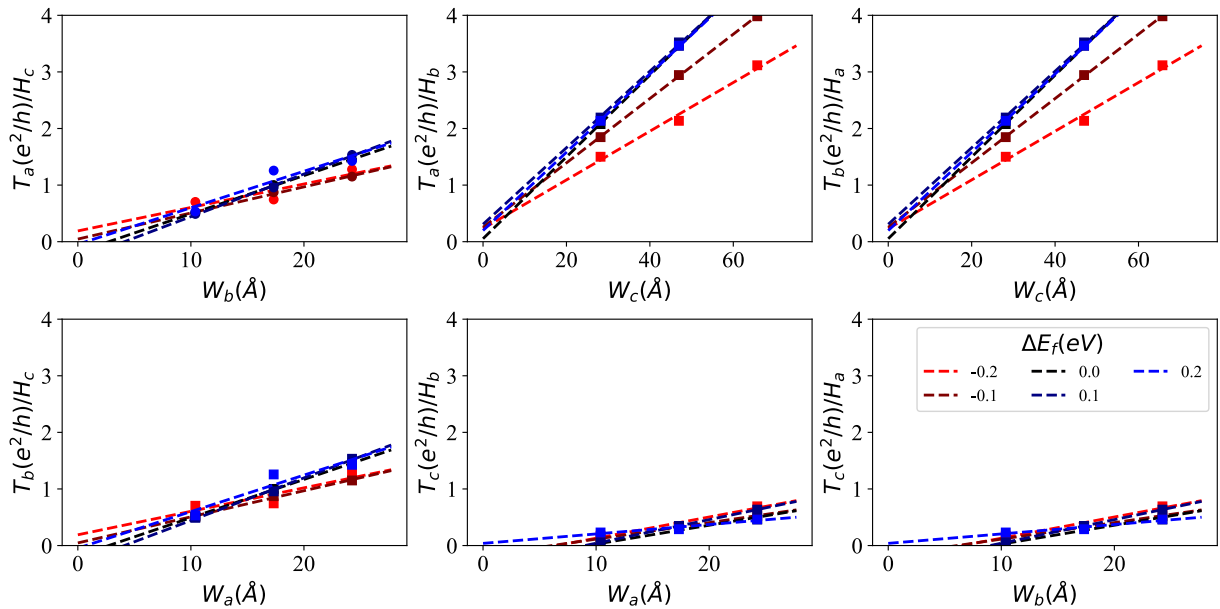

Compound: NbN
Materials Project ID: 2634

Lattice (conventional cell):

| Parameter | Value | Unit |
|-----------|----------|------|
| a | 2.9762 | Å |
| b | 2.9762 | Å |
| c | 2.8987 | Å |
| α (alpha) | 90.0000 | ° |
| β (beta) | 90.0000 | ° |
| γ (gamma) | 120.0000 | ° |

Crystal structure (conventional cell):

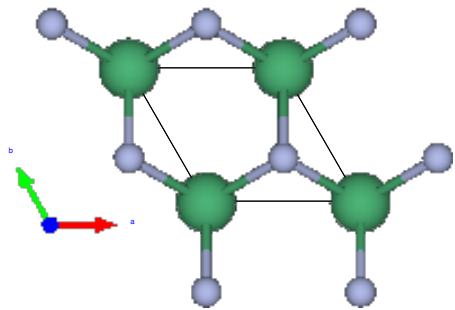

Nanowire Transmission:

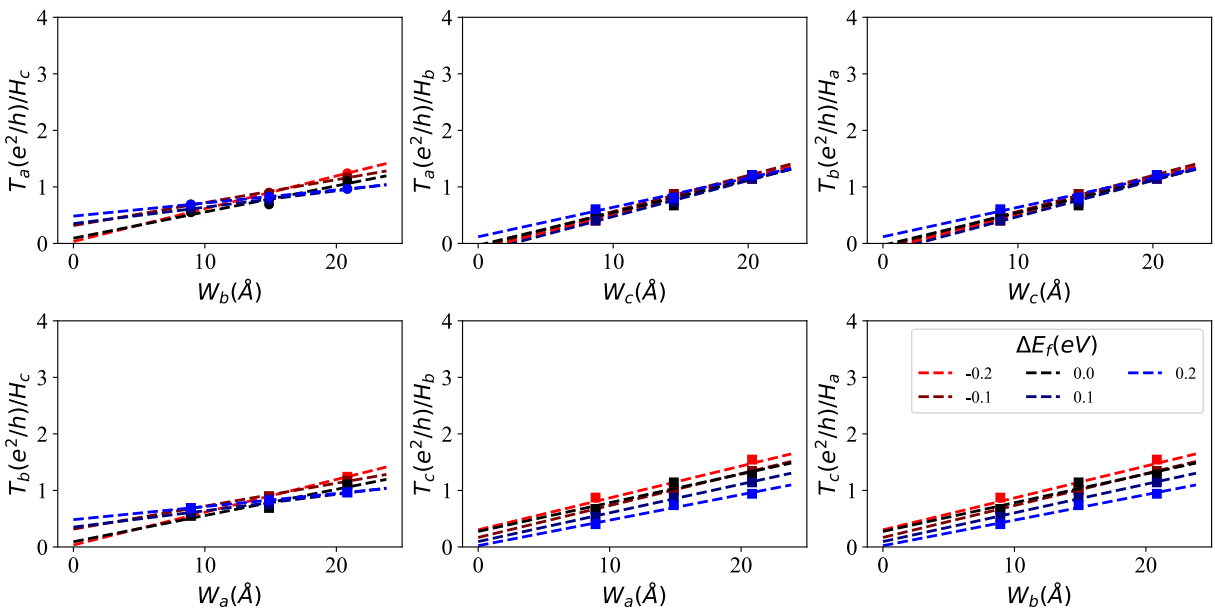

Compound: NbP
Materials Project ID: 9339

Lattice (conventional cell):

| Parameter | Value | Unit |
|-----------|-------|------|
| a | 3.3564 | Å |
| b | 3.3564 | Å |
| c | 11.4365 | Å |
| α (alpha) | 90.0000 | ° |
| β (beta) | 90.0000 | ° |
| γ (gamma) | 90.0000 | ° |

Crystal structure (conventional cell):

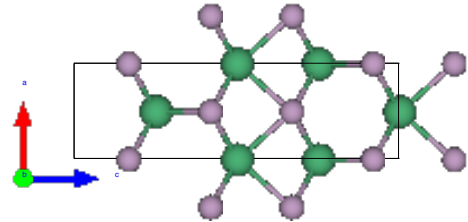

Nanowire Transmission:

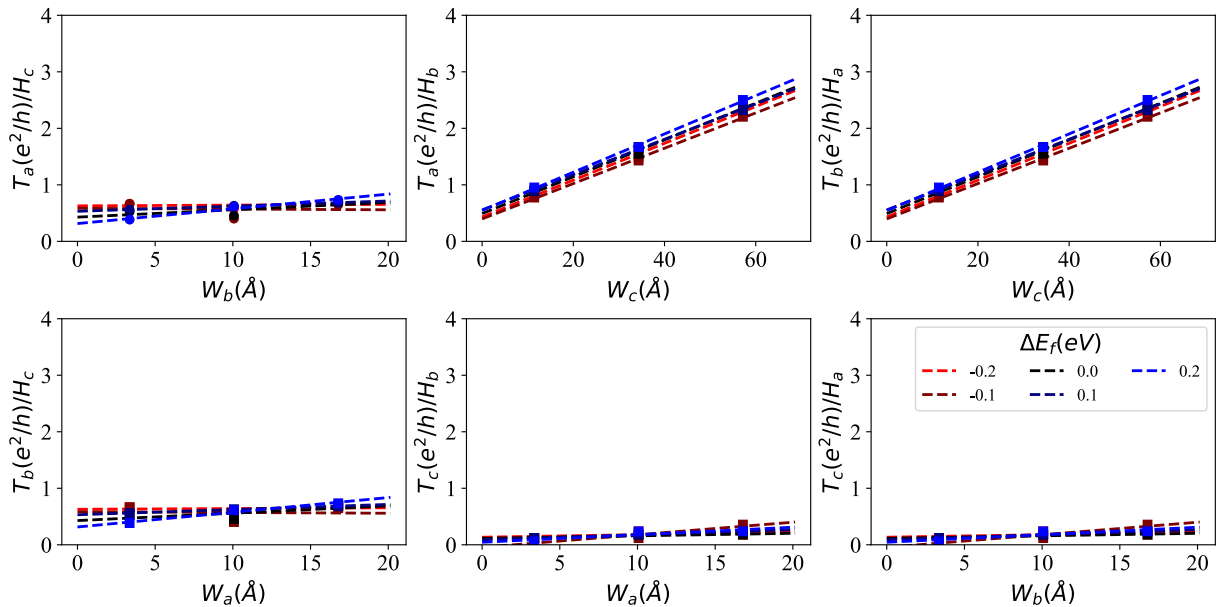

Compound: NbP$_2$
Materials Project ID: 1077116

Lattice (conventional cell):

| Parameter | Value | Unit |
|-----------|---------|------|
| a | 8.9129 | Å |
| b | 3.2925 | Å |
| c | 7.5664 | Å |
| α (alpha) | 90.0000 | ° |
| β (beta) | 118.9943 | ° |
| γ (gamma) | 90.0000 | ° |

Crystal structure (conventional cell):

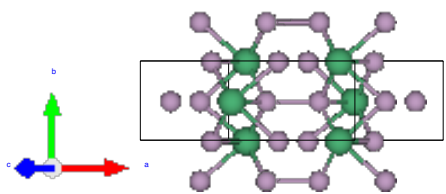

Nanowire Transmission:

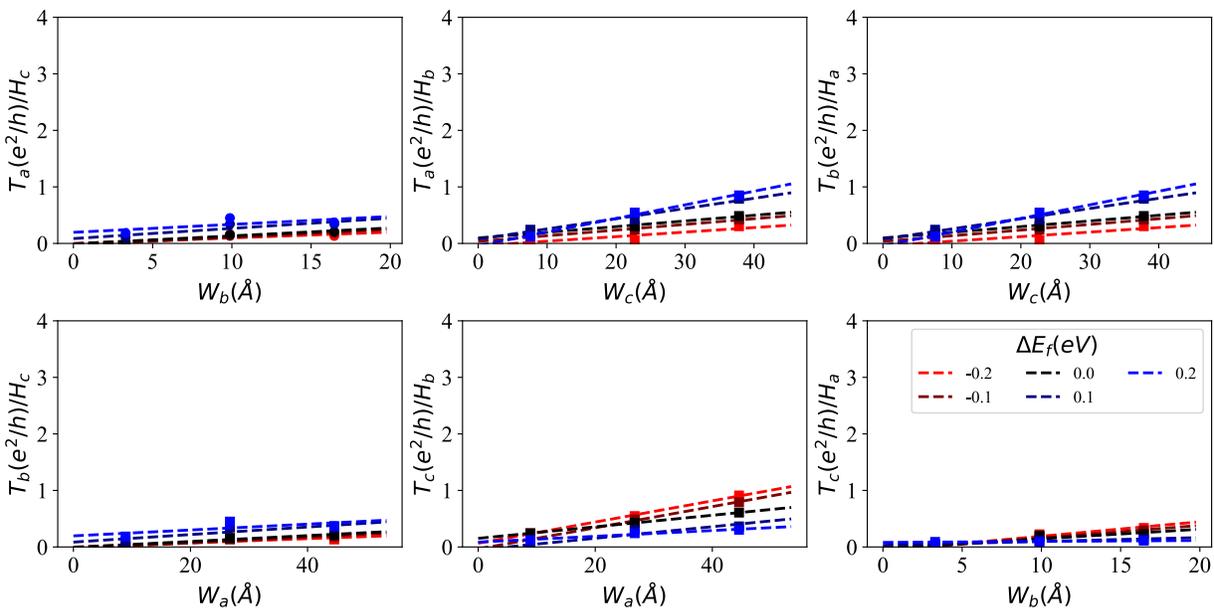

Compound: NbS
Materials Project ID: 2243

Lattice (conventional cell):

| Parameter | Value | Unit |
|---|---|---|
| a | 3.3186 | Å |
| b | 3.3186 | Å |
| c | 3.3243 | Å |
| α (alpha) | 90.0000 | ° |
| β (beta) | 90.0000 | ° |
| γ (gamma) | 120.0000 | ° |

Crystal structure (conventional cell):

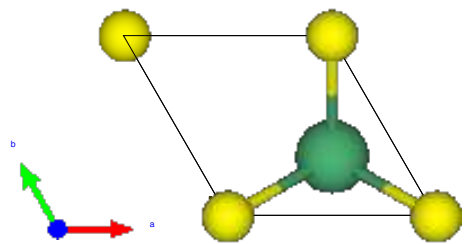

Nanowire Transmission:

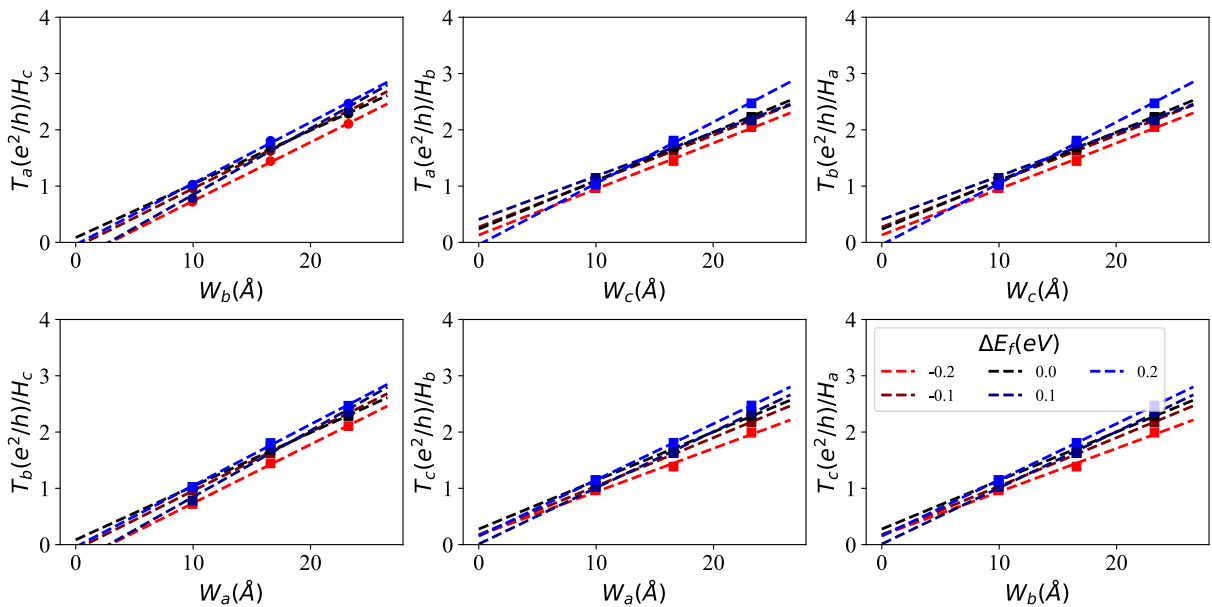

Compound: NbSe$_4$
Materials Project ID: 1078815

Lattice (conventional cell):

| Parameter | Value | Unit |
|-----------|-------|------|
| a | 5.9708 | Å |
| b | 5.9708 | Å |
| c | 6.5262 | Å |
| α (alpha) | 90.0000 | ° |
| β (beta) | 90.0000 | ° |
| γ (gamma) | 90.0000 | ° |

Crystal structure (conventional cell):

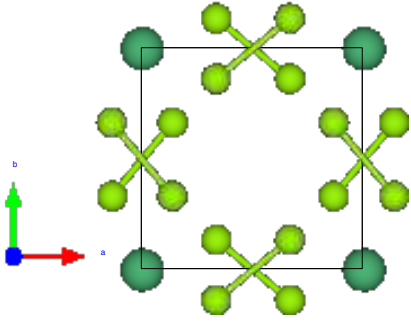

Nanowire Transmission:

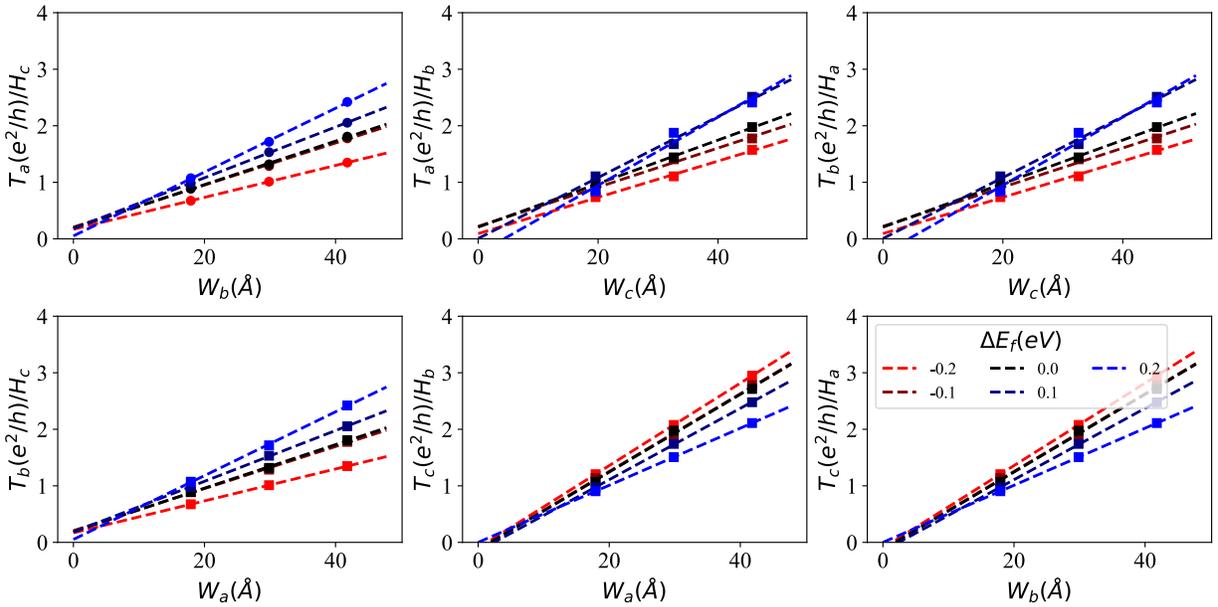

Compound: NbSiAs
Materials Project ID: 21907

Lattice (conventional cell):

Crystal structure (conventional cell):

| Parameter | Value | Unit |
|-----------|-------|------|
| a | 3.5165 | Å |
| b | 3.5165 | Å |
| c | 7.9705 | Å |
| α (alpha) | 90.0000 | ° |
| β (beta) | 90.0000 | ° |
| γ (gamma) | 90.0000 | ° |

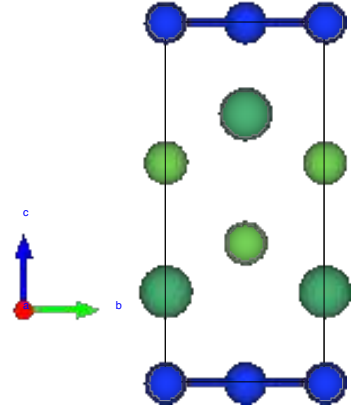

Nanowire Transmission:

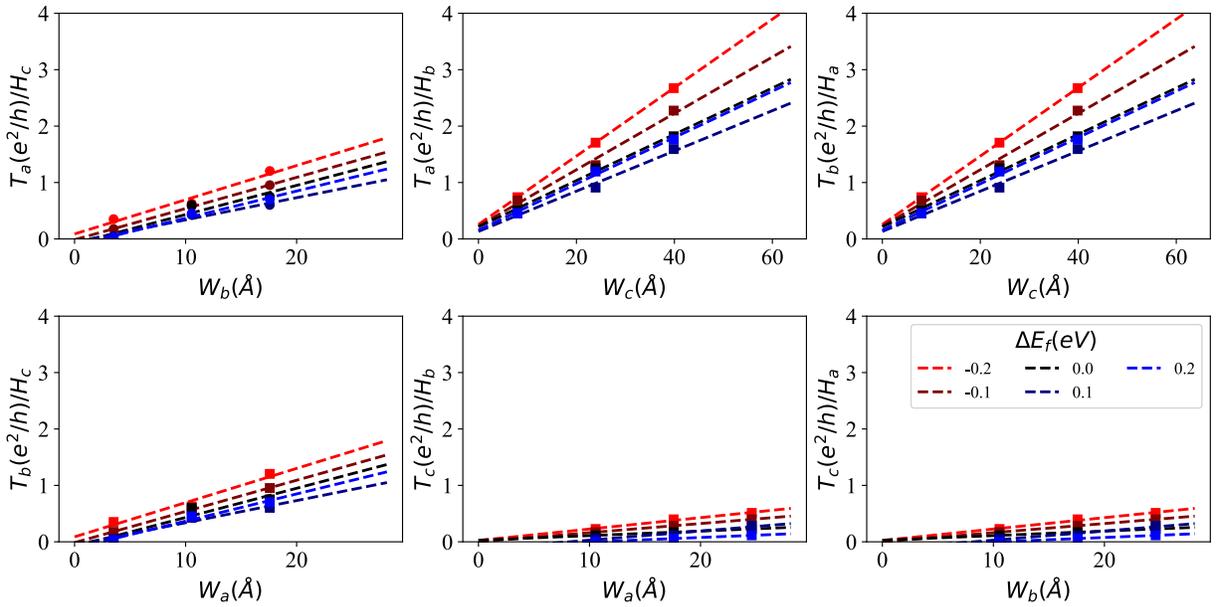

Compound: Ni(BMo)$_2$
Materials Project ID: 9999

Lattice (conventional cell):

| Parameter | Value | Unit |
|-----------|-------|------|
| a | 3.1850 | Å |
| b | 4.5764 | Å |
| c | 7.1383 | Å |
| α (alpha) | 90.0000 | ° |
| β (beta) | 90.0000 | ° |
| γ (gamma) | 90.0000 | ° |

Crystal structure (conventional cell):

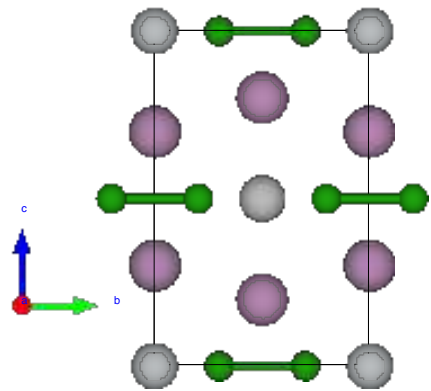

Nanowire Transmission:

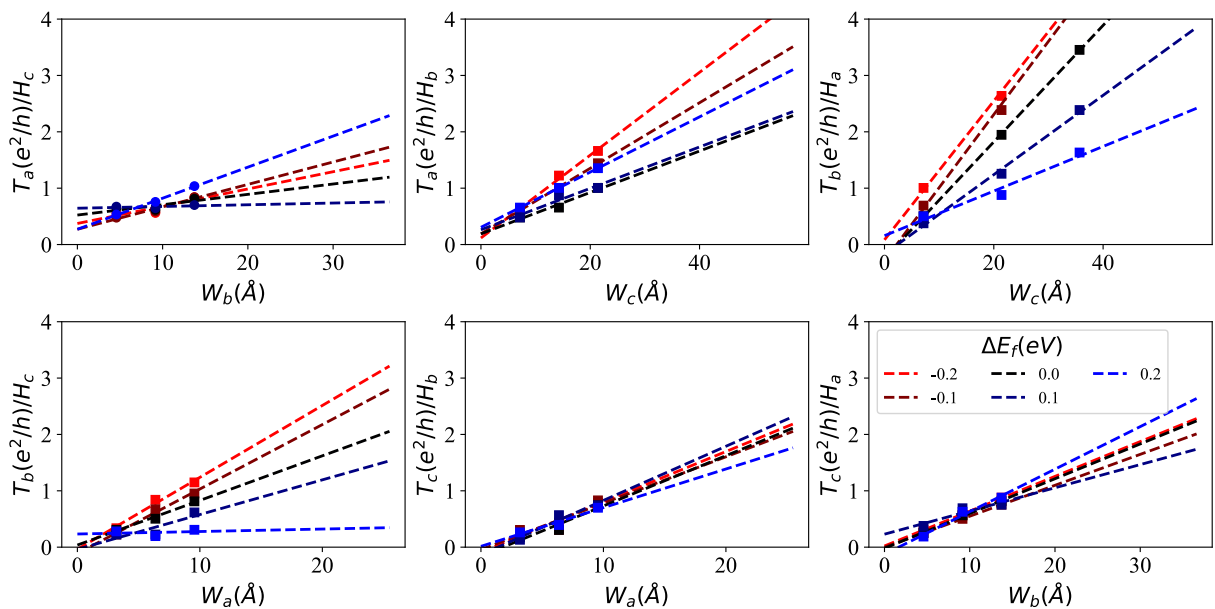

Compound: NiS
Materials Project ID: 1547

Lattice (conventional cell):

| Parameter | Value | Unit |
|-----------|---------|------|
| a | 9.5644 | Å |
| b | 9.5644 | Å |
| c | 3.1260 | Å |
| α (alpha) | 90.0000 | ° |
| β (beta) | 90.0000 | ° |
| γ (gamma) | 120.0000 | ° |

Crystal structure (conventional cell):

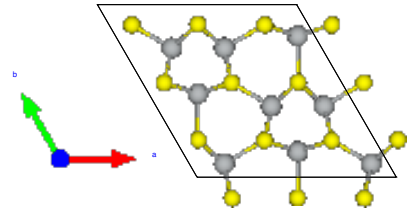

Nanowire Transmission:

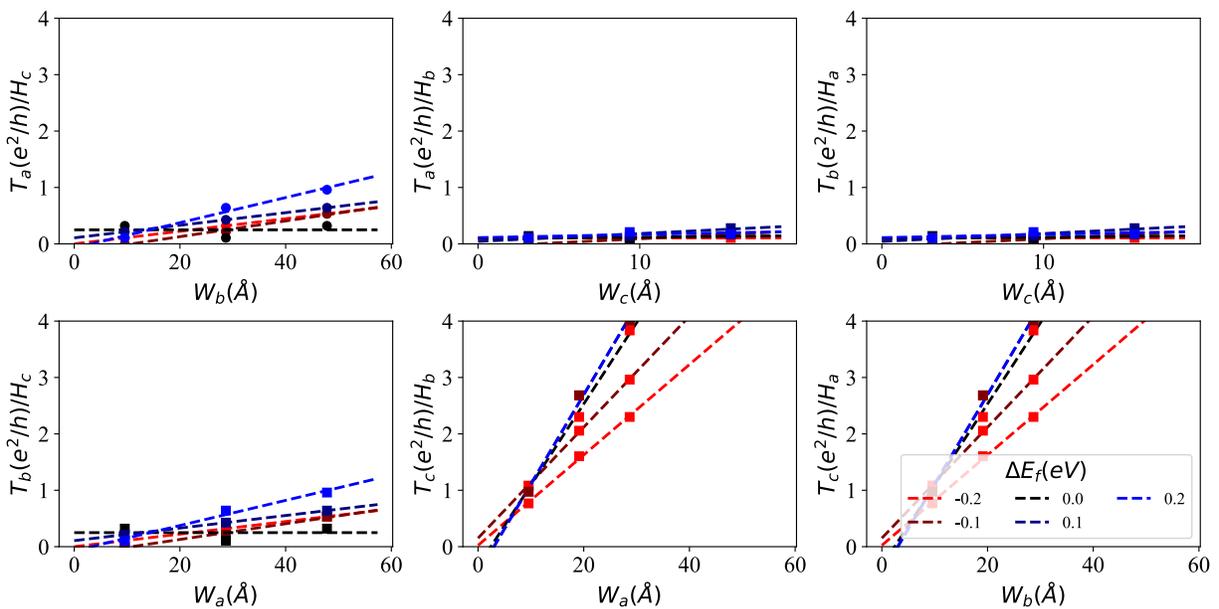

Compound: NiSe
Materials Project ID: 15651

Lattice (conventional cell):

| Parameter | Value | Unit |
|-----------|---------|------|
| a | 10.0013 | Å |
| b | 10.0013 | Å |
| c | 3.3350 | Å |
| α (alpha) | 90.0000 | ° |
| β (beta) | 90.0000 | ° |
| γ (gamma) | 120.0000 | ° |

Crystal structure (conventional cell):

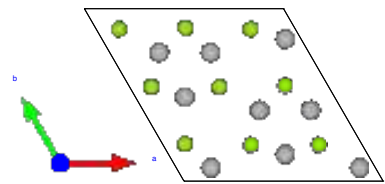

Nanowire Transmission:

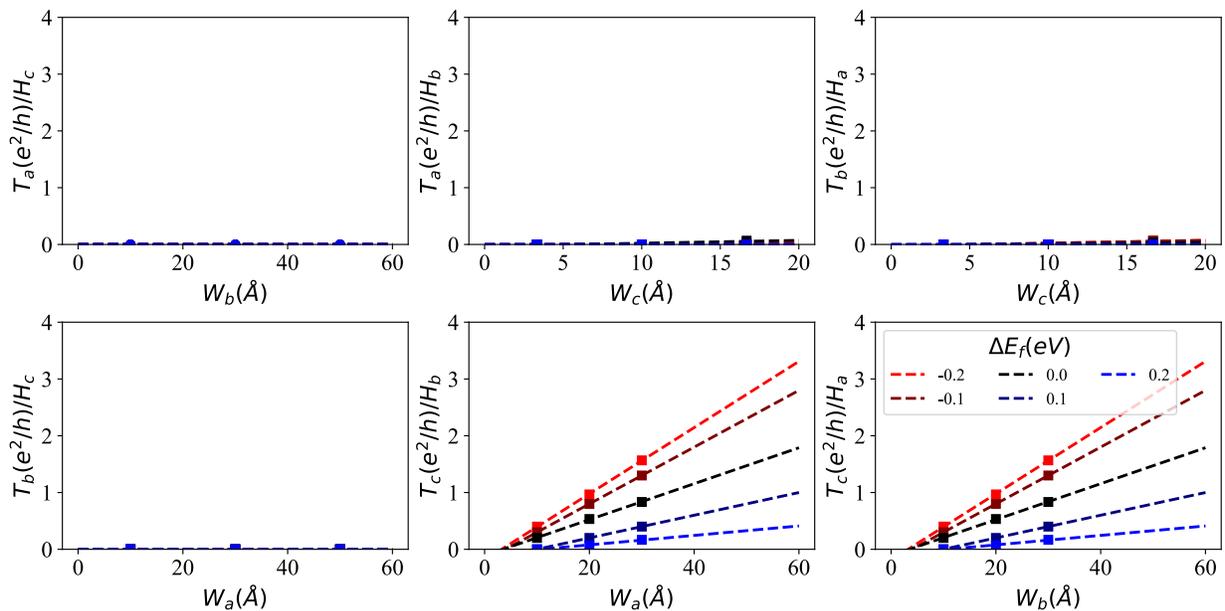

Compound: Pd$_2$N
Materials Project ID: 510087

Lattice (conventional cell):

| Parameter | Value | Unit |
|---|---|---|
| a | 2.9424 | Å |
| b | 4.8726 | Å |
| c | 5.4625 | Å |
| α (alpha) | 90.0000 | ° |
| β (beta) | 90.0000 | ° |
| γ (gamma) | 90.0000 | ° |

Crystal structure (conventional cell):

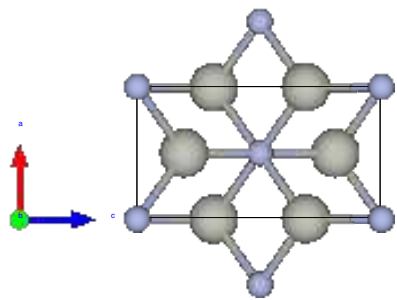

Nanowire Transmission:

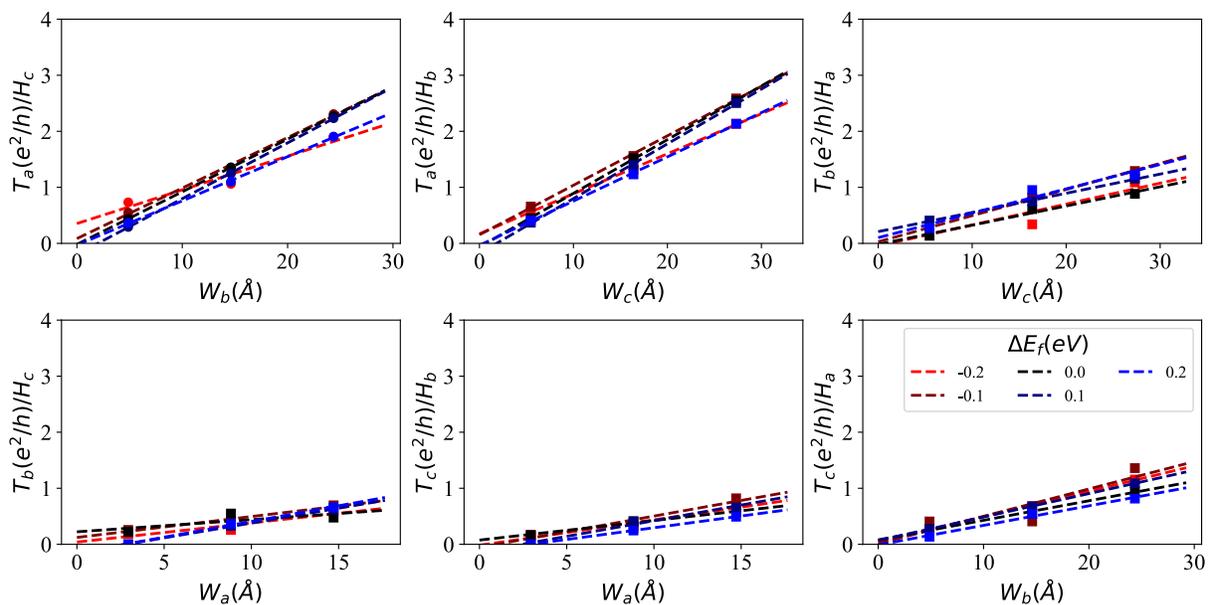

Compound: Re
Materials Project ID: 8

Lattice (conventional cell):

| Parameter | Value | Unit |
|-----------|-------|------|
| a | 2.7811 | Å |
| b | 2.7811 | Å |
| c | 4.4971 | Å |
| α (alpha) | 90.0000 | ° |
| β (beta) | 90.0000 | ° |
| γ (gamma) | 120.0000 | ° |

Crystal structure (conventional cell):

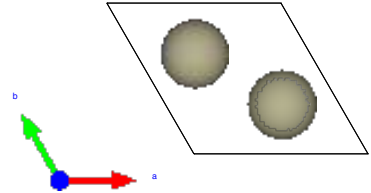

Nanowire Transmission:

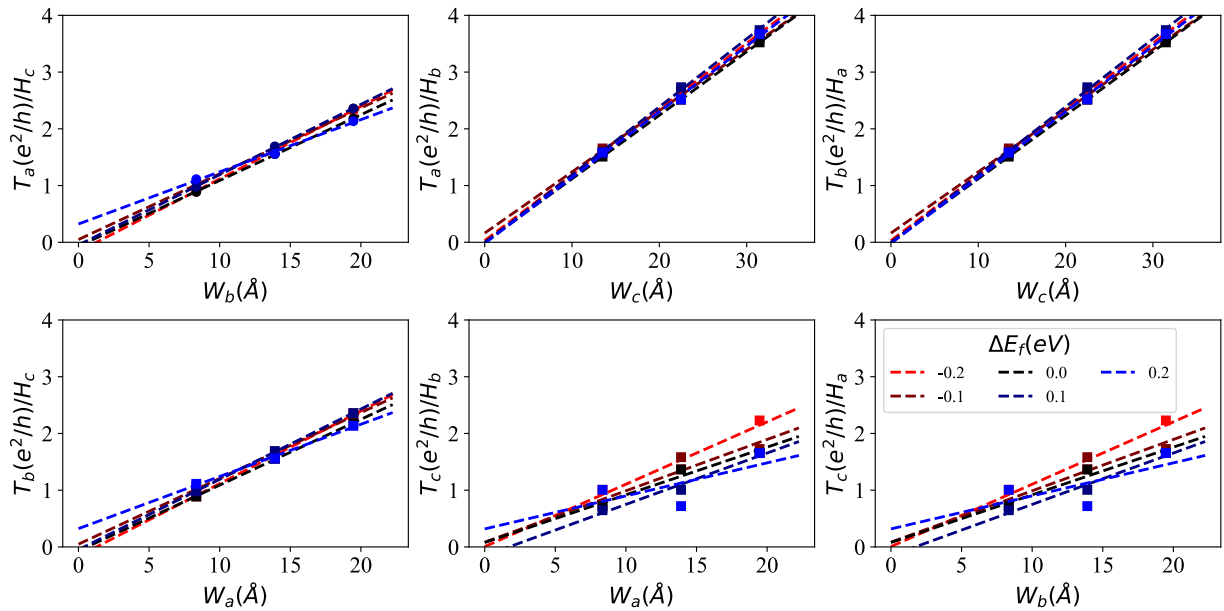

Compound: Ru
Materials Project ID: 33

Lattice (conventional cell):

| Parameter | Value | Unit |
|-----------|-------|------|
| a | 2.7329 | Å |
| b | 2.7329 | Å |
| c | 4.3139 | Å |
| α (alpha) | 90.0000 | ° |
| β (beta) | 90.0000 | ° |
| γ (gamma) | 120.0000 | ° |

Crystal structure (conventional cell):

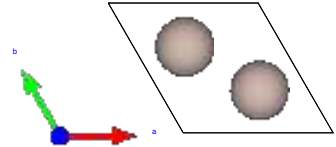

Nanowire Transmission:

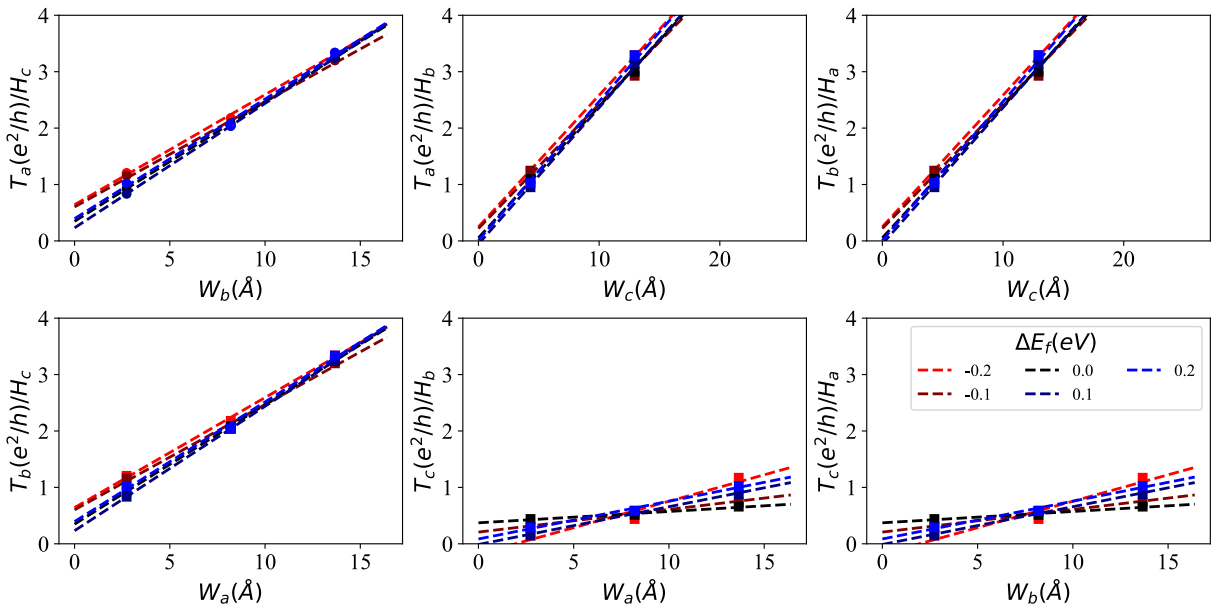

Compound: ScGa$_3$
Materials Project ID: 932

**Lattice (conventional cell):**

| Parameter | Value | Unit |
|-----------|---------|------|
| a | 4.1216 | Å |
| b | 4.1216 | Å |
| c | 4.1216 | Å |
| α (alpha) | 90.0000 | ° |
| β (beta) | 90.0000 | ° |
| γ (gamma) | 90.0000 | ° |

**Crystal structure (conventional cell):**

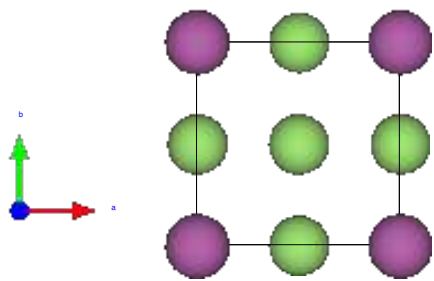

**Nanowire Transmission:**

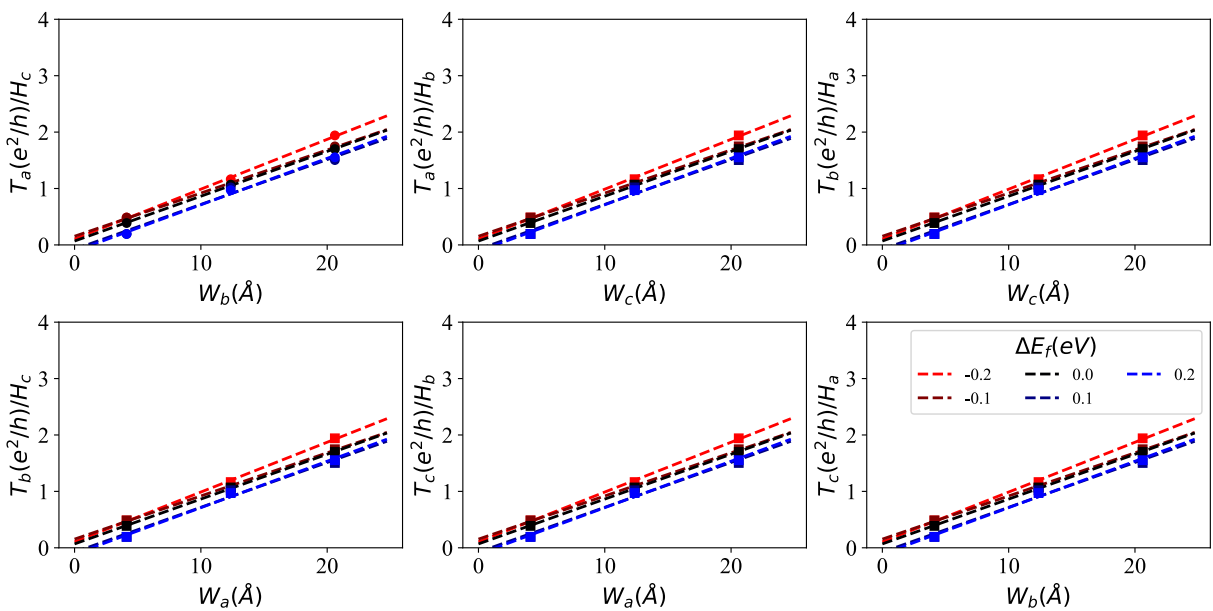

Compound: Se
Materials Project ID: 14

Lattice (conventional cell):

| Parameter | Value | Unit |
|-----------|---------|------|
| a | 4.5195 | Å |
| b | 4.5195 | Å |
| c | 5.0500 | Å |
| α (alpha) | 90.0000 | ° |
| β (beta) | 90.0000 | ° |
| γ (gamma) | 120.0000 | ° |

Crystal structure (conventional cell):

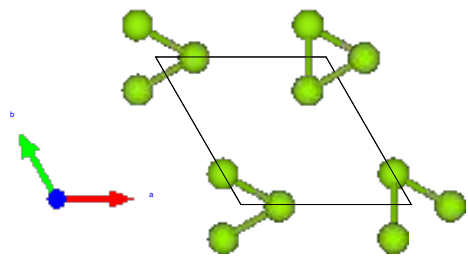

Nanowire Transmission:

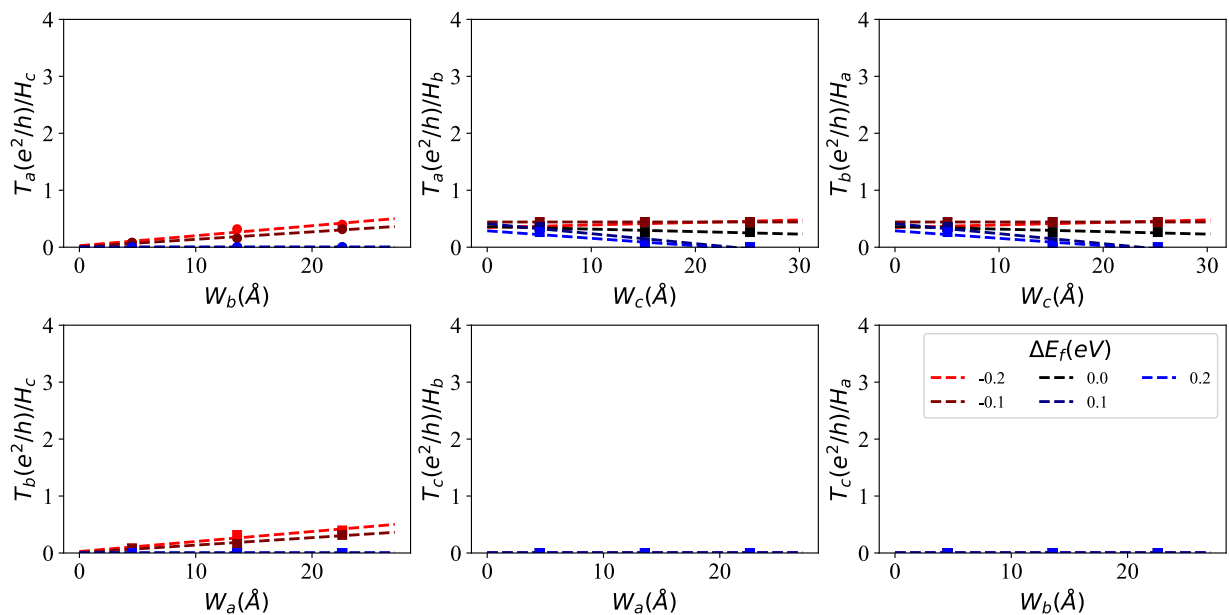

Compound: SiRh
Materials Project ID: 1483

Lattice (conventional cell):

| Parameter | Value | Unit |
|-----------|-------|------|
| a | 4.7295 | Å |
| b | 4.7295 | Å |
| c | 4.7295 | Å |
| α (alpha) | 90.0000 | ° |
| β (beta) | 90.0000 | ° |
| γ (gamma) | 90.0000 | ° |

Crystal structure (conventional cell):

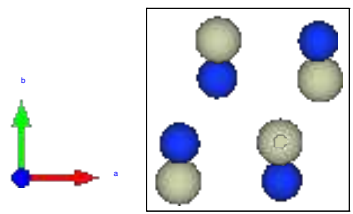

Nanowire Transmission:

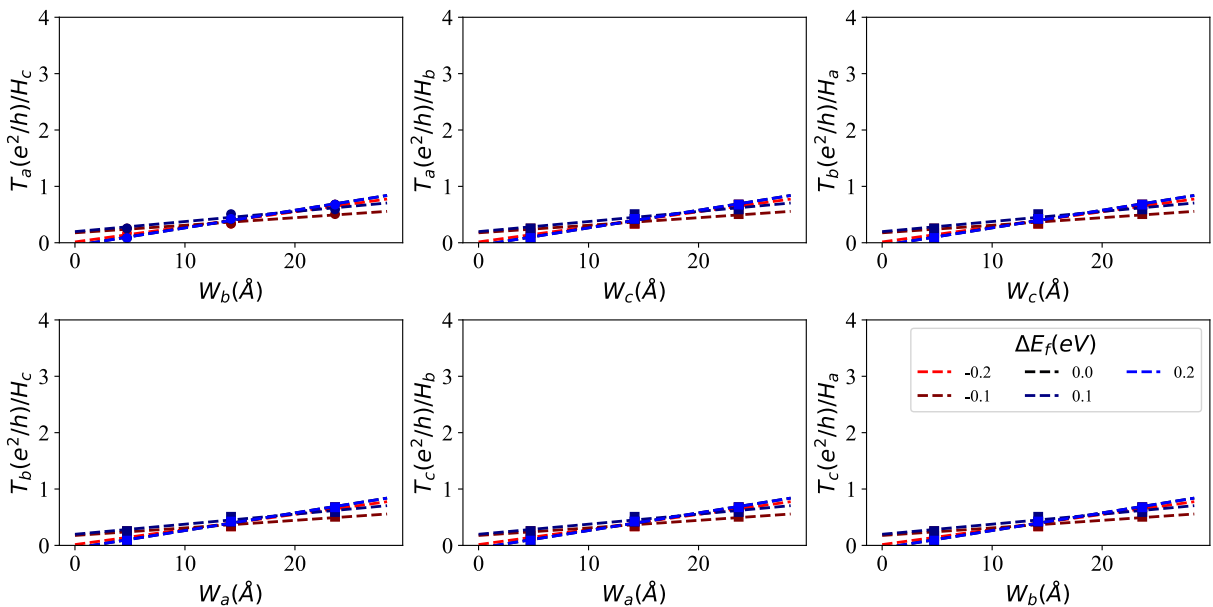

Compound: ScCuGe
Materials Project ID: 1078430

Lattice (conventional cell):

Crystal structure (conventional cell):

| Parameter | Value | Unit |
|-----------|-------|------|
| a | 6.5606 | Å |
| b | 6.5606 | Å |
| c | 3.9977 | Å |
| α (alpha) | 90.0000 | ° |
| β (beta) | 90.0000 | ° |

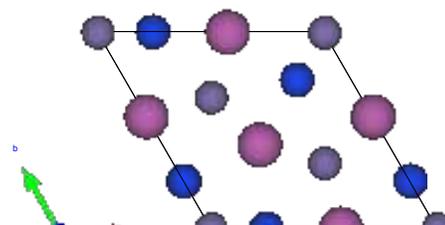

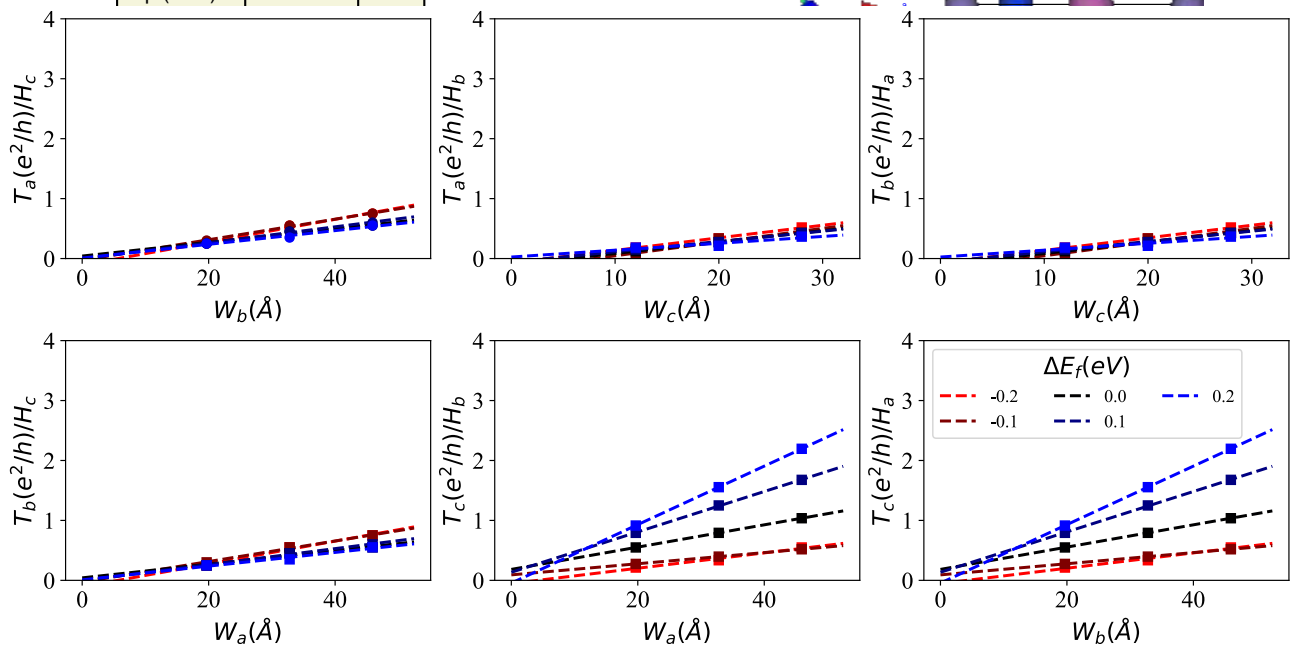

Compound: SnH$_4$
Materials Project ID: 1080725

## Lattice (conventional cell):

| Parameter | Value | Unit |
|-----------|-------|------|
| a | 3.9522 | Å |
| b | 3.9522 | Å |
| c | 6.4992 | Å |
| $\alpha$ (alpha) | 90.0000 | ° |
| $\beta$ (beta) | 90.0000 | ° |
| $\gamma$ (gamma) | 120.0000 | ° |

## Crystal structure (conventional cell):

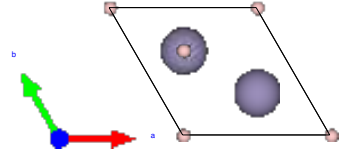

## Nanowire Transmission:

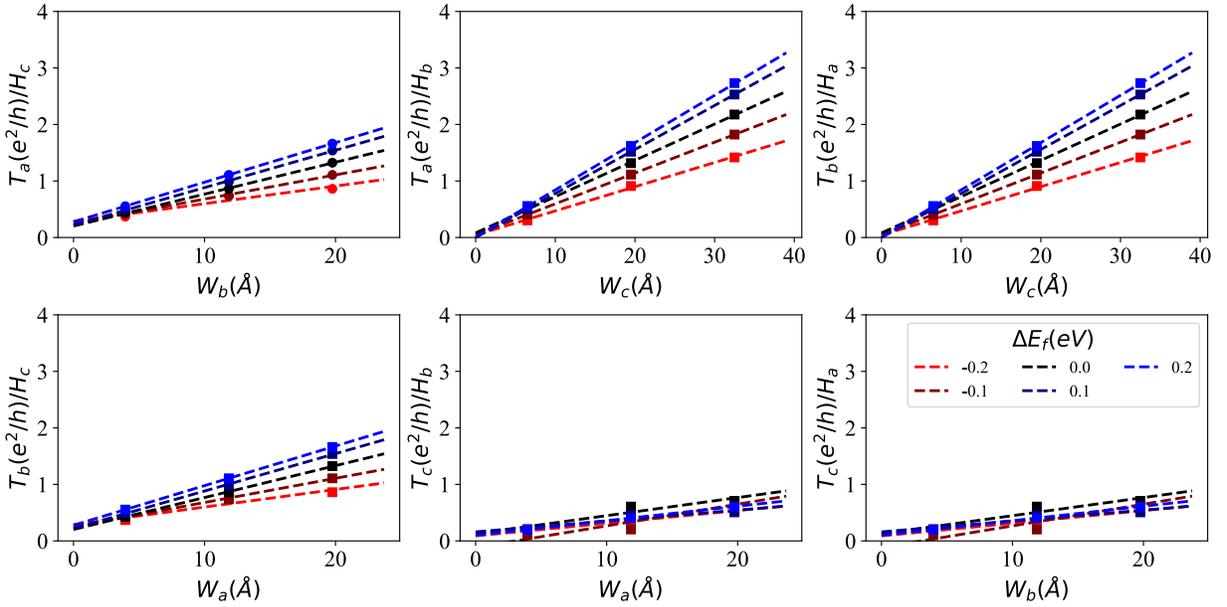

Compound: Sn(HgSe$_2$)$_2$
Materials Project ID: 10955

Lattice (conventional cell):

| Parameter | Value | Unit |
|---|---|---|
| a | 5.9365 | Å |
| b | 5.9365 | Å |
| c | 11.9940 | Å |
| α (alpha) | 90.0000 | ° |
| β (beta) | 90.0000 | ° |
| γ (gamma) | 90.0000 | ° |

Crystal structure (conventional cell):

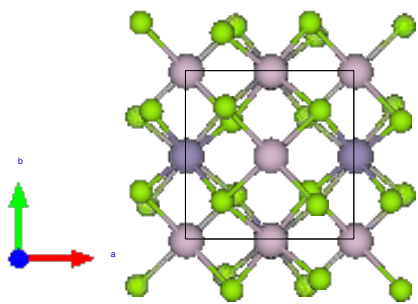

Nanowire Transmission:

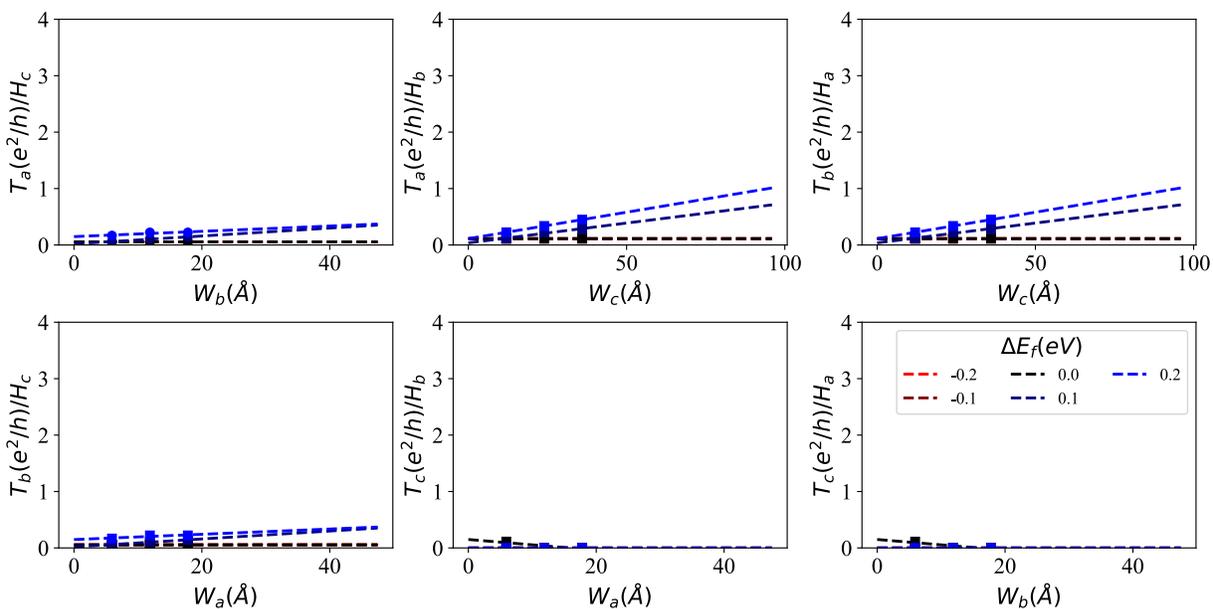

Compound: Sr(CdSb)$_2$
Materials Project ID: 7432

## Lattice (conventional cell):

| Parameter | Value | Unit |
|---|---|---|
| a | 4.7972 | Å |
| b | 4.7972 | Å |
| c | 7.9056 | Å |
| α (alpha) | 90.0000 | ° |
| β (beta) | 90.0000 | ° |
| γ (gamma) | 120.0000 | ° |

## Crystal structure (conventional cell):

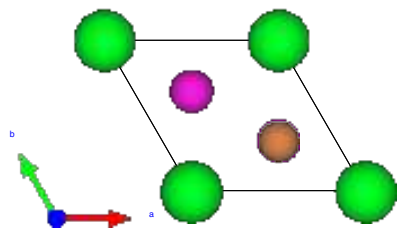

## Nanowire Transmission:

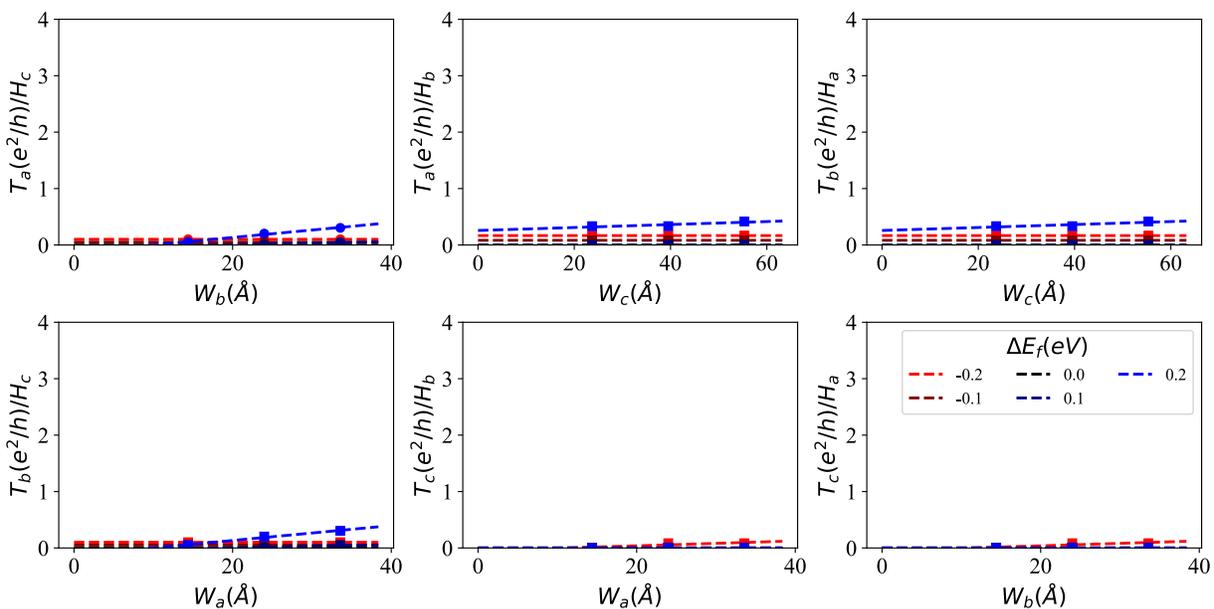

Compound: Sr$_3$(AlGe)$_2$
Materials Project ID: 571216

Lattice (conventional cell):

| Parameter | Value | Unit |
|-----------|-------|------|
| a | 12.7411 | Å |
| b | 4.2166 | Å |
| c | 8.9551 | Å |
| α (alpha) | 90.0000 | ° |
| β (beta) | 110.1782 | ° |
| γ (gamma) | 90.0000 | ° |

Crystal structure (conventional cell):

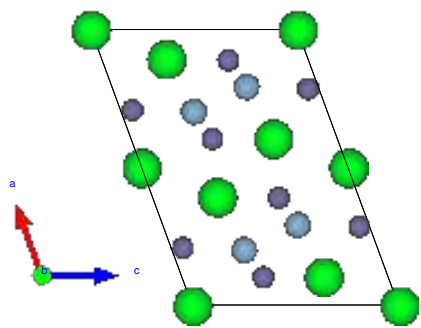

Nanowire Transmission:

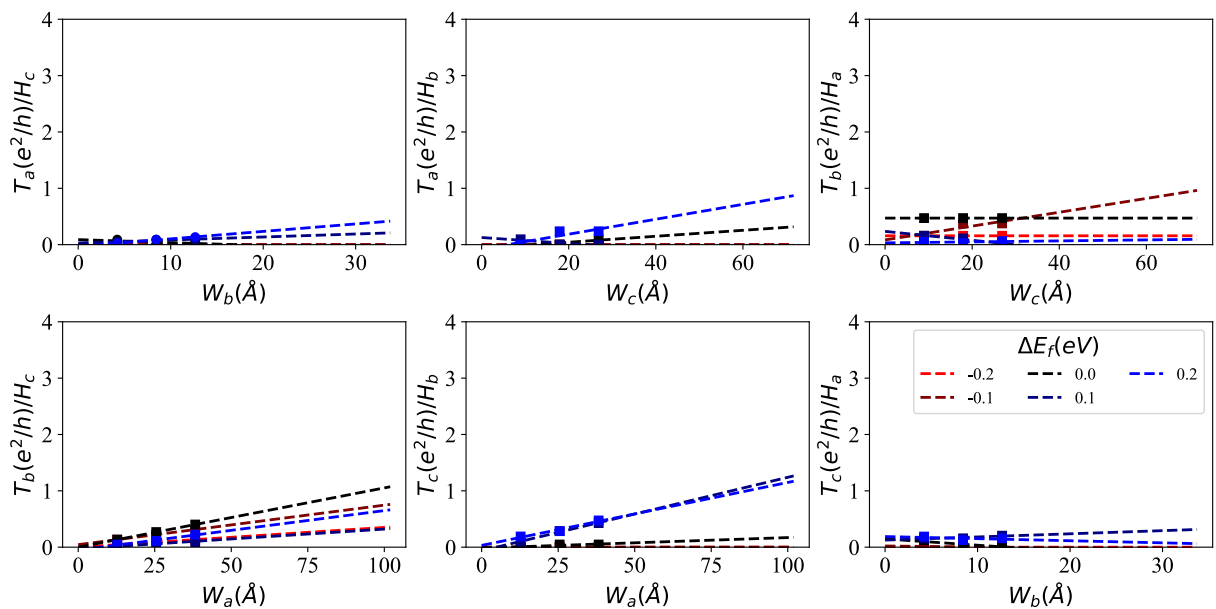

Compound: SrAs$_3$
Materials Project ID: 9907

Lattice (conventional cell):

Crystal structure (conventional cell):

| Parameter | Value | Unit |
|---|---|---|
| a | 9.7449 | Å |
| b | 7.7486 | Å |
| c | 5.9535 | Å |
| α (alpha) | 90.0000 | ° |
| β (beta) | 112.9053 | ° |
| γ (gamma) | 90.0000 | ° |

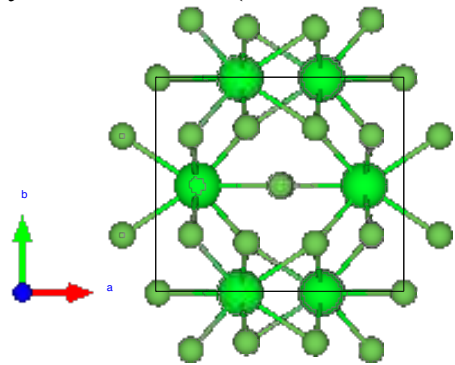

Nanowire Transmission:

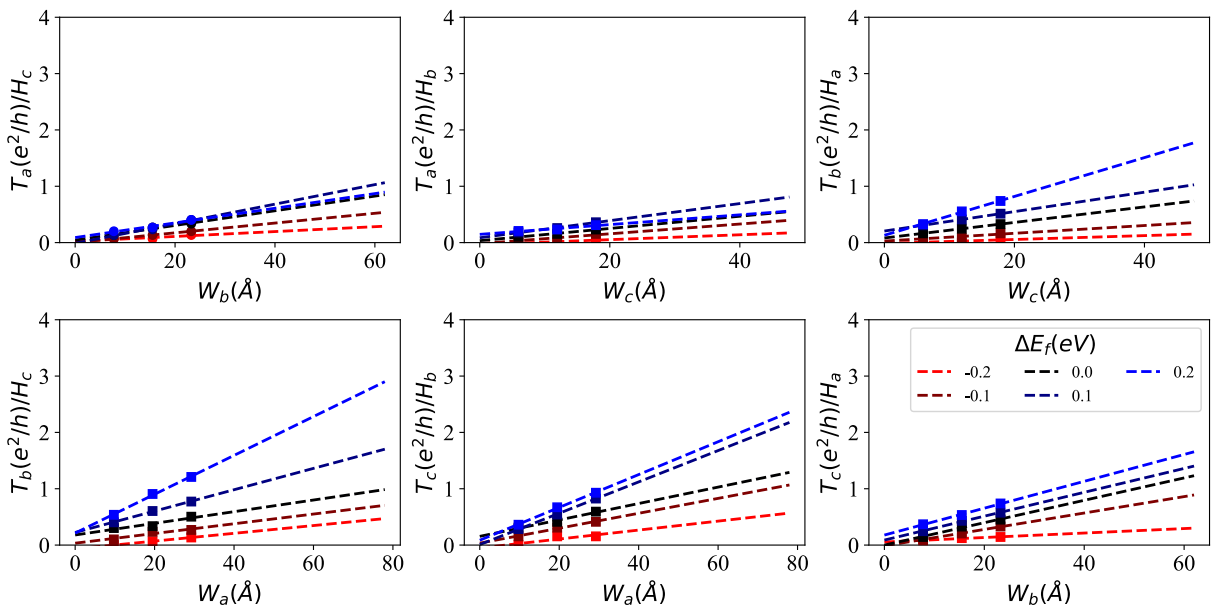

Compound: Sr$_2$Bi
Materials Project ID: 29619

Lattice (conventional cell):

| Parameter | Value | Unit |
|-----------|---------|------|
| a | 5.1526 | Å |
| b | 5.1526 | Å |
| c | 18.1556 | Å |
| α (alpha) | 90.0000 | ° |
| β (beta) | 90.0000 | ° |
| γ (gamma) | 90.0000 | ° |

Crystal structure (conventional cell):

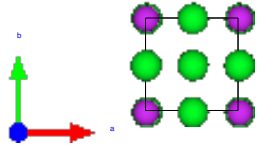

Nanowire Transmission:

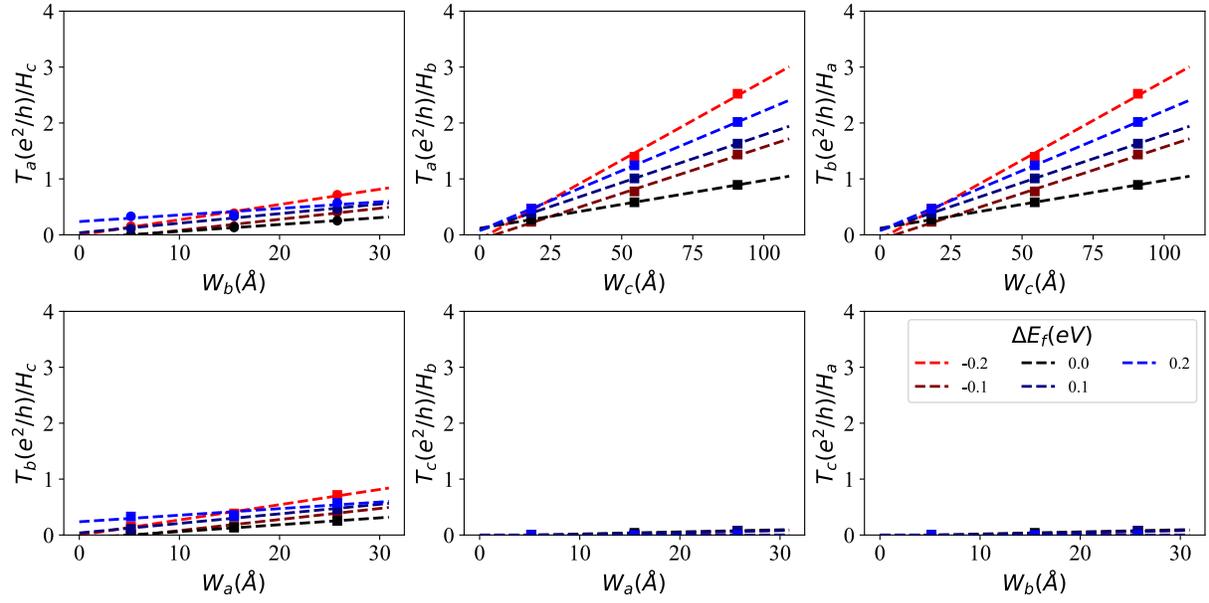

Compound: SrGe$_2$
Materials Project ID: 7390

Lattice (conventional cell):

| Parameter | Value | Unit |
|-----------|---------|------|
| a | 4.1699 | Å |
| b | 4.1699 | Å |
| c | 5.1991 | Å |
| α (alpha) | 90.0000 | ° |
| β (beta) | 90.0000 | ° |
| γ (gamma) | 120.0000 | ° |

Crystal structure (conventional cell):

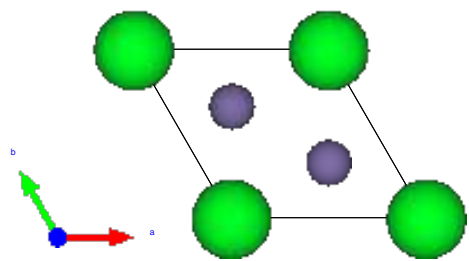

Nanowire Transmission:

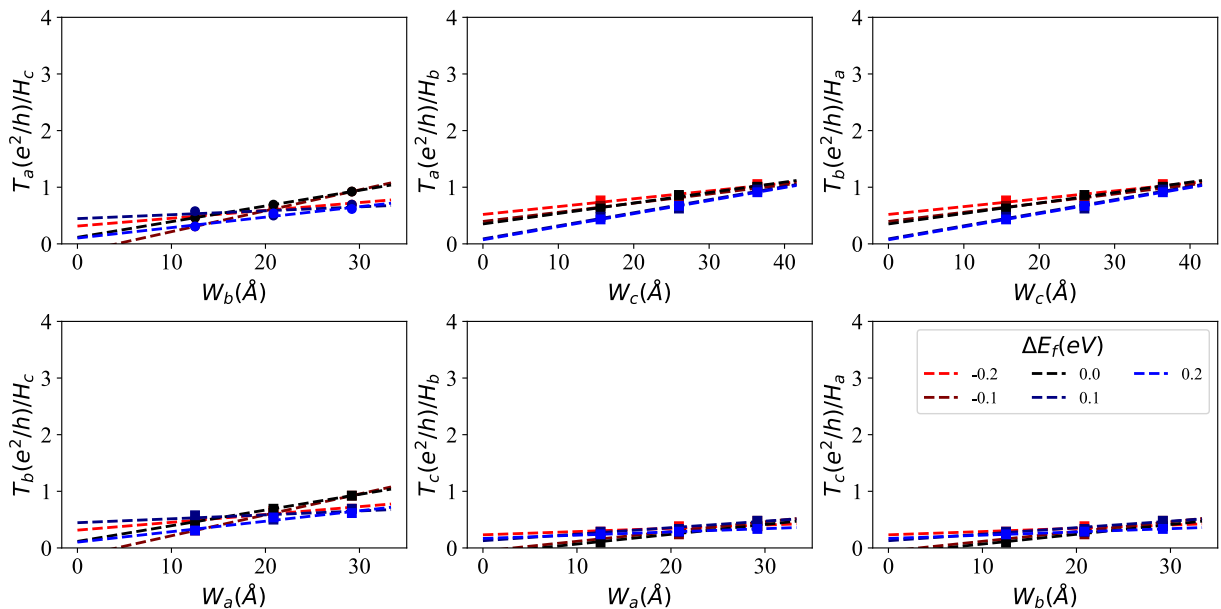

Compound: SrMnSb$_2$
Materials Project ID: 1079721

Lattice (conventional cell):

| Parameter | Value | Unit |
|---|---|---|
| a | 4.4170 | Å |
| b | 4.4170 | Å |
| c | 11.7201 | Å |
| α (alpha) | 90.0000 | ° |
| β (beta) | 90.0000 | ° |
| γ (gamma) | 90.0000 | ° |

Crystal structure (conventional cell):

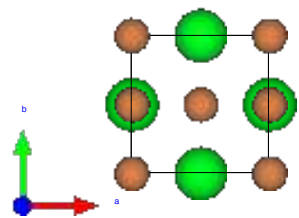

Nanowire Transmission:

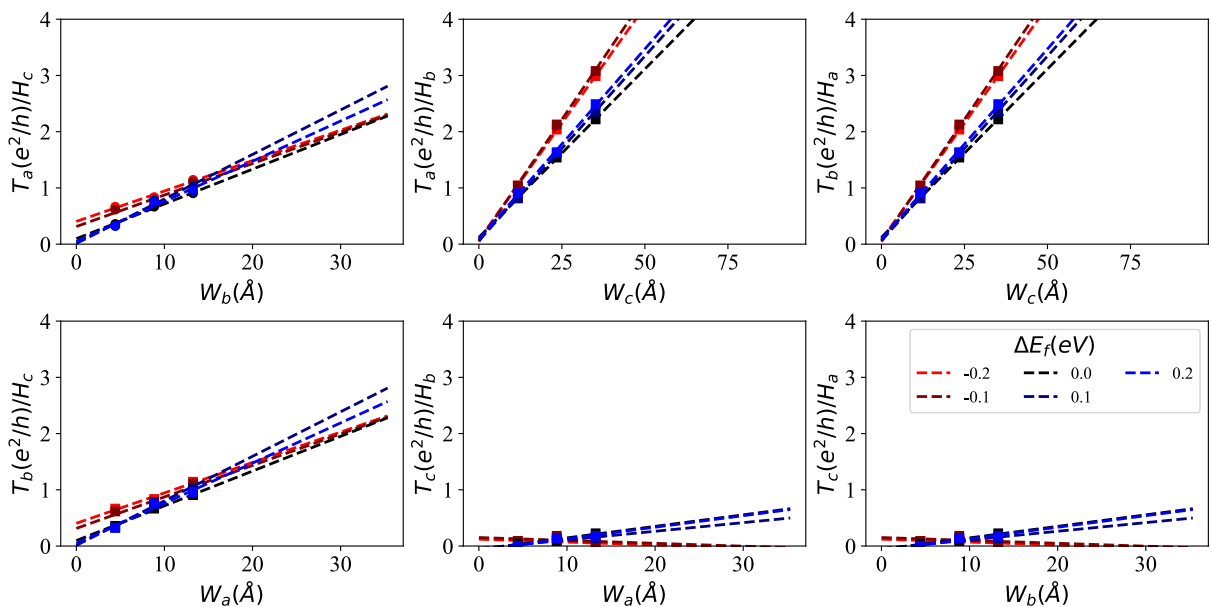

Compound: SrSi$_2$
Materials Project ID: 1727

Lattice (conventional cell):

| Parameter | Value | Unit |
|-----------|-------|------|
| a | 4.4377 | Å |
| b | 4.4377 | Å |
| c | 13.9661 | Å |
| α (alpha) | 90.0000 | ° |
| β (beta) | 90.0000 | ° |
| γ (gamma) | 90.0000 | ° |

Crystal structure (conventional cell):

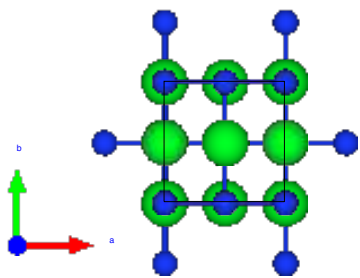

Nanowire Transmission:

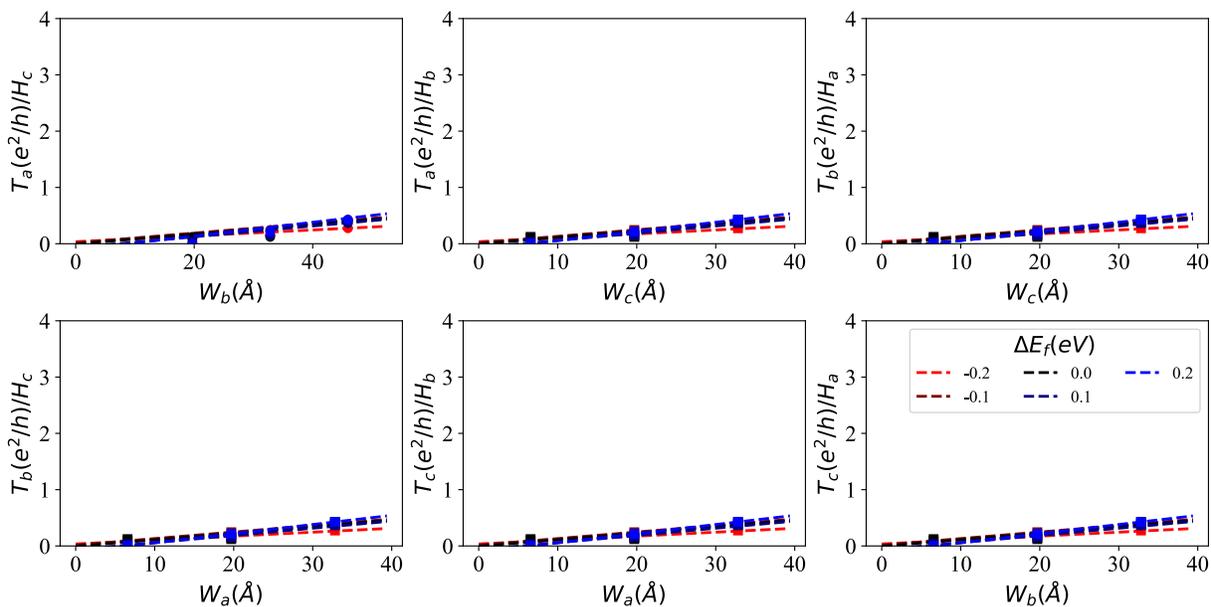

Compound: SrSnHg
Materials Project ID: 1019259

### Lattice (conventional cell):

| Parameter | Value | Unit |
|---|---|---|
| a | 5.0060 | Å |
| b | 5.0060 | Å |
| c | 8.3087 | Å |
| α (alpha) | 90.0000 | ° |
| β (beta) | 90.0000 | ° |
| γ (gamma) | 120.0000 | ° |

### Crystal structure (conventional cell):

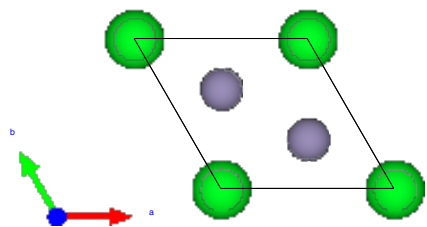

### Nanowire Transmission:

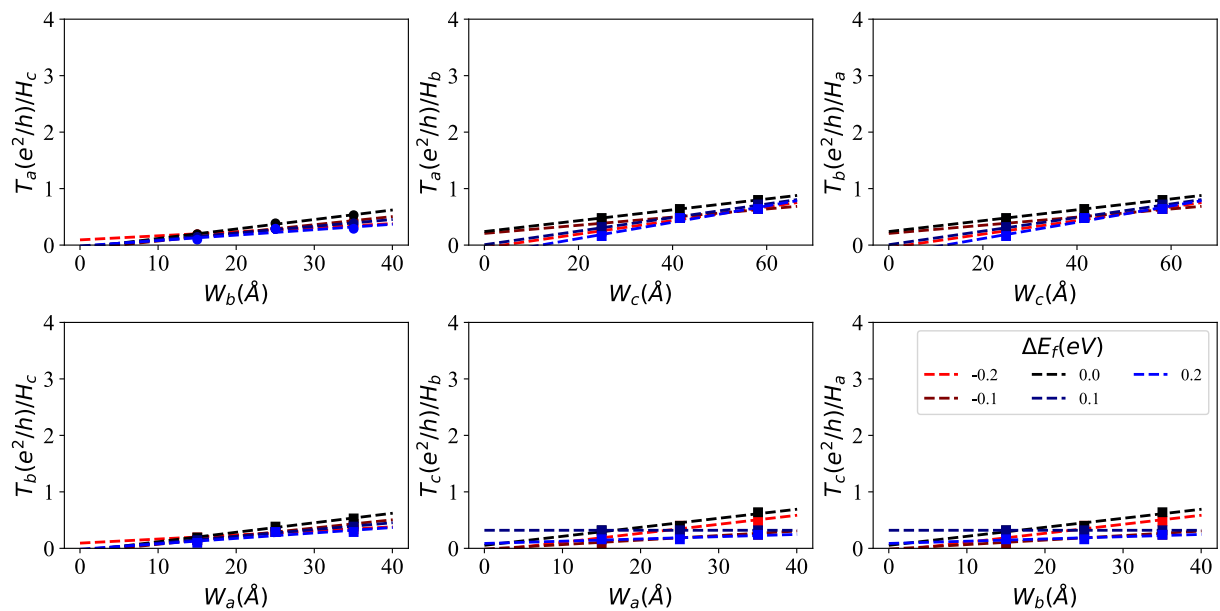

Compound: TaAs
Materials Project ID: 1936

Lattice (conventional cell):

| Parameter | Value | Unit |
|-----------|-------|------|
| a | 3.4696 | Å |
| b | 3.4696 | Å |
| c | 11.7349 | Å |
| $\alpha$ (alpha) | 90.0000 | ° |
| $\beta$ (beta) | 90.0000 | ° |
| $\gamma$ (gamma) | 90.0000 | ° |

Crystal structure (conventional cell):

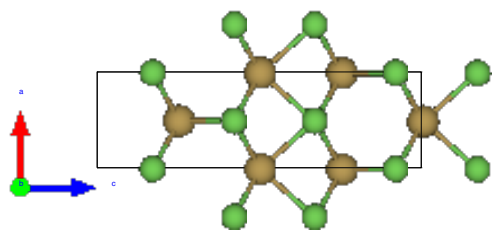

Nanowire Transmission:

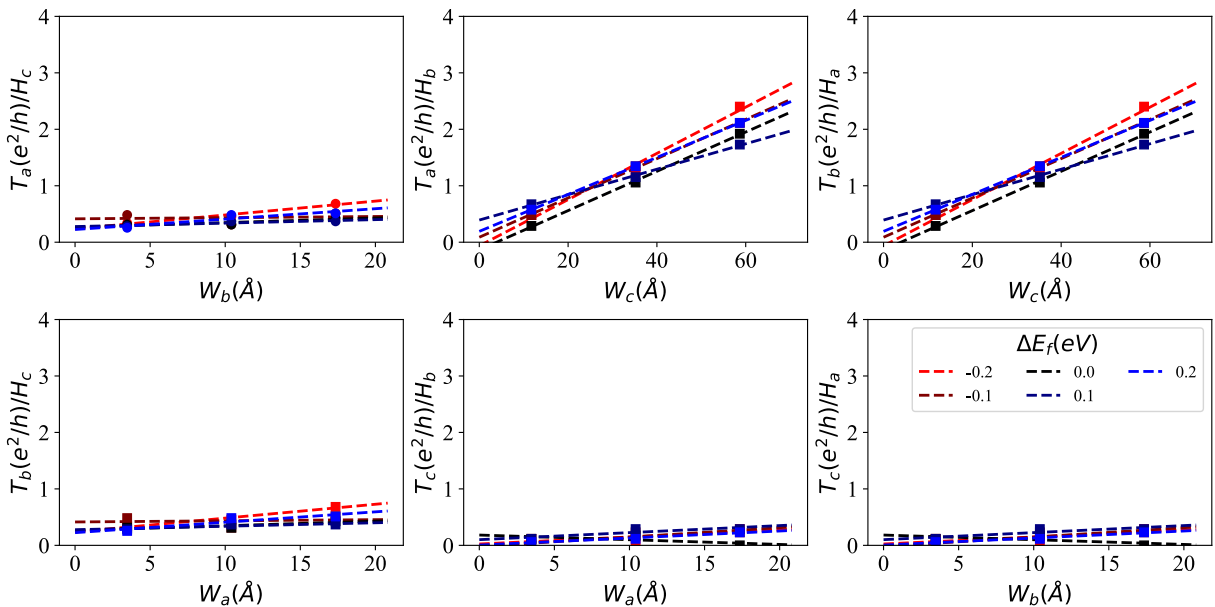

Compound: TaAs$_2$
Materials Project ID: 12561

Lattice (conventional cell):

| Parameter | Value | Unit |
|---|---|---|
| a | 9.4399 | Å |
| b | 3.4259 | Å |
| c | 7.8478 | Å |
| α (alpha) | 90.0000 | ° |
| β (beta) | 119.7016 | ° |
| γ (gamma) | 90.0000 | ° |

Crystal structure (conventional cell):

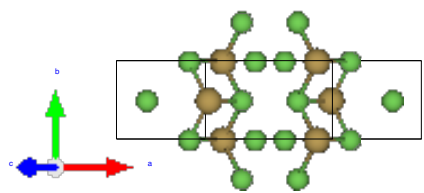

Nanowire Transmission:

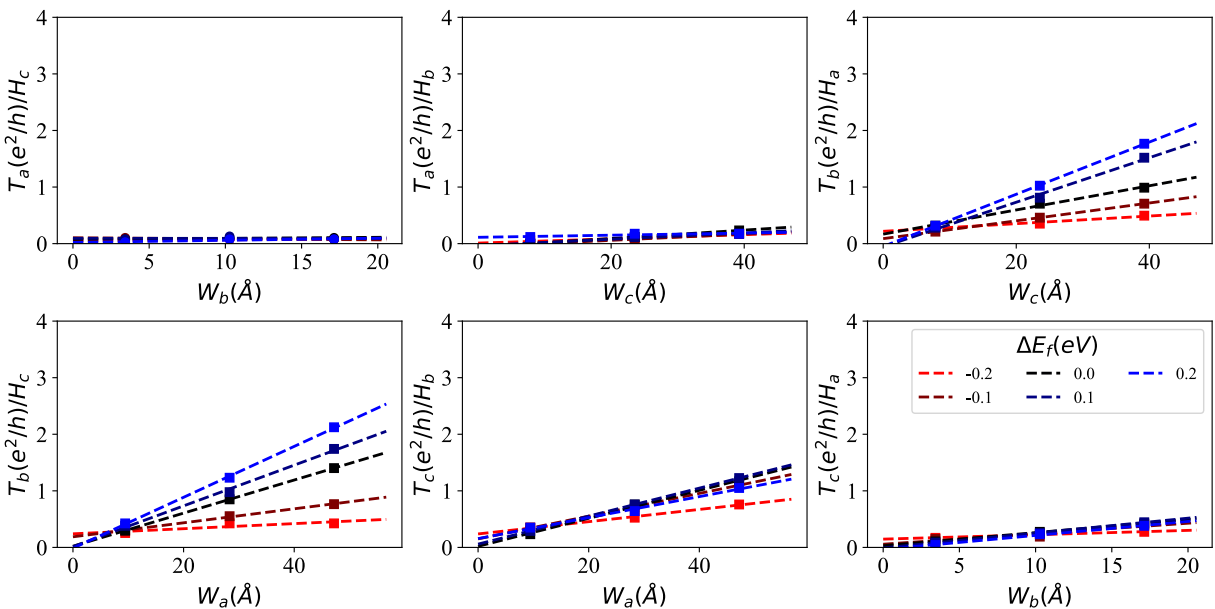

Compound: TaInS$_2$
Materials Project ID: 22332

## Lattice (conventional cell):

| Parameter | Value | Unit |
|-----------|-------|------|
| a | 3.3249 | Å |
| b | 3.3249 | Å |
| c | 8.9364 | Å |
| α (alpha) | 90.0000 | ° |
| β (beta) | 90.0000 | ° |
| γ (gamma) | 120.0000 | ° |

## Crystal structure (conventional cell):

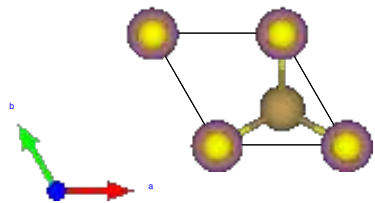

## Nanowire Transmission:

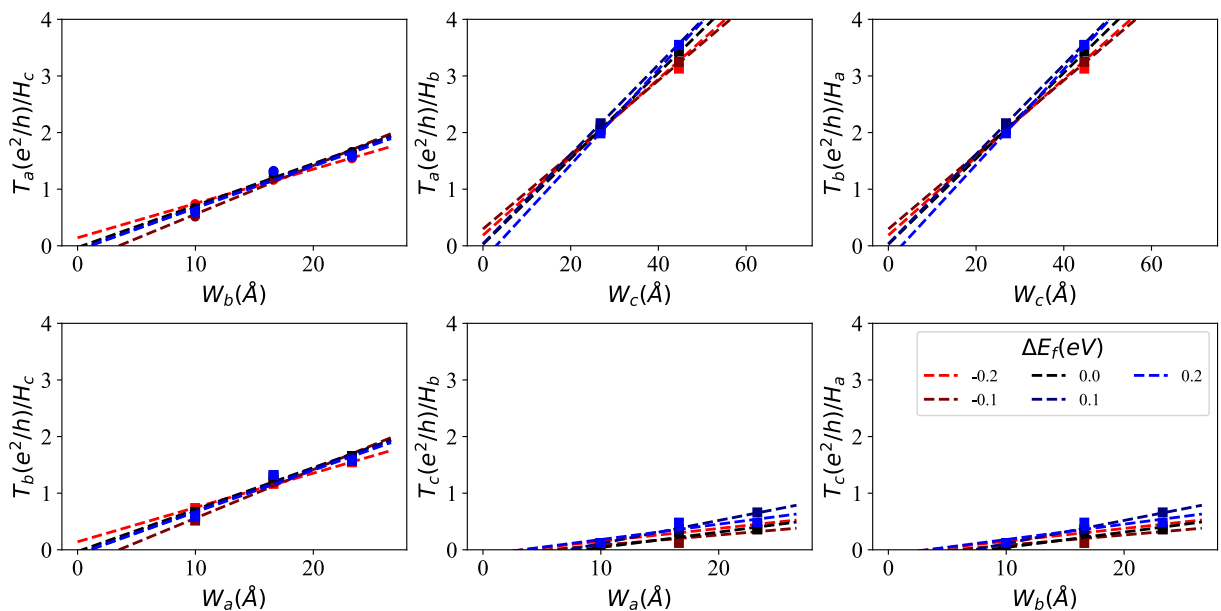

Compound: TaN
Materials Project ID: 1459

Lattice (conventional cell):

| Parameter | Value | Unit |
|-----------|-------|------|
| a | 2.9540 | Å |
| b | 2.9540 | Å |
| c | 2.9015 | Å |
| α (alpha) | 90.0000 | ° |
| β (beta) | 90.0000 | ° |
| γ (gamma) | 120.0000 | ° |

Crystal structure (conventional cell):

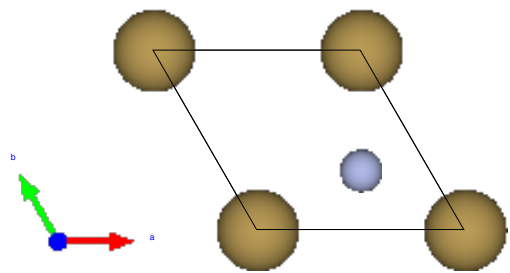

Nanowire Transmission:

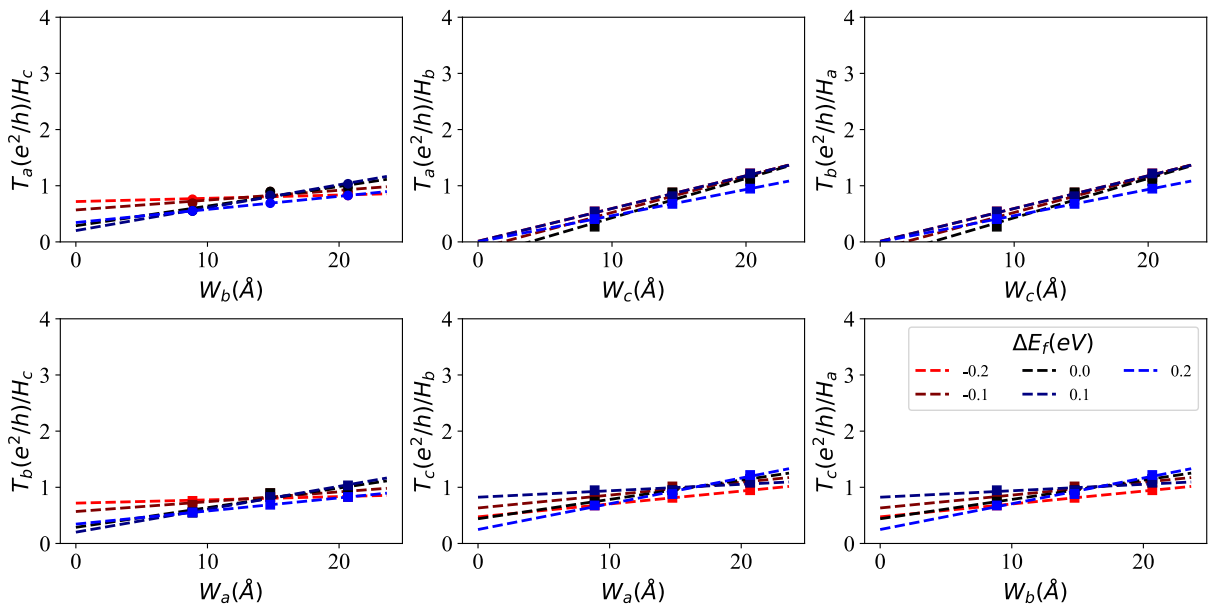

Compound: TaP
Materials Project ID: 1067587

Lattice (conventional cell):

| Parameter | Value | Unit |
|-----------|-------|------|
| a | 3.3402 | Å |
| b | 3.3402 | Å |
| c | 11.4036 | Å |
| α (alpha) | 90.0000 | ° |
| β (beta) | 90.0000 | ° |
| γ (gamma) | 90.0000 | ° |

Crystal structure (conventional cell):

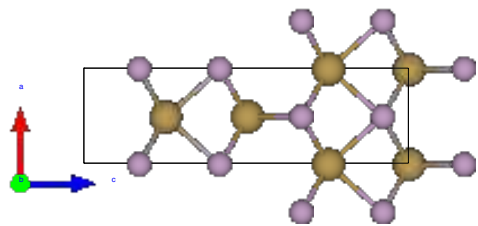

Nanowire Transmission:

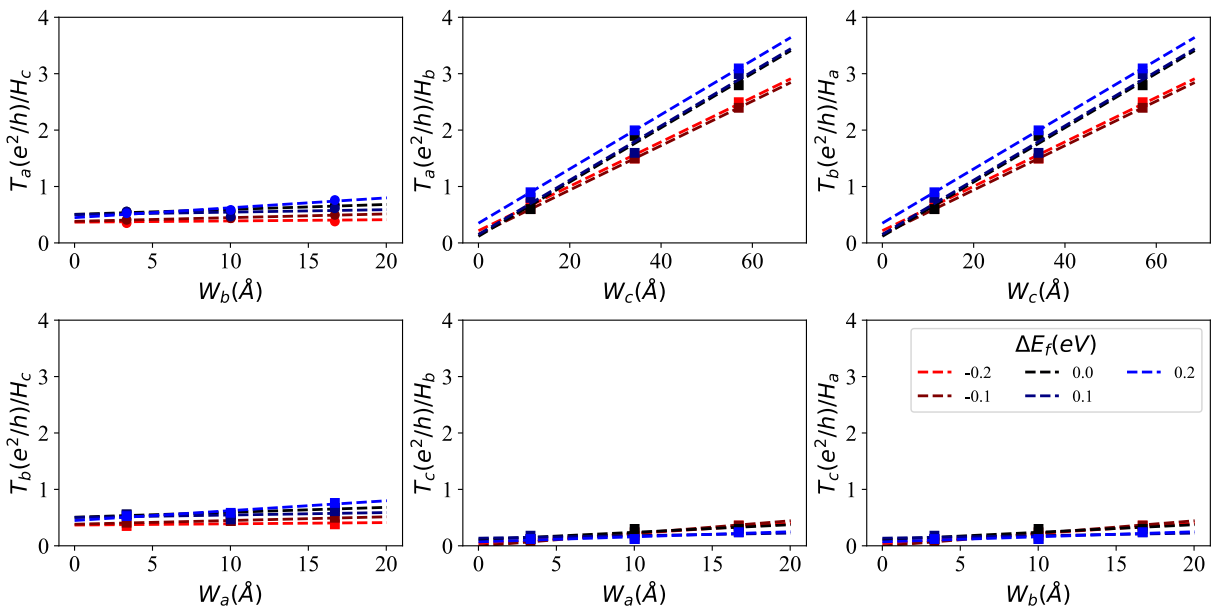

Compound: TaPbSe$_2$
Materials Project ID: 567736

Lattice (conventional cell):

| Parameter | Value | Unit |
|---|---|---|
| a | 3.4835 | Å |
| b | 3.4835 | Å |
| c | 9.4755 | Å |
| α (alpha) | 90.0000 | ° |
| β (beta) | 90.0000 | ° |
| γ (gamma) | 120.0000 | ° |

Crystal structure (conventional cell):

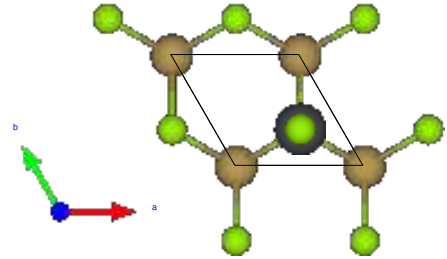

Nanowire Transmission:

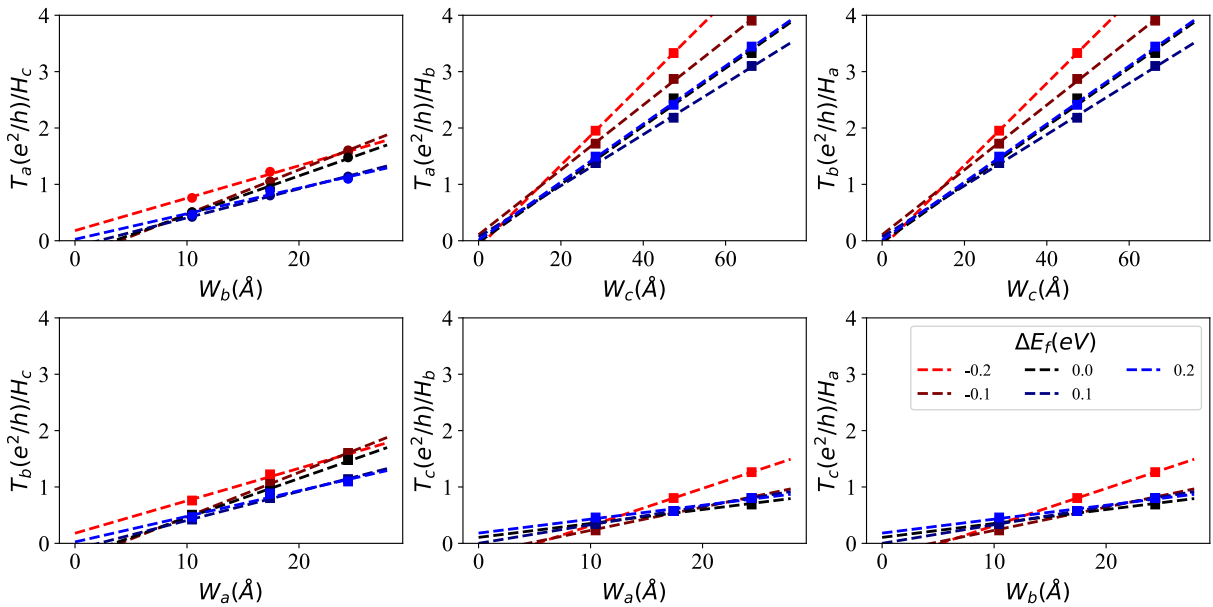

Compound: TaSiAs
Materials Project ID: 81

Lattice (conventional cell):

| Parameter | Value | Unit |
|---|---|---|
| a | 3.5206 | Å |
| b | 3.5206 | Å |
| c | 7.9142 | Å |
| α (alpha) | 90.0000 | ° |
| β (beta) | 90.0000 | ° |
| γ (gamma) | 90.0000 | ° |

Crystal structure (conventional cell):

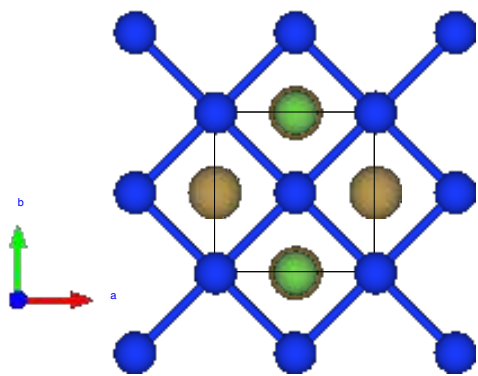

Nanowire Transmission:

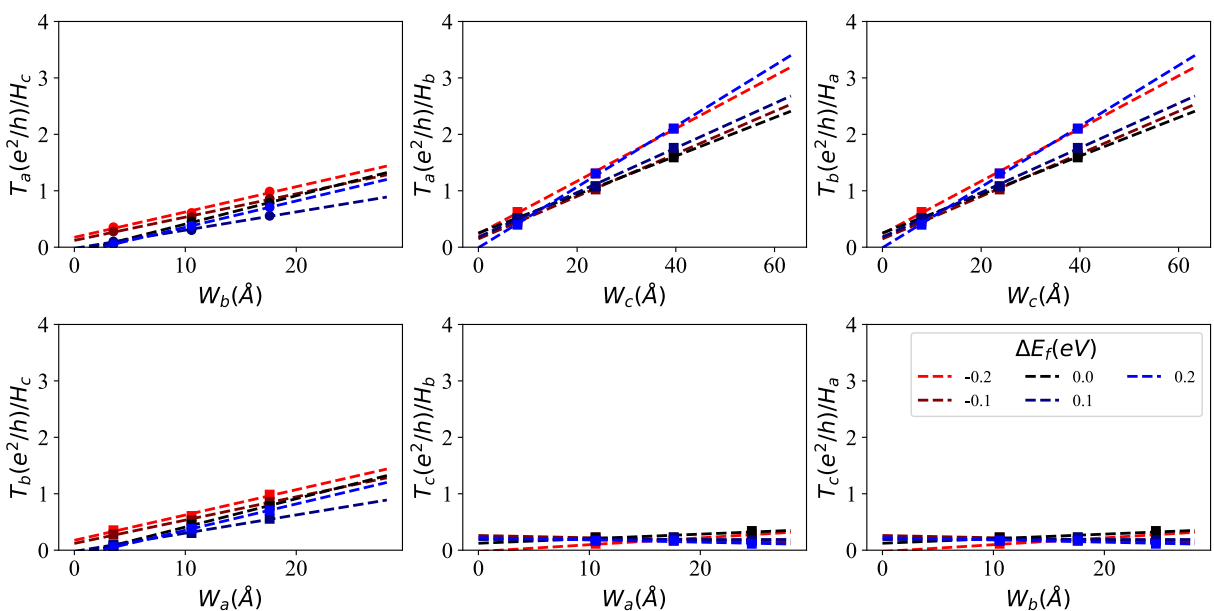

Compound: Te
Materials Project ID: 19

Lattice (conventional cell):

| Parameter | Value | Unit |
|-----------|-------|------|
| a | 4.5124 | Å |
| b | 4.5124 | Å |
| c | 5.9599 | Å |
| α (alpha) | 90.0000 | ° |
| β (beta) | 90.0000 | ° |
| γ (gamma) | 120.0000 | ° |

Crystal structure (conventional cell):

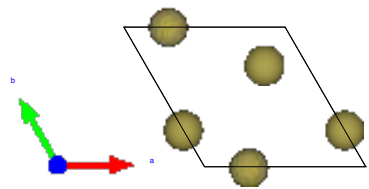

Nanowire Transmission:

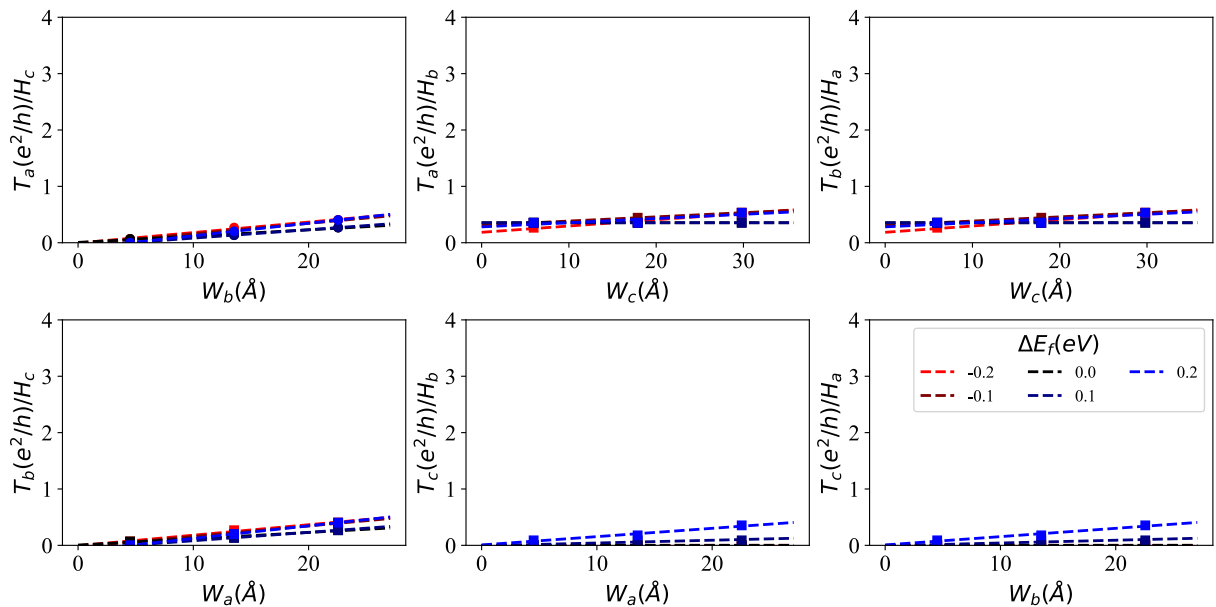

Compound: Te₂Ir

Materials Project ID: 2285

Lattice (conventional cell):

| Parameter | Value | Unit |
|-----------|-------|------|
| a | 3.9878 | Å |
| b | 3.9878 | Å |
| c | 5.5433 | Å |
| α (alpha) | 90.0000 | ° |
| β (beta) | 90.0000 | ° |
| γ (gamma) | 120.0000 | ° |

Crystal structure (conventional cell):

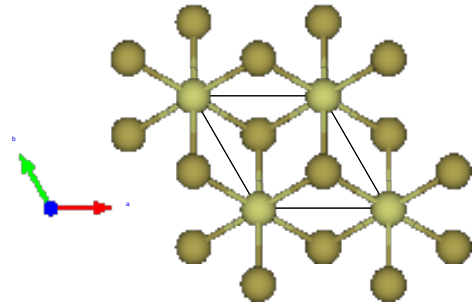

Nanowire Transmission:

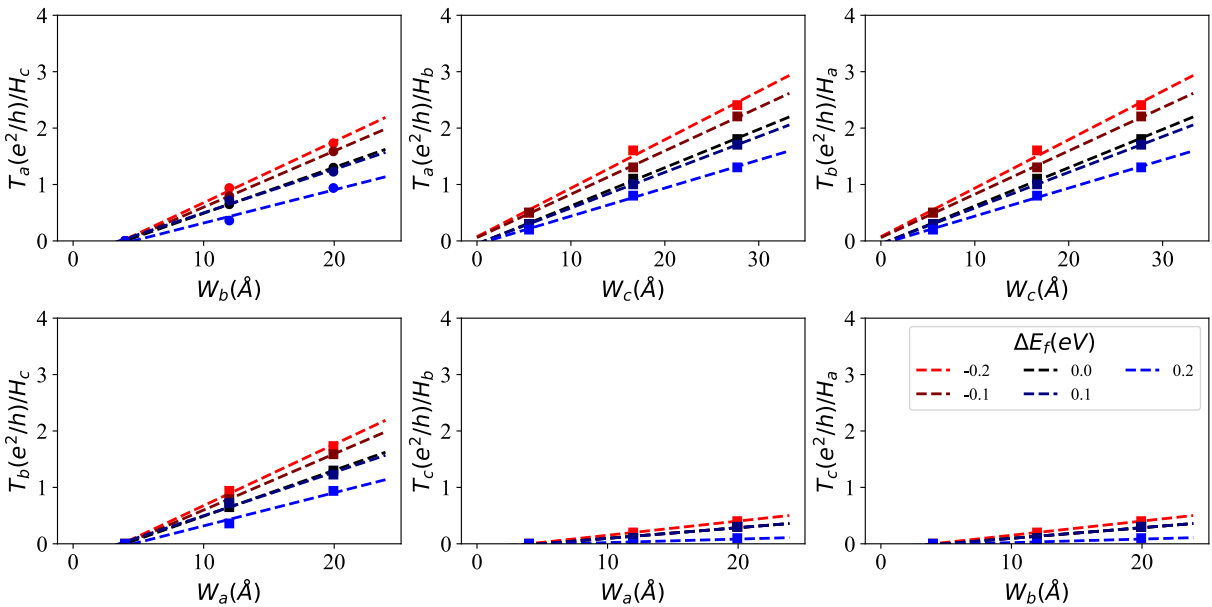

Compound: Te$_2$Mo
Materials Project ID: 7459

## Lattice (conventional cell):

| Parameter | Value | Unit |
|---|---|---|
| a | 6.3662 | Å |
| b | 3.4869 | Å |
| c | 15.5479 | Å |
| α (alpha) | 90.0000 | ° |
| β (beta) | 92.3879 | ° |
| γ (gamma) | 90.0000 | ° |

## Crystal structure (conventional cell):

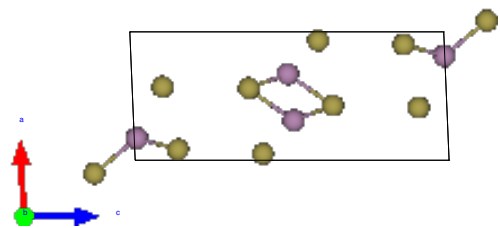

## Nanowire Transmission:

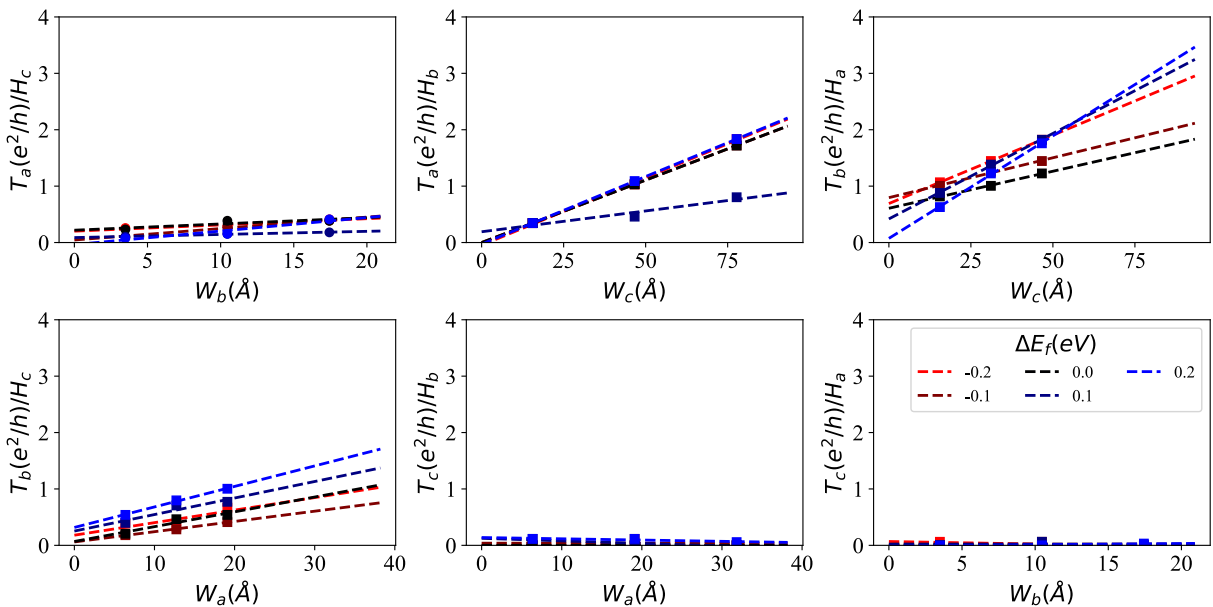

Compound: Te$_2$W
Materials Project ID:22693

Lattice (conventional cell):

| Parameter | Value | Unit |
|-----------|-------|------|
| a | 3.4979 | Å |
| b | 6.3383 | Å |
| c | 15.4319 | Å |
| α (alpha) | 90.0000 | ° |
| β (beta) | 90.0000 | ° |
| γ (gamma) | 90.0000 | ° |

Crystal structure (conventional cell):

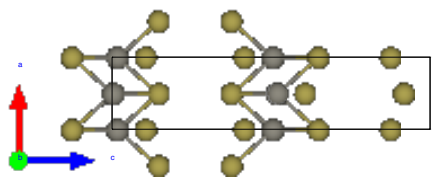

Nanowire Transmission:

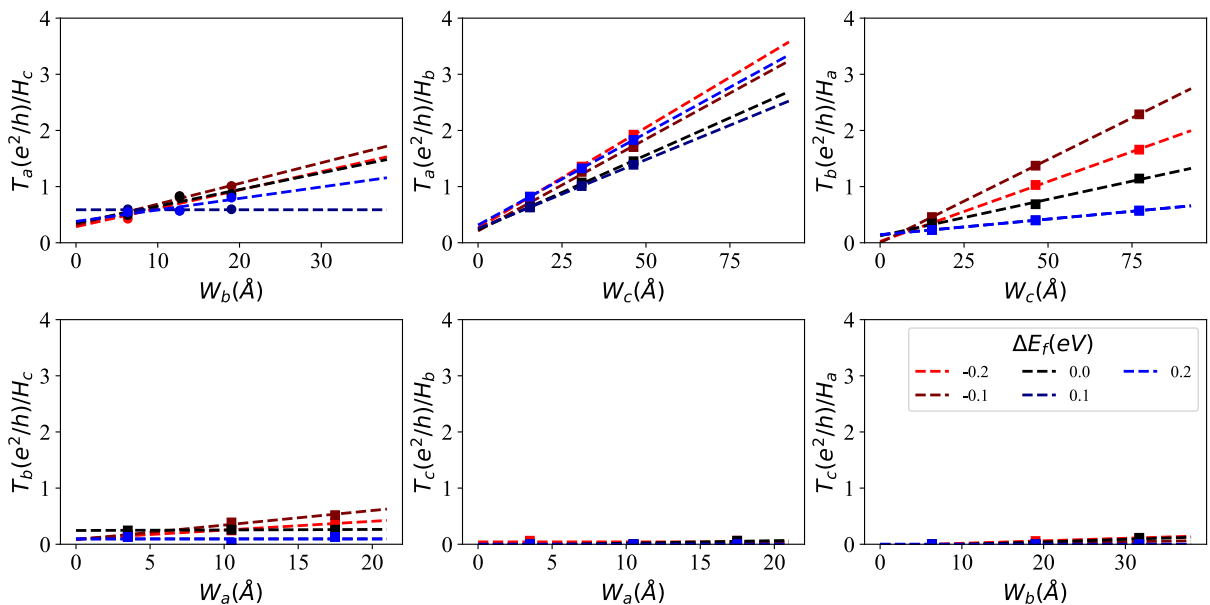

Compound: TiB$_2$
Materials Project ID: 1145

Lattice (conventional cell):

| Parameter | Value | Unit |
|-----------|---------|------|
| a | 3.0351 | Å |
| b | 3.0351 | Å |
| c | 3.2232 | Å |
| α (alpha) | 90.0000 | ° |
| β (beta) | 90.0000 | ° |
| γ (gamma) | 120.0000 | ° |

Crystal structure (conventional cell):

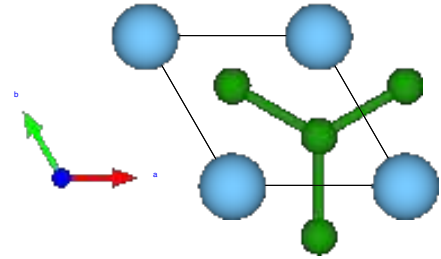

Nanowire Transmission:

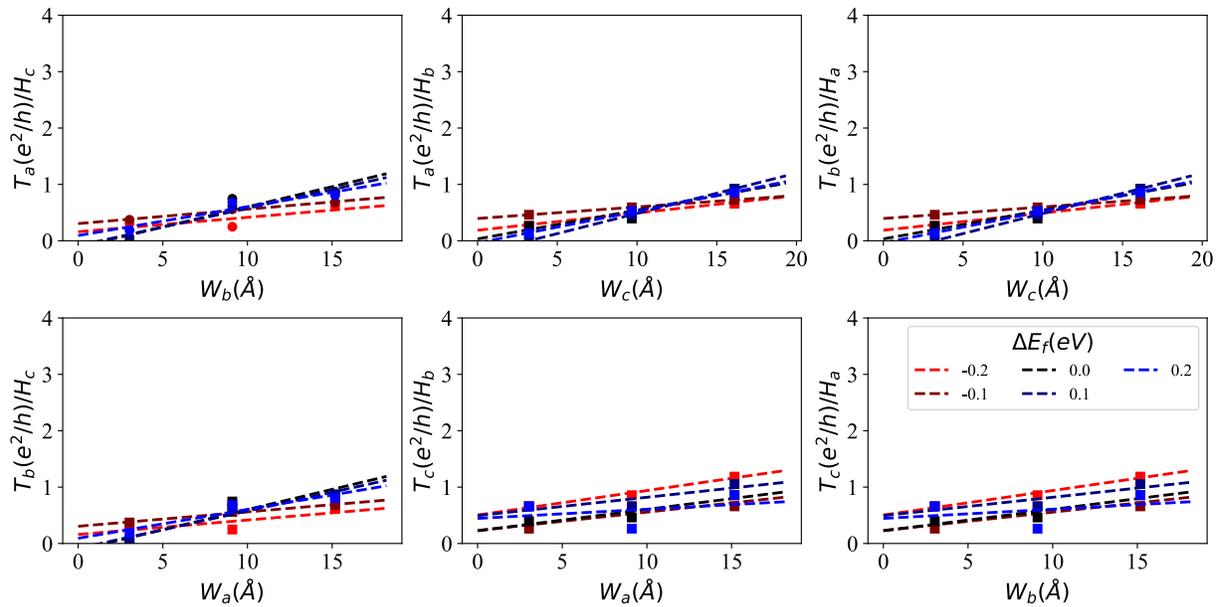

Compound: TiCu$_2$
Materials Project ID: 1077023

Lattice (conventional cell):

| Parameter | Value | Unit |
|-----------|-------|------|
| a | 4.5123 | Å |
| b | 7.9862 | Å |
| c | 4.4641 | Å |
| α (alpha) | 90.0000 | ° |
| β (beta) | 90.0000 | ° |
| γ (gamma) | 90.0000 | ° |

Crystal structure (conventional cell):

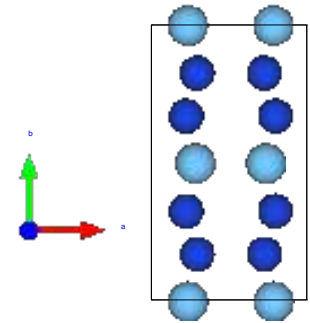

Nanowire Transmission:

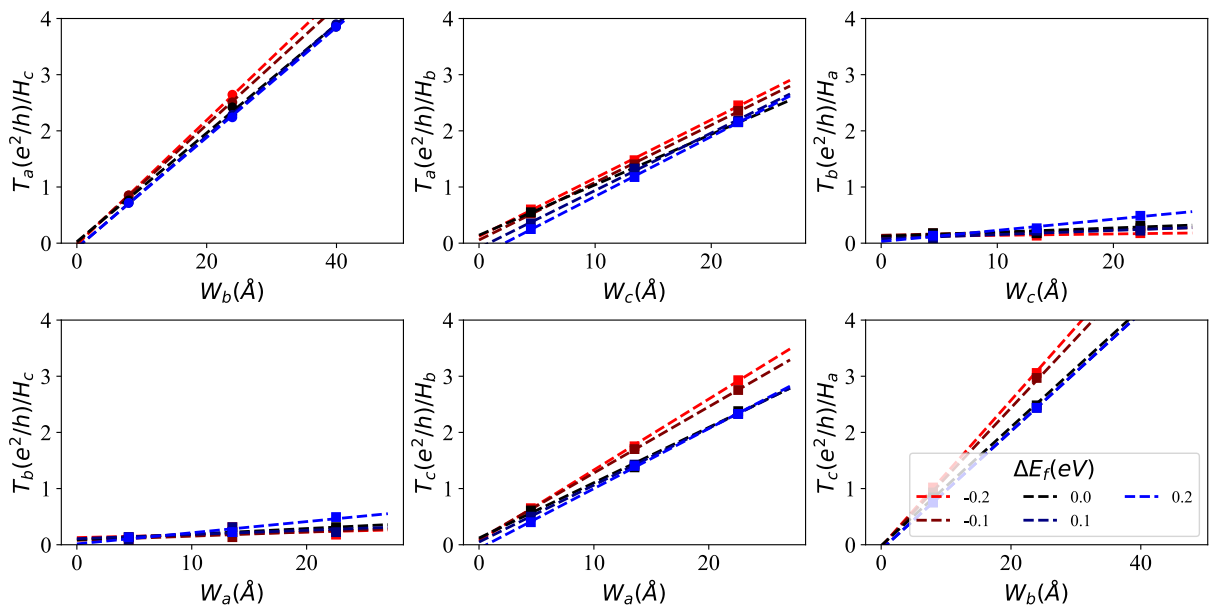

Compound: TiS
Materials Project ID: 1018028

Lattice (conventional cell):

| Parameter | Value | Unit |
|-----------|---------|------|
| a | 3.2739 | Å |
| b | 3.2739 | Å |
| c | 3.2191 | Å |
| α (alpha) | 90.0000 | ° |
| β (beta) | 90.0000 | ° |
| γ (gamma) | 120.0000 | ° |

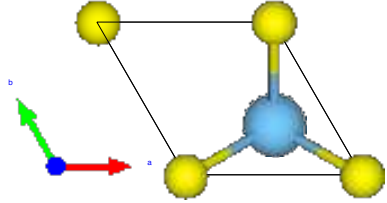

Nanowire Transmission:

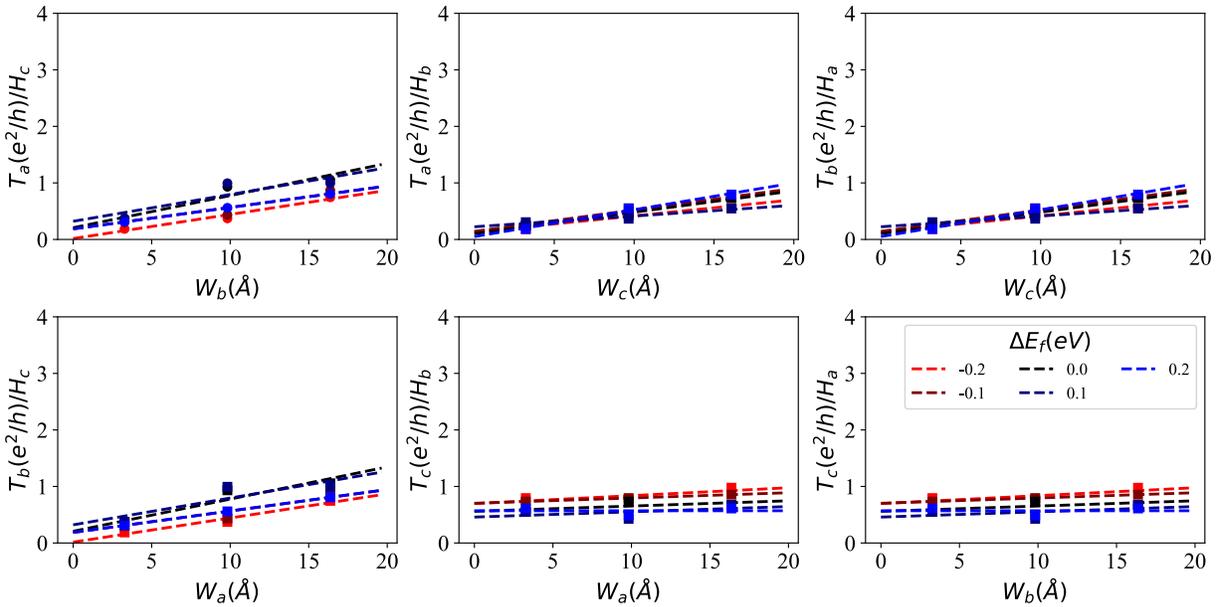

Compound: WC
Materials Project ID: 1894

Lattice (conventional cell):

| Parameter | Value | Unit |
|---|---|---|
| a | 2.9283 | Å |
| b | 2.9283 | Å |
| c | 2.8529 | Å |
| α (alpha) | 90.0000 | ° |
| β (beta) | 90.0000 | ° |
| γ (gamma) | 120.0000 | ° |

Crystal structure (conventional cell):

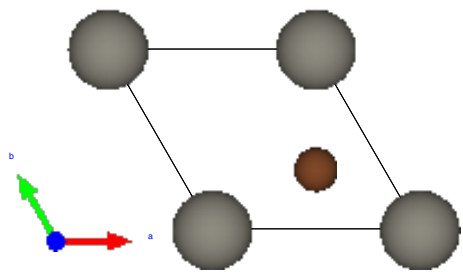

Nanowire Transmission:

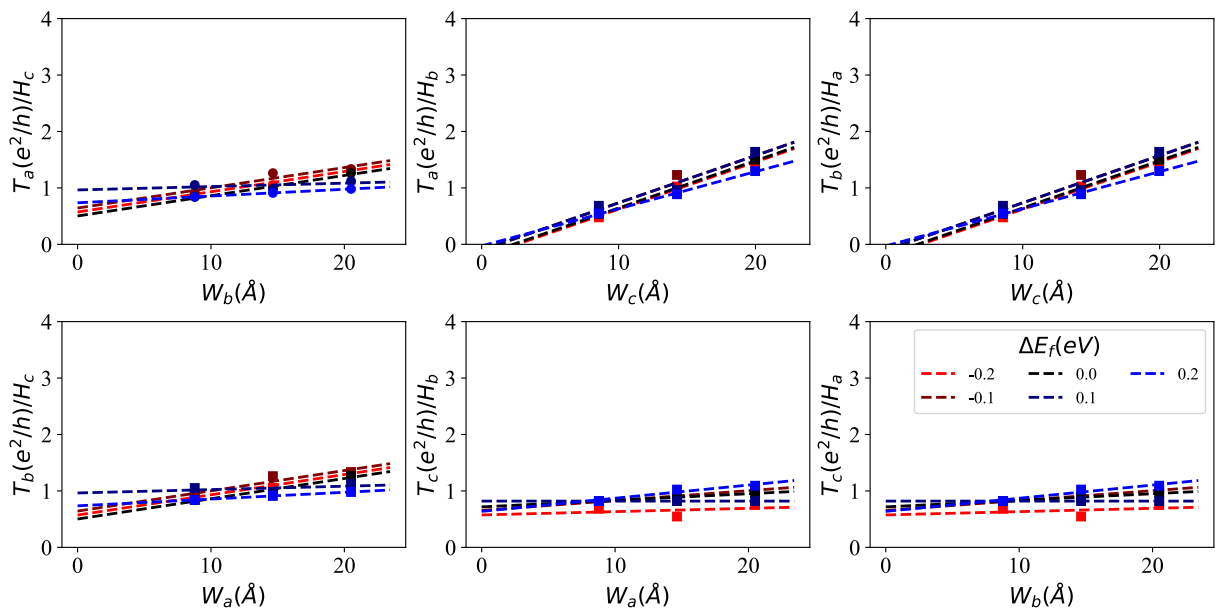

Compound: WN
Materials Project ID: 991

Lattice (conventional cell):

| Parameter | Value | Unit |
|-----------|---------|------|
| a | 2.8760 | Å |
| b | 2.8760 | Å |
| c | 2.9080 | Å |
| α (alpha) | 90.0000 | ° |
| β (beta) | 90.0000 | ° |
| γ (gamma) | 120.0000 | ° |

Crystal structure (conventional cell):

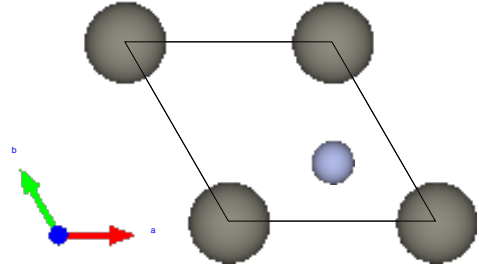

Nanowire Transmission:

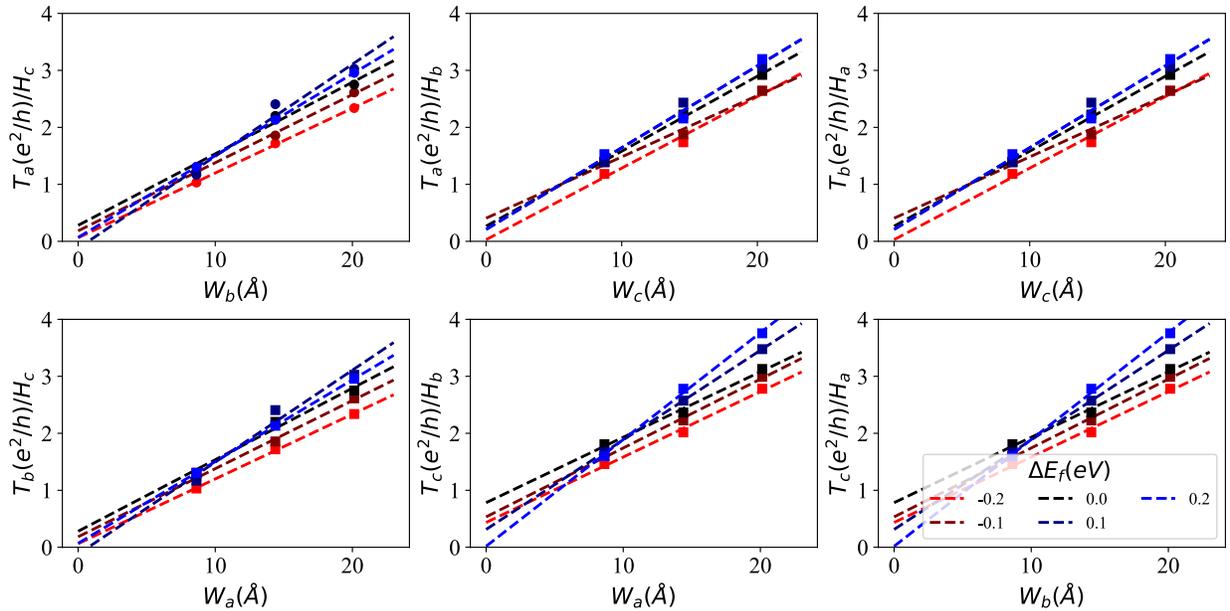

Compound: YAgPb

Materials Project ID: 21505

Lattice (conventional cell):

| Parameter | Value | Unit |
|-----------|-------|------|
| a | 7.5818 | Å |
| b | 7.5818 | Å |
| c | 4.5605 | Å |
| α (alpha) | 90.0000 | ° |
| β (beta) | 90.0000 | ° |
| γ (gamma) | 120.0000 | ° |

Crystal structure (conventional cell):

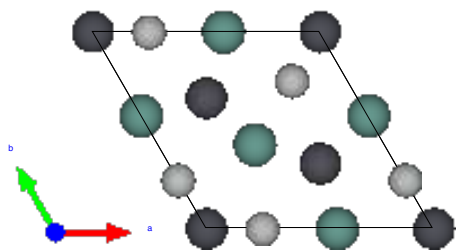

Nanowire Transmission:

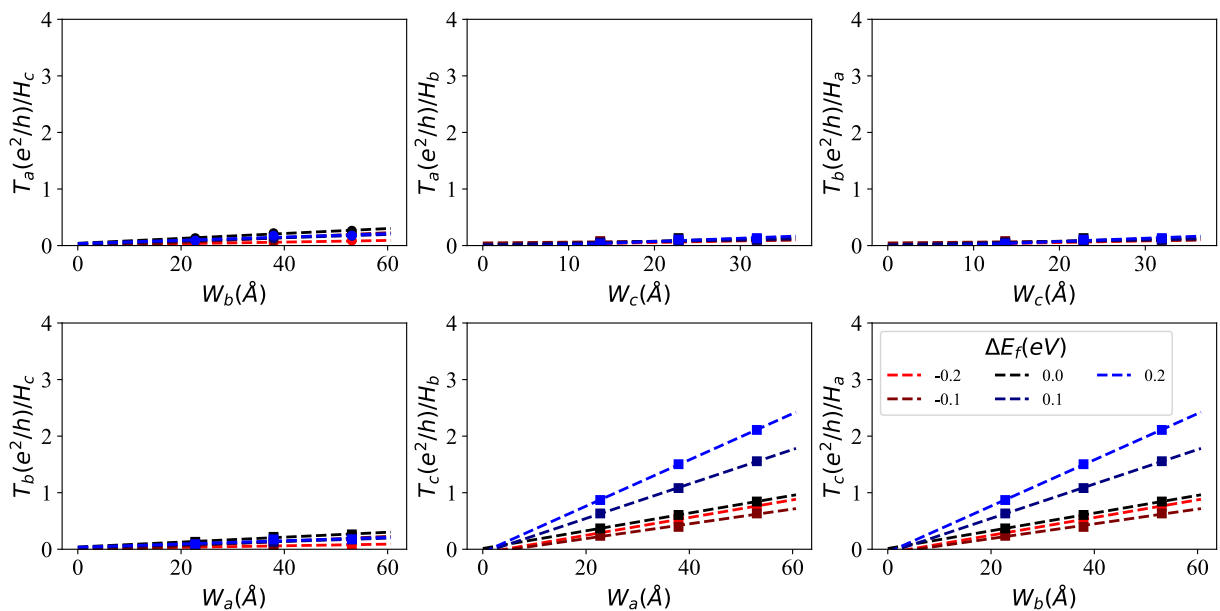

Compound: ZrB$_2$
Materials Project ID: 1472

Lattice (conventional cell):

| Parameter | Value | Unit |
|---|---|---|
| a | 3.1805 | Å |
| b | 3.1805 | Å |
| c | 3.5455 | Å |
| α (alpha) | 90.0000 | ° |
| β (beta) | 90.0000 | ° |
| γ (gamma) | 120.0000 | ° |

Crystal structure (conventional cell):

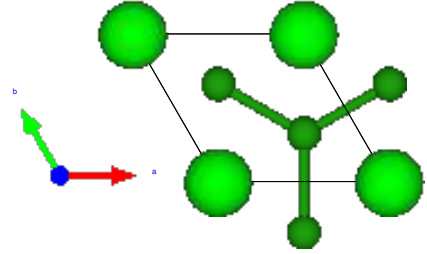

Nanowire Transmission:

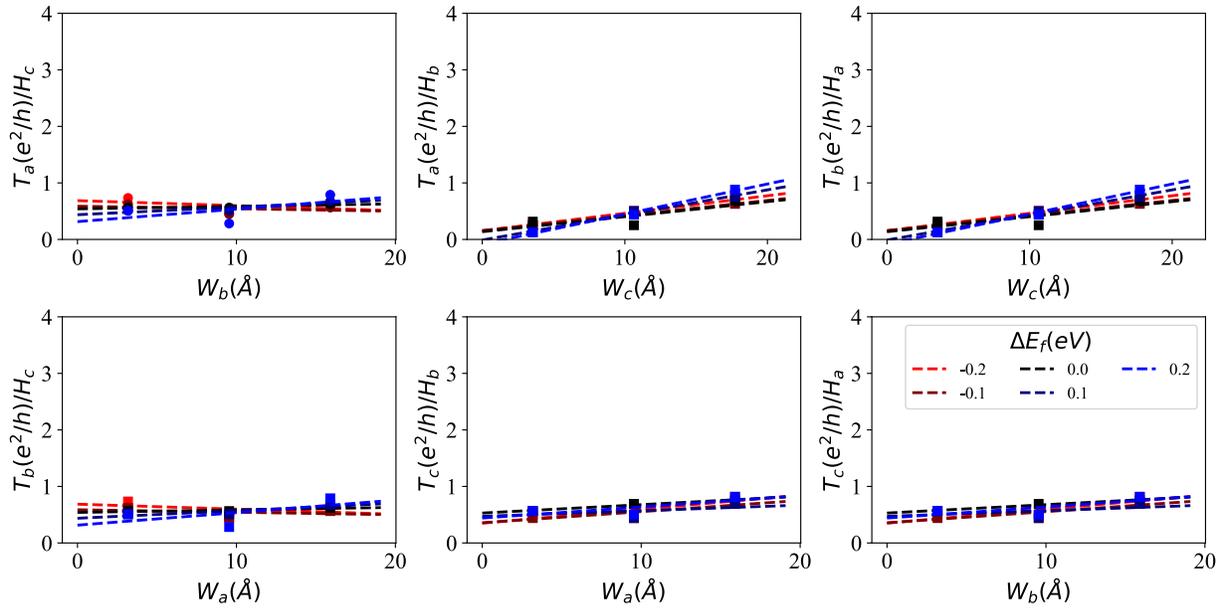

Compound: ZrGePt
Materials Project ID: 1095610

Lattice (conventional cell):

| Parameter | Value | Unit |
|-----------|-------|------|
| a | 4.0158 | Å |
| b | 6.7301 | Å |
| c | 7.7560 | Å |
| $\alpha$ (alpha) | 90.0000 | ° |
| $\beta$ (beta) | 90.0000 | ° |
| $\gamma$ (gamma) | 90.0000 | ° |

Crystal structure (conventional cell):

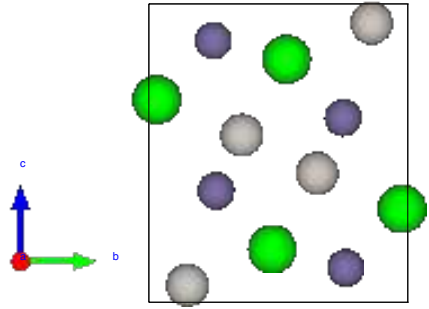

Nanowire Transmission:

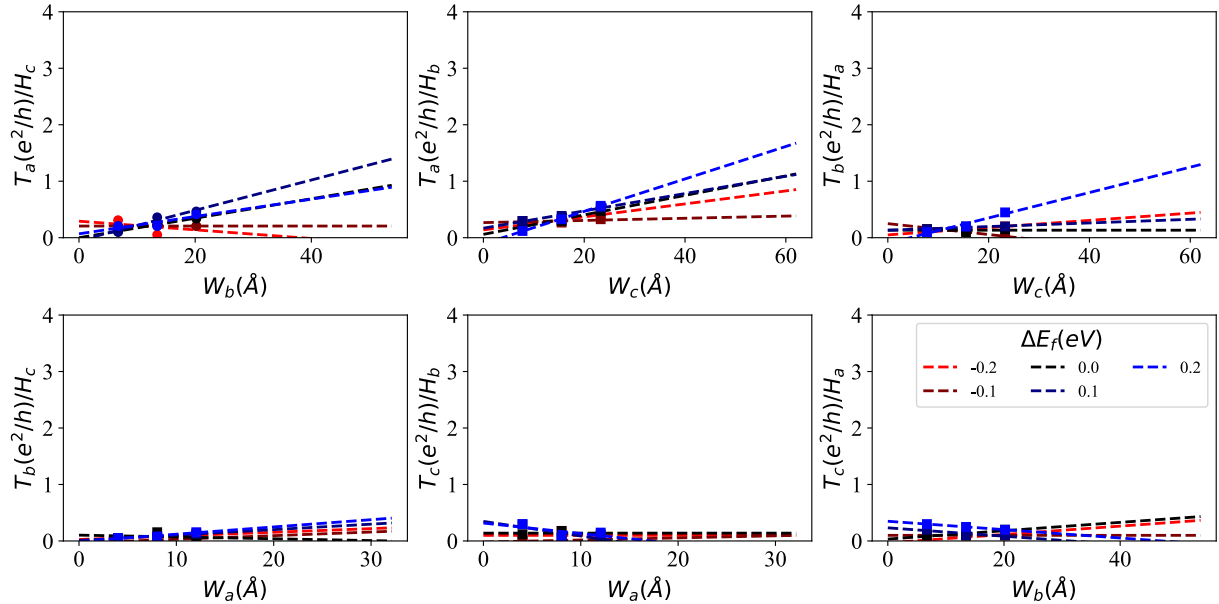

Compound: ZrSiPt
Materials Project ID: 972187

Lattice (conventional cell):

| Parameter | Value | Unit |
|-----------|-------|------|
| a | 3.9323 | Å |
| b | 6.6654 | Å |
| c | 7.6233 | Å |
| α (alpha) | 90.0000 | ° |
| β (beta) | 90.0000 | ° |
| γ (gamma) | 90.0000 | ° |

Crystal structure (conventional cell):

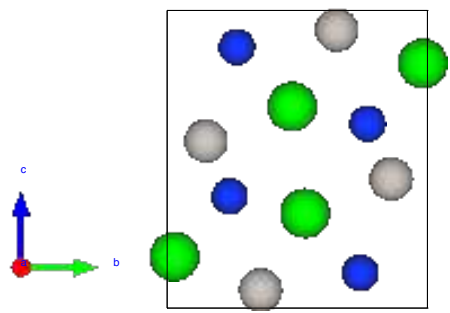

Nanowire Transmission:

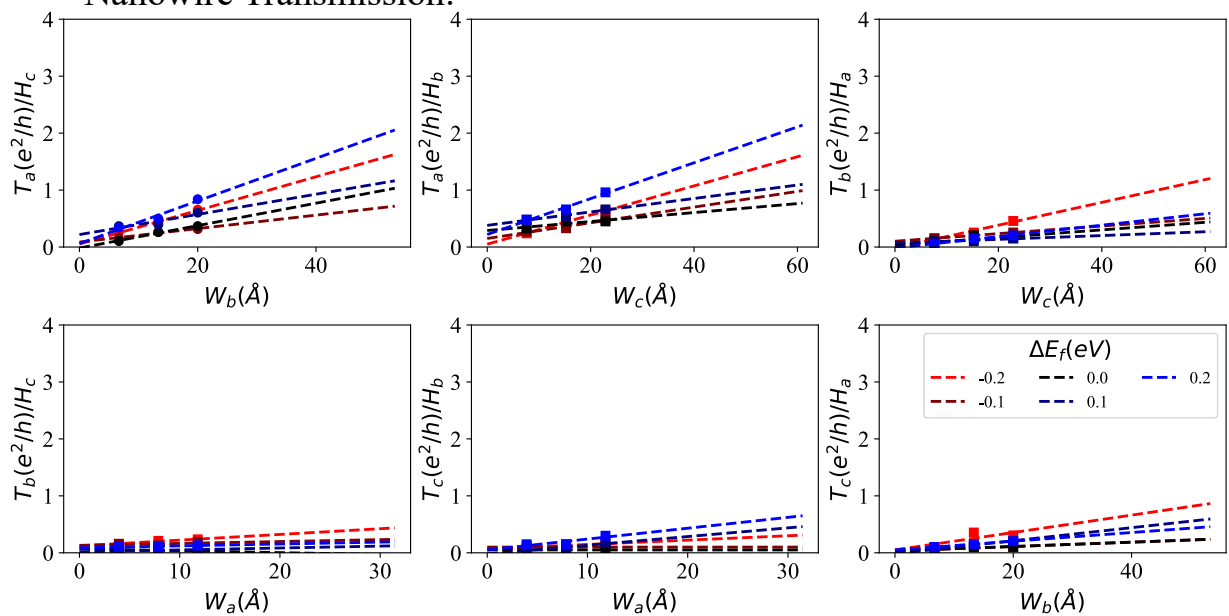

Compound: ZrSiS
Materials Project ID: 3938

## Lattice (conventional cell):

| Parameter | Value | Unit |
|-----------|-------|------|
| a | 3.6168 | Å |
| b | 3.6168 | Å |
| c | 8.6761 | Å |
| α (alpha) | 90.0000 | ° |
| β (beta) | 90.0000 | ° |
| γ (gamma) | 90.0000 | ° |

## Crystal structure (conventional cell):

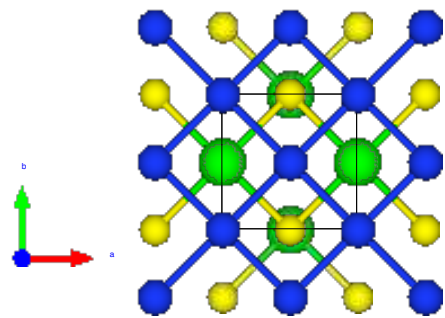

## Nanowire Transmission:

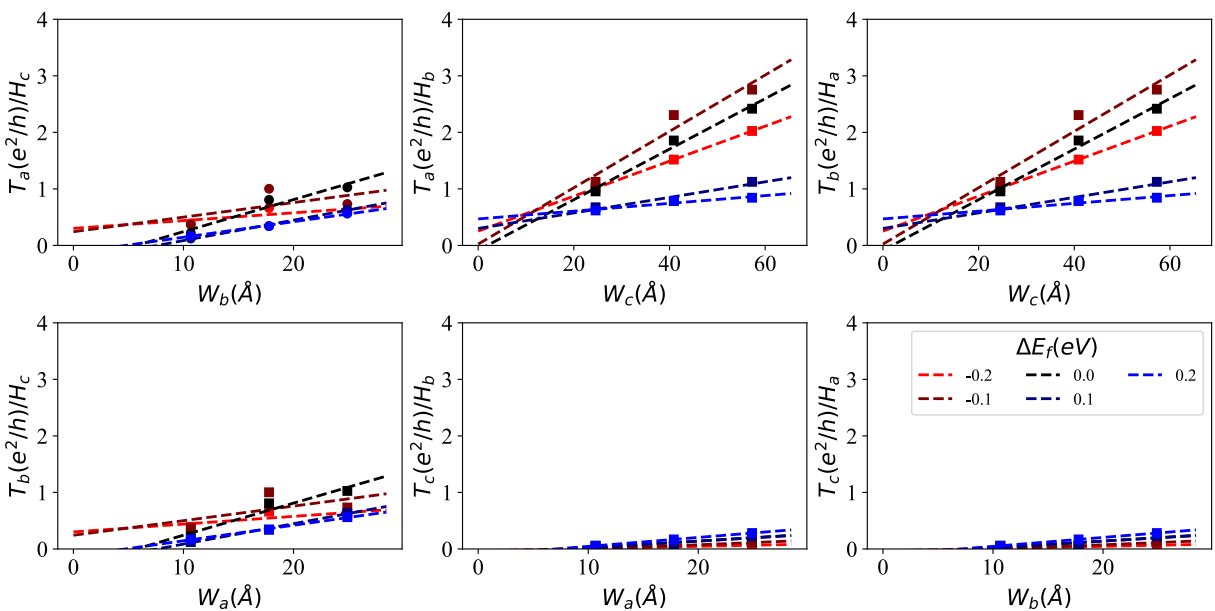

Compound: ZrSiSe

Materials Project ID: 81

**Lattice (conventional cell):**

| Parameter | Value | Unit |
|---|---|---|
| a | 3.6168 | Å |
| b | 3.6168 | Å |
| c | 8.6761 | Å |
| α (alpha) | 90.0000 | ° |
| β (beta) | 90.0000 | ° |
| γ (gamma) | 90.0000 | ° |

**Crystal structure (conventional cell):**

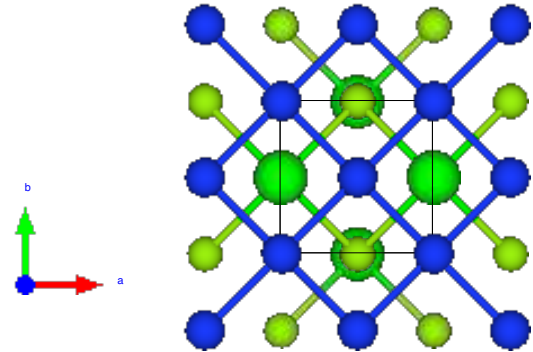

**Nanowire Transmission:**

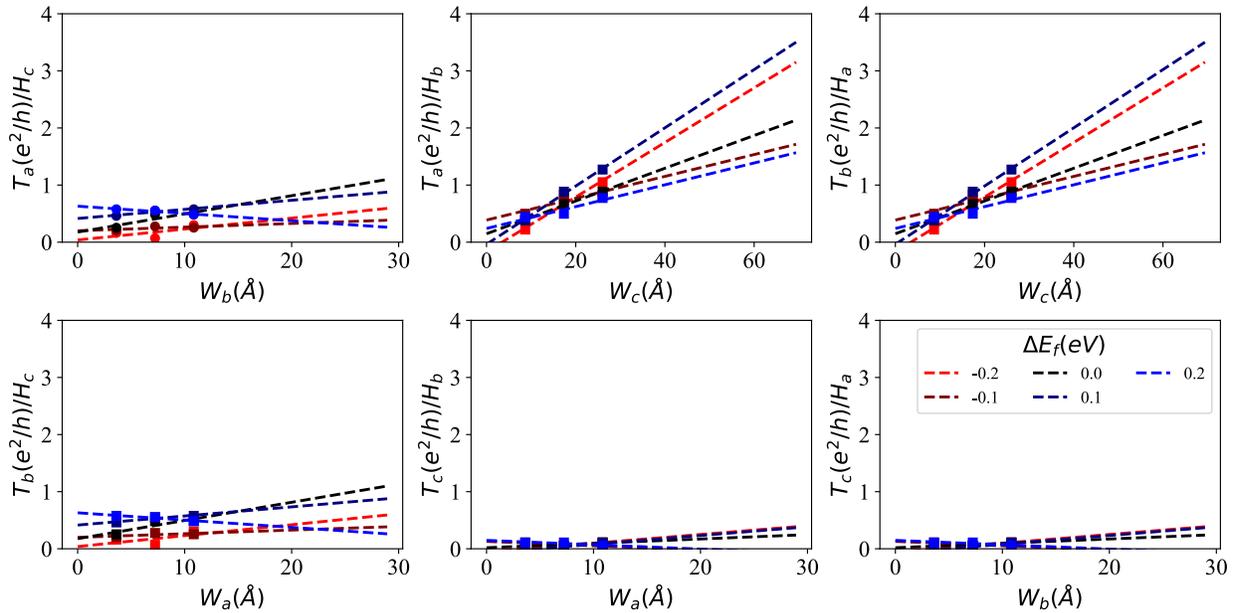

Compound: ZrTe
Materials Project ID: 1539

Lattice (conventional cell):

| Parameter | Value | Unit |
|-----------|-------|------|
| a | 3.7994 | Å |
| b | 3.7994 | Å |
| c | 3.8986 | Å |
| $\alpha$ (alpha) | 90.0000 | ° |
| $\beta$ (beta) | 90.0000 | ° |
| $\gamma$ (gamma) | 120.0000 | ° |

Crystal structure (conventional cell):

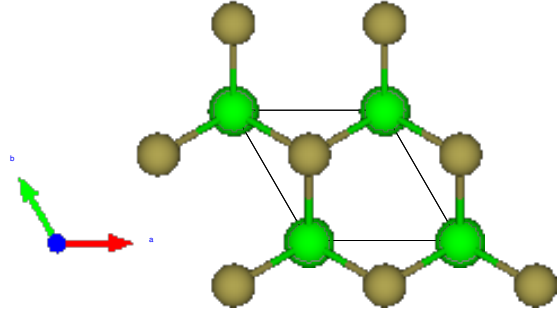

Nanowire Transmission:

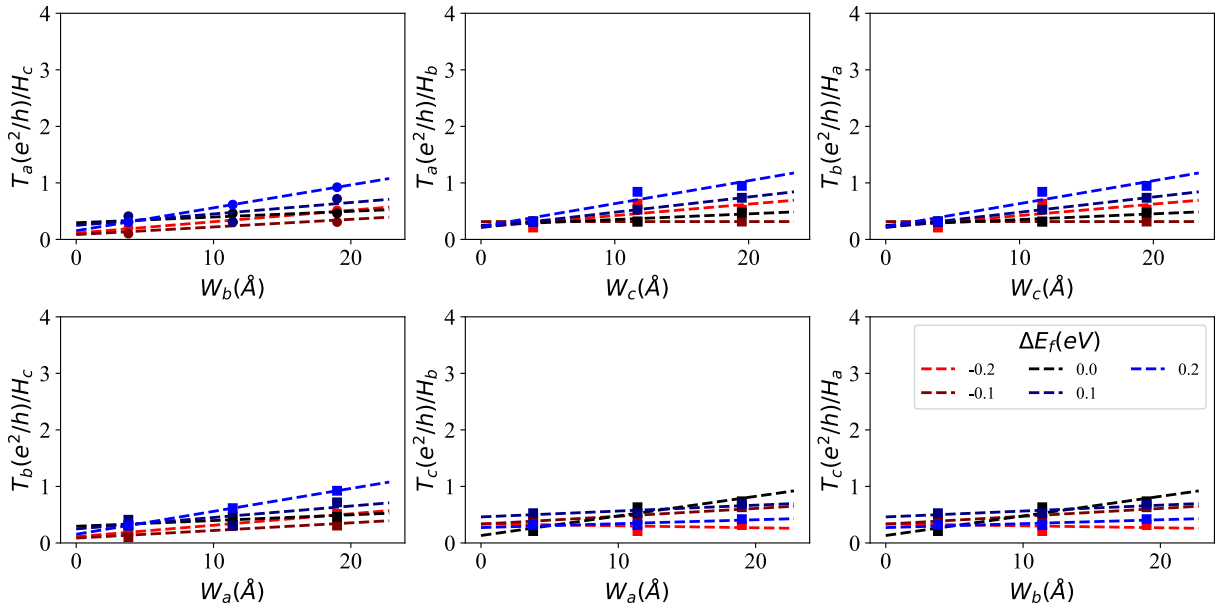



\bibliography{Ref.bib}